\documentclass[aps,rmp,groupedaddress,twocolumn]{revtex4_modcite}

\usepackage[dvipsnames]{xcolor}
\usepackage[utf8]{inputenc}
\usepackage[normalem]{ulem}  
\usepackage{epsfig}
\usepackage{hyperref}

\definecolor{green}{rgb}{0.2,0.5,0.2}

\newcommand{\midtilde}{\raisebox{-0.25\baselineskip}{\textasciitilde}}

\def\roughly#1{\mathrel{\raise.3ex\hbox{$#1$\kern-.75em%
\lower1ex\hbox{$\sim$}}}}
\def\lsim{\roughly<}
\def\gsim{\roughly>}

\begin{document}

\title{Equations of state for supernov{\ae} and compact stars}

\author{M.~Oertel}
\email[]{micaela.oertel@obspm.fr}
\affiliation{Laboratoire Univers et Th\'eories, CNRS, Observatoire de Paris, PSL Research University,
Universit\'e Paris Diderot, Sorbonne Paris Cit\'e, 5 place Jules Janssen, 92195 Meudon, France}
\author{M.~Hempel}
\email[]{matthias.hempel@unibas.ch}
\affiliation{Departement Physik, Universit\"at Basel, Klingelbergstrasse 82, 4056 Basel, Switzerland}
\author{T.~Kl\"ahn}
\email[]{thomas.klaehn@ift.uni.wroc.pl}
\affiliation{Instytut Fizyki Teoretycznej, Uniwersytet Wroc{\l}awski, pl.~M.~Borna~9,  PL-50-204 Wroc{\l}aw, Poland}
\author{S.~Typel}
\email[]{s.typel@gsi.de}
\affiliation{Institut f\"ur Kernphysik, Technische Universit\"{a}t Darmstadt, Schlossgartenstra\ss{}e 9, 64289 Darmstadt, Germany}
\affiliation{GSI Helmholtzzentrum f\"ur Schwerionenforschung, Planckstra{\ss}e 1, 64291 Darmstadt, Germany}
\date{\today}

\begin{abstract}
We review various theoretical approaches for the 
equation of state (EoS) of dense matter, relevant for the description
of core-collapse supernovae, compact stars and compact star
mergers. The emphasis is put on models that are applicable to all of these
scenarios.  Such EoS models have to cover large ranges in baryon
number density, temperature and isospin asymmetry. The
characteristics of matter change dramatically within these ranges,
from a mixture of nucleons, nuclei, and electrons to uniform, strongly
interacting matter containing nucleons, and possibly other particles such as hyperons or
quarks. As the development of an EoS requires joint efforts from many
directions we consider different
theoretical approaches
and discuss relevant experimental and observational constraints which
provide insights for future research.  Finally, results from
applications of the discussed EoS models are summarized.
\end{abstract}


\maketitle
\tableofcontents

\section{Introduction}

Matter under extreme conditions can be found at various places in the
universe. Extremely high densities exist in neutron stars (NSs)
\cite{Glendenning:1997wn,Weber:2004kj,haensel_07,potekhin10,Lattimer:2012nd}.
Densities above nuclear saturation
density \textit{and} high temperatures are reached when the core of a
massive star collapses.  The resulting core-collapse supernova (CCSN)
explosion
\cite{mezzacappa05,Janka_06,kotake06,ott09,janka_12,burrows13} leads
to the formation of a proto-neutron star (PNS)
\cite{Prakash:1996xs,Pons99} and
finally to a NS or a black hole (BH). Similar densities and
temperatures, but higher isospin asymmetries (viz., a higher excess of
neutrons over protons), are involved in the merging of NSs in close
binary systems, NS-NS and NS-BH \cite{Shibata_11,Faber_12,Rosswog_15}.
The dynamical evolution of such violent events and the structure of
the emerging compact stars are determined among others by the equation
of state (EoS) of matter. In addition, the EoS impacts
the conditions for nucleosynthesis and the emerging neutrino spectra.
Hence, the EoS is an essential ingredient in many astrophysical
simulations.\footnote{Within this review we employ the general term
  ``astrophysical simulations'' or ``astrophysical applications''
  synonymously with ``astrophysical simulations of CCSNe, (proto-)NSs,
  and compact binary mergers involving NSs''.}  Many efforts are made
to gain a comprehensive understanding of properties of the involved
matter, which in several aspects are dramatically different from those
in terrestrial experiments.

During the last decades numerous theoretical investigations,
laboratory experiments as well as astronomical observations have been
conducted in order to constrain the thermodynamic properties and
chemical composition of stellar matter for conditions relevant to the
description of compact stars, CCSNe and NS mergers, 
see, e.g., \citet{Klahn_06,Lattimer:2006xb,Lattimer:2012xj}. There is an
intrinsic connection between the macroscopic structure and evolution
of such astrophysical objects and the underlying fundamental
interactions between the constituent particles at the microscopic
level.  This makes the study of the aforementioned systems very
rewarding as they challenge our understanding of nature on both
scales. The aim of this paper is to review existing approaches for the
description of dense matter that can yield EoSs relevant for
compact star astrophysics, from both purely theoretical and
phenomenological perspectives. There is a large
number of different approaches. In many cases the properties of matter
can be provided only for particular thermodynamic conditions,
not always sufficient to describe simultaneously all the astrophysical
systems we address. 
Hence, the main emphasis of this
work is placed on the discussion of approaches to the EoS that
are readily available for use in astrophysical simulations. Such EoSs, 
that cover the full thermodynamic parameter range of 
temperature, density and isospin asymmetry relevant for CCSNe, NSs, and 
compact binary mergers we shall call ``general purpose EoSs''.

Within this review we will not consider the EoS for white dwarfs since, although
being compact stars in astronomical terminology, the underlying
microphysics is quite different. 
Due to the complexity of the topic we will also not discuss pairing 
and related effects of superfluidity and superconductivity, 
see, e.g., \citet{Lombardo:2000ec,Alford:2007xm,Chamel:2008ca,page13}
for corresponding reviews.

In section \ref{sec:eos_astro} some basic thermodynamic
considerations, definitions and requirements for an EoS in
astrophysical applications are discussed. In order to get an idea
about the challenge of constructing such an EoS, it is useful to
state the relevant degrees of freedom and the ranges of the
thermodynamic variables that have to be covered.

It is not an easy task to obtain a reliable description of dense matter that
covers the full range of thermodynamic variables.  In section
\ref{sec:formal} formal approaches to the description of dense matter
are discussed.  The main uncertainties arise from two sources:
\begin{enumerate}
\item the, at least partly, poor knowledge of the interactions, and
\item the treatment of the many-body problem for strongly interacting
  particles.
\end{enumerate} 
Basic considerations about interactions
are presented in subsection~\ref{sec:fewbodyinteractions}.  
In subsections \ref{sec:manybody} and
\ref{sec:clusteredmatter}, we will outline  currently
available techniques that address the many-body problem at finite
densities and temperatures for homogeneous and inhomogeneous matter,
respectively.  
We  will briefly emphasize their
respective advantages and current limitations. 
The description of inhomogeneous matter
is particularly important for CCSNe. 
It is conceptually and computationally very involved
and up to now it has not been possible to apply sophisticated
{\em ab-initio} many-body methods on a grand scale.
Finally, subsection \ref{sec:PT} discusses specific features that
appear in the treatment of phase transitions. 

The inherent uncertainties of any EoS model require a careful analysis
of and comparison with available experimental and observational data.
Therefore, we give an overview of constraints of
the EoS from terrestrial experiments, theoretical considerations, and
astrophysical observations in Sec.~\ref{sec:constraints}.

Section \ref{sec:generalEoS} presents an overview of EoS models for
astrophysical applications. Since EoSs for cold compact stars have been
discussed extensively in the literature, only a summary of available
models and their main features is given. The main emphasis is put on
currently existing general purpose EoS.

Section \ref{sec_application_in_astro}
summarizes the impact of the EoS on the astrophysics of compact stellar objects,
e.g., on (proto)-NSs, binary mergers, CCSNe, and the
formation of BHs. The main aim of this section is to show how
different parts of the EoS and the associated uncertainties are
related to potential astrophysical observations. These considerations
might be useful to identify open questions which
inspire further work for improving EoS models.

An appendix lists freely available resources, data bases
and software, which are related to the EoS and its application in the
astrophysics of compact stellar objects.

Throughout this paper we use units where $k_{B}=\hbar=c=1$.

\section{General remarks on the EoS}
\label{sec:eos_astro}

\subsection{Basic thermodynamic considerations}
\label{sec:thermodynamics}
In its most general form the expression ``equation of state'' is used
for any relation between thermodynamic state variables. Depending on the 
context, often we use it more specifically for the (set of) thermodynamic 
equations that fully specifies the state of matter under a given set of 
physical conditions.

EoSs are typically employed in astrophysical models that use a
hydrodynamic description of the macroscopic system.  In this case, it
is assumed that matter can be considered as a fluid, and explicit
effects from the gravitational field do not have to be included in the
thermodynamic description.  The construction of an EoS supposes that
the local system under consideration is in thermodynamic equilibrium.
This usually means that intensive thermodynamic variables such as
temperature, pressure, or chemical potentials are well defined and
that the conditions of thermal equilibrium (equivalent to a constant
temperature throughout the chosen domain), mechanical equilibrium
(constant pressure) and chemical equilibrium (constant chemical
potentials) hold. Therefore, uniformity of all independent intensive
variables has to be demanded. In order to obtain a thermodynamically
consistent approach, it is most convenient to start from a
thermodynamic potential, chosen according to the set of natural
variables used, and to derive all relevant quantities by standard
thermodynamic relations, see, e.g., \citet{LandauLifshitzV} or the
CompOSE manual \cite{Typel:2013rza,compose}. An example is the
Helmholtz free energy $F(T,\{N_{i}\},V)$, see, e.g.,
\citet{Lattimer:1991nc}, depending on the natural variables
temperature $T$, the set of particle numbers $N_{i}$
($i=1,\dots,N_{\rm part}$) and the volume $V$. It attains a minimum in
the ground state of the system for given values of the thermodynamic
variables.  In the thermodynamic limit, the actual value of $V$ is
irrelevant, and all extensive variables follow the same
scaling. Therefore one can work with ratios of extensive variable such
as the particle number densities $n_i=N_i/V$ that behave as intensive
variables.  In general, the different particle species $i$ are not
inert but can convert to other species by reactions.  If they are in
equilibrium the state of the system is characterized by a number
$N_{\rm cons} \leq N_{\rm part}$ of independent conserved
charges. Thus, in general, the individual particle densities $n_{i}$
are not independent, but connected by conditions of chemical
equilibrium that can be expressed with the help of the particle
chemical potentials $\mu_{i}= \partial F/\partial N_{i}$.  Sometimes a
theoretical description that starts from the grand-canonical potential
$\Omega(T,\{\mu_{i}\},V)= F - \sum_{i} \mu_{i}N_{i}$ is more
convenient.

\subsection{Specific requirements for astrophysical EoS}
\label{subsec:requirements}

\subsubsection{Equilibrium conditions}
\label{sec:equil}

An EoS can be applied only if the system is in
thermodynamic equilibrium. In astrophysical simulations,
this concerns in particular the chemical equilibrium 
since thermal and mechanical equilibrium 
are in general quickly achieved with
$T$ and $p$ as the associated intensive variables.
Then, the use of an EoS in chemical equilibrium
is only justified if the timescales of the corresponding reactions 
are much shorter than the timescales of the system's hydrodynamic evolution.

In contrast, chemical equilibrium among different
nuclear species is not achieved, if
an ensemble of nuclei, nucleons and
electrons is considered
at densities and temperatures as reached in main sequence
stars, in the outer regions of a CCSN
or in explosive nucleosynthesis. Hence the time
evolution of the composition has to be followed with a reaction
network depending on the reaction cross sections of the participating
particle species. Typically it is assumed that a temperature on
the order of 0.5~MeV and above 
is sufficient to reach the so-called nuclear statistical
equilibrium (NSE) \cite{iliadis07}.

A typical set of conserved charges of the system are 
the total baryon number $N_{B}$, the total (electric)
charge number $N_{Q}$, the total electronic lepton number
$N_{L^{(e)}}$, and the total strangeness number $N_{S}$. The
  quantities $N_{i}$ are thereby defined as net particle numbers.
Correspondingly, for every particle the chemical potential is given by
\begin{equation}
\label{eq:mu_general}
 \mu_{i} = B_{i} \mu_{B} + Q_{i} \mu_{q} + L_{i}^{(e)} \mu_{le} +
 S_{i} \mu_{s} 
\end{equation}
with the baryon ($B_{i}$), charge ($Q_{i}$), electronic lepton
($L_{i}^{(e)}$), and strangeness number ($S_{i}$) of the individual
particle.  Hence, the specification of the baryon chemical potential
$\mu_{B}$, the charge chemical potential $\mu_{q}$, the electronic
lepton chemical potential $\mu_{le}$, the strangeness chemical
$\mu_{s}$, is sufficient to obtain
the chemical potential of every constituent.  In particular, in
  NSE, the chemical potential of each nucleus $a$ with neutron number
$N_a$ and proton number $Z_a$ is given by
\begin{equation}
\label{eq_nse}
 \mu_a=(N_a + Z_a) \,\mu_B + Z_a\,\mu_q \equiv N_a \mu_n + Z_a \mu_p \: , 
\end{equation}
where $\mu_n$ ($\mu_p$) is the chemical potential of neutrons (protons).
Conditions on (electric) charge neutrality and weak equilibrium can
further reduce the number of independent particle numbers or
chemical potentials. 

Weak interactions, for instance the electron capture reaction $p +
e^{-} \to n + \nu_{e}$, cannot be considered in equilibrium in
general, since the relevant timescales can exceed the dynamical
timescale of the astrophysical object of interest. In particular in
CCSNe, except at the highest densities roughly
above $n_B = N_B/V = 10^{-3}$ fm$^{-3}$, no weak equilibrium is
obtained. In addition, neutrinos are not necessarily in equilibrium, 
neither thermal nor chemical. Usually
they are not considered within the EoS, but treated via a transport
approach. The neutrino transport
equations, together with the employed weak interaction rates,
are then coupled via energy, momentum and lepton number
conservation to the hydrodynamic evolution of the system and to the
EoS. They determine the electron number densities, which remain a
degree of freedom of the EoS.
Concerning strangeness changing weak
interactions, in the temperature and density range where strange
particles have non-negligible abundances, the timescales estimated for
the relevant processes are of the order of $10^{-6}$ s or below, see,
e.g., \citet{Brown92}. 
Therefore, in general strangeness changing weak
equilibrium is assumed, i.e., $\mu_s = 0$. Hence, 
strangeness is not a conserved charge and does not represent 
an independent thermodynamic variable.

The situation is different for the highest
densities reached in CCSNe, i.e., in hot (proto)-NSs.  At the
prevailing high temperatures and densities, neutrinos are trapped and
equilibrium with respect to weak reactions is achieved.  They can be
treated as part of the EoS, parameterized by the neutrino fraction
$Y_{\nu_e}=n_{\nu_e}/n_B$ or the lepton fraction
$Y_{L_e}=Y_{\nu_e}+Y_e$ with the electron fraction $Y_e=n_e/n_B$.  At
a later cooling stage neutrinos become untrapped, i.e., their mean
free path becomes longer than the system size and $\beta$-equilibrium
without neutrinos is established.  This condition can be expressed by
setting the electronic lepton chemical potential
$\mu_{le}$ to zero in Eq.~(\ref{eq:mu_general})  as for cold NSs. Together
with charge neutrality it implies that $n_e$ or $Y_e$ are fixed 
by $n_B$ and are no longer free variables of the EoS.

Assuming lepton flavor conversion via neutrino oscillations to be
negligible, the heavy flavor lepton numbers are conserved
independently of the electronic lepton number. For the moment no
simulation has been performed that includes heavy charged leptons
explicitly. The influence 
of heavy flavor leptons is expected to be
small due to their high rest masses.
Nevertheless, in EoSs of cold NSs muons are usually included.

\subsubsection{Charge neutrality and inhomogeneity effects}
\label{sec:charge_neutrality}
In all astrophysical scenarios considered in this review, 
the system can be regarded as infinitely large on
the length scales of the microscopic model. The
thermodynamic limit is reached and electric charge neutrality 
is required to avoid instabilities due to the occurrence 
of strong electric fields.

In its simplest form, charge neutrality can be formulated as a local
condition, $n_{Q} = \sum_{i} Q_{i}  n_{i} = 0$.  
Thus $n_{Q}$ is not an
independent thermodynamic degree of freedom and 
it is convenient to introduce the hadronic charge density
$n_{q}=\sum_{i}'Q_{i}n_{i}$ where the primed sum runs over all
hadrons (and/or quarks, if present).  If electrons are the only leptonic component, this
implies $n_{q}=n_{e}$.

In the case of inhomogeneous matter, the charge distribution
can be imbalanced locally. The resulting competition between nuclear surface 
and Coulomb energies, causes the formation of clusters or more 
complicated structures such as ''pasta phases'' 
\cite{Ravenhall:1983uh,HaSeYa1984,WiKo1985}.
Charge neutrality is maintained only globally. 
In a simple approximation, this occurrence of finite-size 
structures with low and high baryon number densities can be treated as a 
coexistence of phases, however, surface and Coulomb effects are neglected 
in this case, see Sec.\ \ref{sec:PT} for details.

\subsubsection{Range of thermodynamic variables}
\label{sec:range}

The most general case we are interested in, is an EoS depending on the
temperature $T$, total baryon number density $n_{B}$ and the total
hadronic charge density $n_q$ or a set of equivalent thermodynamic
variables.  Instead of $n_{q}$, the corresponding fraction
$Y_{q}=n_{q}/n_{B}$ may be used. The baryon number density $n_{B}$ is
sometimes replaced by the mass density $\varrho = m_{B} n_{B}$ with
the mass unit $m_{B}$, which is often taken to be either the atomic
mass unit $m_u$ or the neutron mass $m_n$. As an
order of magnitude estimate, a baryon number density of
$0.1$~fm$^{-3}$ corresponds to a mass density of $\varrho \approx 1.66
\cdot 10^{14}$~g/cm$^3$.

The observations of compact stars with masses of $2$~M$_{\odot}$
\cite{demorest_10,Antoniadis_13,Fonseca:2016tux} 
imply that the maximum
baryon number density in NSs can approach approximately
ten times the nuclear saturation density $n_{\rm sat}\approx
0.16$~fm$^{-3}$ \cite{Lattimer:2010uk}.  The densities  in
CCSNe and PNSs are generally lower. An
exception is the case of so-called ``failed''CCSNe
leading to BH formation, see Sec.~\ref{sec_bh_formation}.  
During the final collapse to a BH, densities well above
$10~n_{\rm sat}$ can be reached before the formation of an event horizon
\cite{sumiyoshi_07,O'Connor:2010tk,Hempel_11a, Peres_13}. However,
these extremely high densities occur only for less than a millisecond.
The free-fall like collapse to a BH is largely dominated by gravity
and the EoS is not expected to influence its dynamics.

For the gross structure of NSs, the state of matter below
$10^{-11}$~fm$^{-3}$ is practically irrelevant, 
as it makes up only the few outermost centimeters of the star. In CCSNe, and to
some extent also in NS mergers, the situation is
different. Here, low-density matter plays an important role. 
In both cases one is interested in the ejecta. Their densities
decrease continuously during their expansion. Typically, a simple ideal gas EoS
is used for the description of matter under such conditions, however,
full thermodynamic equilibrium, see Sec.~\ref{sec:equil}, cannot
always be assumed. Instead a network of time-dependent 
nuclear reactions has to be considered. For CCSNe, 
the ongoing burning processes, which contribute to the final explosion 
energies \cite{Yamamoto_13,Perego_15}, occur in low-density matter.
To follow the evolution of a
CCSNe for several seconds, it is unavoidable to
include a description of low-density matter out of NSE. 
From a practical point of view, the connection of such a region to a
tabulated EoS in the higher-density and temperature regime can be
quite intricate.

\begin{figure}
\centering
\includegraphics[width=1.0\columnwidth]{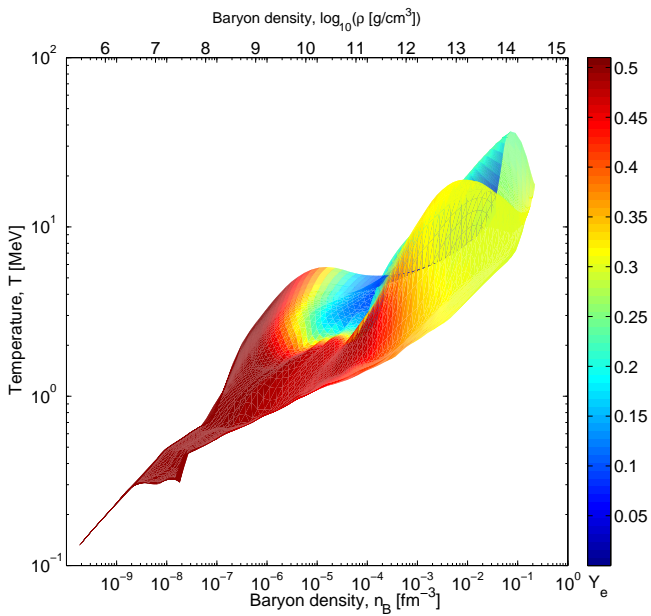}
\caption{(color online)  
Temperatures and densities (lower scale shows the baryon number density and the upper scale shows the mass density) reached during a CCSN simulation at $1$~s post bounce. The color-coding shows the electron fraction $Y_e$. Figure taken from \citet{fischer_11}.}
\label{fig:phasediagram_sn}
\end{figure}

The temperature of a typical NS older than a few minutes is small
(below 1~MeV) on nuclear energy scales~\cite{Pons99,Suwa:2013mva} 
and can be considered as zero in most
applications. However, these objects are born in CCSNe which can be
extremely hot events. The same holds, e.g., for NS-NS and NS-BH
mergers. Typical temperatures in CCSNe and PNS
are in the range from a fraction to a few tens of MeV. This can be
inferred from Fig.~\ref{fig:phasediagram_sn} which shows the
temperatures and densities reached during a 
CCSN simulation for the 15~M$_{\odot}$
progenitor of \citet{woosley95} within the first second after bounce.
It is a typical example for the
core-collapse of an intermediate-mass progenitor which is expected
to lead to an explosion.  The temperatures
obtained correspond to entropies per baryon in the range from $1$
to $5$ at the stage of collapse and up to 20 in the shock heated matter.
For other progenitors that are also expected to lead to
explosions, the range of temperatures is similar. 
Scenarios with BH formation set
the upper limits for density and temperature which a general purpose
EoS has to cover. The temperature in such an event can rise above
$100$~MeV, see, e.g., \citet{O'Connor:2010tk}. Therefore the
temperature domain to be covered by a general purpose EoS 
is $0~\mbox{MeV}\lsim T\lsim150$ MeV.

The color coding in Fig.~\ref{fig:phasediagram_sn} illustrates the
electron fractions $Y_{e}$ reached during the early evolution
of a CCSN. The core of the
supernova progenitor has almost an equal number of electrons, protons
and neutrons, i.e., $Y_{e} \approx 0.5$.  During the collapse,
electron capture reactions lead to a strong neutronization of matter,
decreasing $Y_e$. The presence of trapped neutrinos, which acquire a
finite chemical potential, limits the lowest electron fractions
$Y_{e}$ which are reached in the core \cite{fischer_11}.
In its later evolution, the cooling PNS approaches $\beta$-equilibrium with
neutrinos freely leaving the system. In the cold, final equilibrium
state of a NS, the lowest electron fractions are found to be very
close to zero.  In some parts of the supernova ejecta $Y_{e}$ can rise
to values above $0.5$ corresponding to a proton-rich
environment. Consequently, the range of a general purpose EoS to
be covered is $0<Y_e \lsim\, 0.6$. 

\begin{figure*}
\centering
\includegraphics[width=0.8\textwidth]{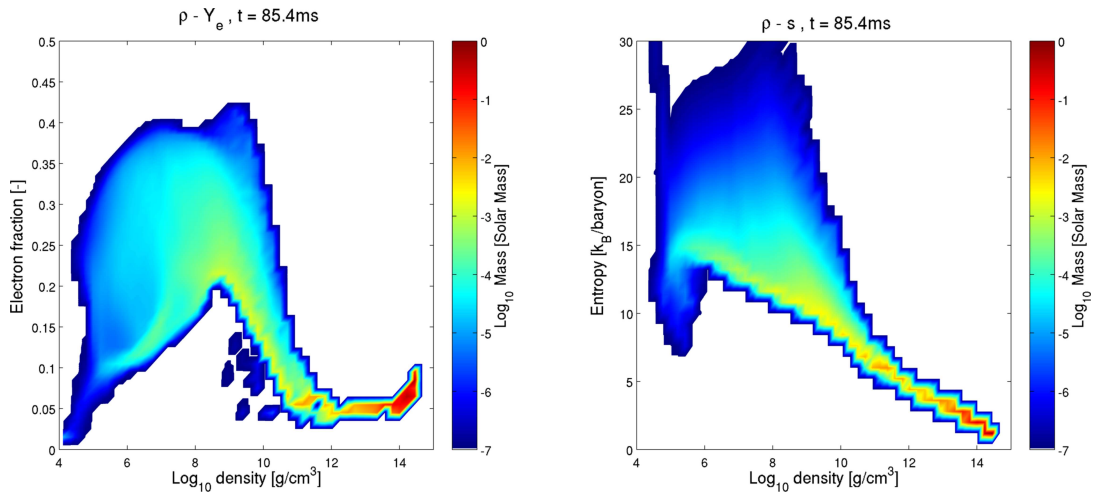}
\caption{(color online) 2D mass histograms 
for mass density $\rho$ and electron fraction $Y_e$ (left panel) 
or entropy $s$ 
(right panel), of a NS merger remnant
at a time of $85.4$~ms after the first contact. Color
coded is a measure of the amount of matter experiencing the specific 
thermodynamical conditions. Figures adapted from \citet{perego14}.}
\label{fig:phasediagram_nsmerger}
\end{figure*}

The conditions in NS mergers are quite diverse. In general they depend on the 
masses of the merging NSs and the EoS, and also on the magnetic fields and NS 
spins. Typical temperatures in the core of a 
post-merger remnant NS are in the range from 20 to 60 MeV \citep{bauswein10}. 
These temperatures can be well exceeded in the contact layers in 
the early stage of the merger, where extremely high temperatures up to
150~MeV can occur locally \citep{bauswein10,rosswog13}.
The highest densities in the hot and rotating remnant NS are 
typically between 2 and 6 $n_{\rm sat}$ \citep{hotokezaka13b}. 
In case the remnant collapses to a BH, similar arguments as for failed SNe 
apply: during the collapse much higher densities and correspondingly higher 
temperatures are reached, but are probably not important dynamically.

The dynamic ejecta of NS mergers originate from the crust and outer core 
of the merging NSs. Initially, this material has very low $Y_e$ in the 
range from $0.0$ to $0.2$ \citep{rosswog13,sekiguchi15}. 
Depending on the temperatures reached, the
degeneracy of electrons is lifted and $Y_e$ increases to higher values. In the 
subsequent evolution, neutrino absorptions also influence $Y_e$,
resulting in final values in the range of roughly $0.1$ to $0.4$ 
\citep{Wanajo:2014wha,sekiguchi15}. 
See also \citet{foucart15} for a comparison of the thermodynamic 
conditions for different EoS.
Fig.~\ref{fig:phasediagram_nsmerger} shows the thermodynamic conditions reached
in the remnant in the aftermath of a neutron star merger. 
For the later ejecta that 
appear in form of a neutrino-driven wind, extremely high entropies above 50 are 
found (located mostly in the polar regions), whereas most of the matter has 
entropies below 7, and the entropy tends to decrease with 
increasing density \citep{perego14,sekiguchi15}.

To conclude, we summarize in Table~\ref{tab:ranges} the overall ranges 
that have to be covered by a general purpose EoS 
to describe NS mergers, CCSNe and cold NSs. 

\begin{table}[b]
\caption{Approximate ranges of temperature, baryon number density and electron fraction a general purpose EoS has to cover to be able to describe cold NSs, NSs in binary mergers, and CCSNe.}
\begin{center}
\begin{tabular}{c|c}
 \hline \hline
 quantity & range \\
 \hline
 temperature & $0$~MeV $\leq T < 150$~MeV \\
 baryon number density & $10^{-11}$~fm$^{-3}$ $ < n_{B} < 10$~fm$^{-3}$ \\
 electron fraction & $ 0 < Y_{e} < 0.6$ \\
 \hline \hline
\end{tabular}
\end{center}
\label{tab:ranges}
\end{table}

\subsubsection{Particle degrees of freedom} 

Within the ranges of the thermodynamic variables given above, the
composition of matter changes dramatically. 
In cold NSs, heavy nuclei are present in the inner and outer crust
\cite{Chamel:2008ca}. The surface layer of the outer crust is made of $^{56}$Fe
ions immersed in a sea of electrons. With increasing density, the
nuclei become more massive and neutron rich, reaching nuclei of
the neutron drip line at the boundary to the inner crust, c.f.,
section \ref{sec:crust} for details.  

At low densities and finite temperatures, a plasma is expected
with a mixture of nuclei, nucleons and electrons. 
In the shock-heated matter
of CCSNe, light nuclear clusters, such
as $\alpha$-particles, deuterons or tritons, are found to be the
dominant baryonic particle degrees of freedom besides nucleons
\cite{Sumiyoshi08a}.  At densities just below nuclear saturation
or sufficiently high temperatures, nuclei dissolve and one is left
with strongly interacting matter composed of nucleons and electrons. 
At high densities and/or temperatures, additional particle
species are expected to occur
\cite{Glendenning:1997wn,Weber:2004kj}, such
as nuclear resonances, e.g., $\Delta$-baryons \cite{drago14}, or
mesons, e.g., pions. Also strange degrees of freedom such as
hyperons \cite{glendenning82,chatterjee15} or kaons can be
present.  There is the possibility that
the mesons form condensates at low temperatures
\cite{Glendenning:1997wn}. Even a transition to deconfined quark
matter is possible at high densities and temperatures.  The
occurrence of antiparticles is relevant at high temperatures, 
in particular for light particle species.  
Besides electrons, muons are relevant leptonic degrees of
freedom, and so are electron-, muon-, and tau-flavor neutrinos and
anti-neutrinos.  Neutrinos are not
necessarily in equilibrium with matter, see Sec.~\ref{sec:equil}.  At
finite temperatures, thermal photons complete the composition.  While
leptons and photons can be mostly treated as free
gases, this does not hold for hadrons or quarks. Their
contribution to the EoS is mainly governed by the strong interaction
deeply inside the non-perturbative regime.

\section{Formal approaches to the description of dense matter}
\label{sec:formal}

The theoretical description of strongly interacting matter requires
methods which capture the essential thermodynamic properties of the
many-body system. The challenges are multifaceted. First of all, the
relevant degrees of freedom have to be identified. Approaches for
nuclear matter that are based on nucleons are the predominant choice
and may suffice in many cases. It will be necessary to consider other
degrees of freedom for certain thermodynamic conditions, e.g., nuclei
at low temperatures and densities and hyperons or even quarks at high
temperatures and densities.  Furthermore, large isospin asymmetries of
the system can shift the dominating degrees of freedom to more exotic
particles. Secondly, the interactions between the constituents have to
be specified.  This is a nontrivial task due to the very complex
nature of the strong interaction.  In addition, the representation
depends on the chosen degrees of freedom. In principle, one would like
to describe matter directly within the well-founded theory of QCD.
However, there are no {\em ab initio} QCD calculations of dense matter
available at the thermodynamic conditions that are
characteristic for compact stars or CCSNe. Even a
derivation of the ``true'' interaction between nucleons or other
strongly interacting hadrons from QCD remains a very complex task
despite intensive efforts. Hence, calculations have to rely on model
interactions, which are partially constrained by laboratory
measurements.
  Modern theoretical approaches aim at a derivation of input
  interactions that are systematically improvable with controlled
  uncertainties.

In the next step, an appropriate  method  is applied
to find the actual state of the system 
incorporating few- and many-body correlations.
In the case of a phase transition, additional thermodynamical considerations 
have to be applied, see Sec.~\ref{sec:PT}.

The choice of the degrees of freedom, the
choice of the interaction and the selection
of the many-body method are not independent and many different
  approaches exist. Here, we divide them into two
  categories:
\begin{enumerate}
\item {\em Ab-initio many-body methods} start from ``realistic''
  few-body interactions (mainly two- and three-nucleon
  forces),i.e. interactions that are fitted to observables in
  nucleon-nucleon scattering in vacuum and properties of bound
  few-nucleon systems. The many-body problem is then treated using
  different techniques, e.g., Green's function methods,
  (Dirac-)Brueckner-Hartree Fock calculations, coupled
    cluster, variational and Monte-Carlo methods, see section
  \ref{sec:abinitio}.  Some of these many-body methods are limited
    by technical problems, such as the Monte-Carlo methods, others
    introduce approximations, such as for example, Brueckner-type
  approaches, that consider only a subclass of all possible
  diagrams.
\item {\em Phenomenological approaches} use effective interactions
  that often have a more simple structure than realistic
  interactions used in {\em ab initio} approaches. They depend on
  a small number of parameters, usually of the order of 10 to 15,
  which are fitted, in the ideal case, to different properties of
  several nuclei all over the chart of nuclei and nuclear matter
  properties.  Typical representatives of these effective interactions
  are the Skyrme and Gogny forces in non-relativistic calculations and
  meson-exchange forces in relativistic mean-field models, see section
  \ref{sec:pheno}.  Nowadays, these phenomenological approaches are
    interpreted in terms of energy density functional (EDF) theory.
  Applying simple many-body methods, mostly on the mean-field level,
  results already in a rather precise description of nuclei and
  nuclear matter.  The extrapolation to exotic conditions has to be
  considered with caution, nevertheless phenomenological approaches
  are the most widely used methods to construct EoSs for astrophysical
  applications.  
\end{enumerate}

\subsection{Basic few-body interactions}
\label{sec:fewbodyinteractions}

Basic few-body interactions are the starting point of any calculation
of dense matter with {\em ab initio} many-body methods. The two-body
interaction is largely dominant, but interactions beyond the two-body
level become important in matter at high densities.  For example, it
is well known that the nuclear three-body force is essential to
reproduce the saturation properties of nuclear matter. Forces among
four or more nucleons are difficult to construct and can be neglected
in many circumstances.  Realistic two- and three-body
forces between nucleons and hyperons, as discussed in this section,
should not be confused with effective temperature and density
dependent interactions used in phenomenological models, see
Sec.~\ref{sec:pheno}.  A more detailed survey of modern theories for
nuclear forces can be found, e.g., in the review of
\citet{Epelbaum:2008ga}.

Historically, Yukawa~\cite{Yukawa_35} proposed the first model of the
$NN$-interaction based on the exchange of a massive particle, the
pion. His model successfully explained the range of the nuclear
interaction.  Since then,
many phenomenological models have been
developed either based on Yukawa's idea of meson exchange 
or by constructing potentials with appropriate operator structure. 
With the advent of QCD as the theory of the strong interaction in the 1970's,
phenomenological quark models became very fashionable, describing
baryons as quark clusters. They, however, suffer from the missing
confinement and connection with QCD. 
Only recently, with chiral effective field theories ($\chi$EFTs) and with
lattice gauge theory, considerable progress has been achieved to 
link baryonic few-body forces to QCD. 

\subsubsection{Experimental data}
\label{sec:exdatafewbody}

Any theoretical model for baryonic forces 
can be tested by comparing predictions to
experimental data. This concerns scattering
and the structure of light nuclei and hypernuclei.
In the nuclear sector, many thousands of high precision
data points are available. A complete partial wave analysis of
nucleon-nucleon ($NN$) scattering data can be performed,
see \citet{Stoks_93,Arndt_94,Arndt_07,Perez:2013mwa} 
and the corresponding online data bases. 
Deuteron properties, among others its binding energy
and electric quadrupole moment, are an
important input for $NN$-forces.
Owing to the huge amount of data,
today's $NN$-interactions, phenomenological or based on effective
field theories, have reached a very high degree of precision. 

The binding energies of other light nuclei and
nucleon-deuteron scattering data cannot be described satisfactorily on
the basis of a two-nucleon interaction and provide thus valuable
information on the nuclear three-body force,
see~\citet{KalantarNayestanaki_12} for a review. Recently, properties
of very neutron rich nuclei have attracted attention 
since they provide additional constraints on the three-body
force, see, e.g., \citet{Wienholtz_13}. 

For the hyperonic sector data are scarce, see \citet{Gal:2016boi} 
for a detailed review of strangeness in nuclear physics.
Hyperon-nucleon ($YN$) scattering 
experiments are difficult to perform because hyperons have very short lifetimes 
of the order $10^{-10}$~s. Data on hyperon-hyperon ($YY$) 
scattering are not available.  In the fitting procedure for the parameters of 
hyperonic two-body forces, in general only 35 data points
from the 1960's~\cite{Engelmann_66,Alexander_68,SechiZorn_69,Eisele_71}
for low-energy total cross sections in reactions involving $\Lambda$
and $\Sigma$-hyperons are included. First low-energy data on $\Xi^{-} p$
elastic and $\Xi^{-} p \to \Lambda\Lambda$ scattering have been obtained
at KEK~\cite{Ahn_06}.

In addition to scattering experiments, hypernuclear spectroscopy can
provide valuable information. Since the first events 
recorded by \citet{Danysz_53a, Danysz_53b}, many
hypernuclei have been produced. These are mainly
single-$\Lambda$-hypernuclei,
see, e.g., the reviews by~\citet{Hashimoto_06,Gal:2016boi}. 
Some events with double-$\Lambda$ hypernuclei have been detected,
see \citet{Aoki:1991ip,Nakazawa10}. 
The absence of $\Sigma$-bound states, except for an
s-wave $\Sigma$ bound state in $^{4}_{\Sigma} \mathrm{He}$ \cite{Nagae98},
indicates a repulsive $\Sigma N$ interaction
\cite{Bart_99,Saha04,Kohno06}.
Only few events for  $\Xi$-hypernuclei have been observed 
up to now \cite{Aoki_95,Fukuda98,Khaustov99}.
Considerable experimental efforts are underway to improve hypernuclear
data, see, e.g., \citet{Agnello_12, Sugimura_14}.
Three-body forces 
are not yet well explored for hyperons, see, however,
the recent work in~\citet{lonardoni2013a}. 
Hyperonic single-particle potentials in symmetric nuclear matter
are often used to determine the effective hyperon-nucleon
interactions in phenomenological models, see the discussion in, e.g.,
\citet{Ellis:1990qq,glendenning_91,Schaffner:1993qj,Glendenning:1997wn,Balberg97,Vidana01,Oertel:2012qd}.
   
\subsubsection{Phenomenological forces}

Approaches to obtain phenomenological forces can be divided into three
main categories: models based on meson exchange and potential models.

\paragraph{Meson-exchange models}
Meson exchange models follow the original idea of Yukawa that the
$NN$-interaction is mediated by meson exchange. Additional mesons have
been added to capture
the complex dependence of the nuclear interaction on spin, isospin and
spatial coordinates.  Some models have been extended to include
strange mesons in order to describe the $YN$ and in some cases the
$YY$ interaction.  The general idea is that the pion, as the lightest
particle, describes the long-range attractive part of the interaction
and that scalar mesons are responsible for the intermediate range
attraction whereas vector mesons govern the short-range repulsive
contribution. The so-called $\sigma$-meson in the scalar-isoscalar
channel often represents the mid-range attraction, however, its status
as a particle is very ambiguous.  Many models use instead correlated
and uncorrelated two-pion exchange to describe the intermediate range
$NN$-interaction, see, e.g., \citet{Machleidt_94,Donoghue_06} for a
discussion. The various models differ mainly in the mesonic content,
the treatment of two-meson exchange and approximations made in order
to obtain practically applicable potentials from the basic amplitudes,
for instance the form factors used at the meson-baryon interaction
vertices.  From a phenomenological point of view, the latter are
introduced to account for the substructure of baryons. They serve as
regulators in solving the scattering equation in order to avoid any
divergent contributions.

Below, we mention some well-known models that describe $NN$
scattering data and the deuteron satisfactorily. The classical
versions of the Nijmegen interaction for the $NN$ system
\cite{Nagels_77,Nagels_78} are based on the one-meson exchange
picture. They were extended to include $YN$, $YY$ interactions
\cite{maessen89,Rijken99} as well as additional one- and two-meson
exchanges \cite{Rijken_06a,Rijken_06b,Nagels_14}.  The Paris $NN$
potential \cite{Cottingham_73,lacombe80} uses an ad-hoc
parameterization at very short distances arguing that the meson
exchange picture is no longer valid there due to the substructure of
nucleons. The Bonn potential \cite{Machleidt_87,Machleidt_01} for the
$NN$-interaction is given in relativistic form in momentum space to
avoid the local approximation of non-relativistic models, as, e.g., in
the Nijmegen potentials. The J\"{u}lich group has extended the Bonn
model to the $YN$-interaction~\cite{Holzenkamp_89,Haidenbauer_05}.

\paragraph{Potential models}
\label{sec:phenoforces}

In addition to the well-established long-range one-pion exchange,
potential models adopt a sum of local operators, where the
essential ones are central, tensor and spin-orbit terms. The
parameters are fitted to deuteron properties and $NN$-scattering data.
The Urbana~\cite{Lagaris_81} and Argonne
potentials~\cite{Wiringa_84,Wiringa_95} are examples of such
high-quality potential models. The latest version of the Argonne
potential~\cite{Wiringa_95}, called $v_{18}$, not only contains
isoscalar operators but includes an electromagnetic part and isovector
operators such that the charge dependence
of the $NN$-force is successfully described.  
Also, some models with $\Lambda$-hyperons are available, see,
e.g., \citet{Bodmer_84a} but they are much less sophisticated due to
the small amount of hyperonic data. 

Considering only two-nucleon interactions, it is well known that light
nuclei, in particular the triton, $^3$H, are underbound and the
saturation density of nuclear matter is overestimated. This shows the
importance of repulsive many-body  
forces to correctly describe nuclear systems. A major contribution to the
three-nucleon force is the two-step pion exchange between two nucleons
via a third nucleon that can be excited, e.g., to a
$\Delta$-baryon, \cite{Fujita:1957zz}. 
This feature is incorporated for instance in the
Tucson-Melbourne model~\cite{Coon_81,Friar_99,Coon_01}. Such an
interaction is attractive and helps to cure the underbinding problem
in light nuclei, whereas it worsens nuclear matter saturation
properties. Therefore the Urbana group proposed a series of
phenomenological three-nucleon forces, adding to the attractive
two-pion exchange contribution a parameterized repulsive
part~\cite{Carlson_83,Pudliner_95,Pieper_01,Pieper_08}. The latter
interaction is adjusted to the properties of light nuclei. The problem
of such a procedure is  that the three-body force is not
independent of the two-body force  employed in the fits.
More recently, consistent two- and three-nucleon forces have
been derived within $\chi$EFT, see the next section. 

\subsubsection{Interactions from chiral effective field theory and lattice QCD} 
\label{sec:chiralBBforces}

Since the seminal papers by Weinberg in the early
1990's~\cite{Weinberg_91a, Weinberg_91b} many efforts have been
devoted to the derivation of nuclear forces from a $\chi$EFT.  Within
a chiral theory pions emerge naturally as the relevant degrees of
freedom at low energies to describe the interaction of nucleons since
they appear as Goldstone bosons of the theory if the chiral symmetry
of QCD is spontaneously broken.  The systematic framework of effective
field theories has allowed to establish a classification of different
contributions to the interaction and to make the link with QCD. The
starting point is the most general effective chiral Lagrangian that
respects the required symmetries. It is expanded in powers of a small
quantity $p \sim (m_\pi/\Lambda_\chi, |\vec{k}|/\Lambda_\chi)$, where
$m_\pi$ denotes the pion mass, $\vec{k}$ is a (soft) external momentum
and $\Lambda_\chi \sim 1$~GeV specifies the scale of chiral symmetry
breaking.  In addition to dynamical pion contributions, nucleonic
contact operators appear at each order. They contain the unresolved
short-range physics. Their strength is controlled by so-called
low-energy constants (LECs) that are fitted to experimental
data. Pionless effective theories are applicable at very low
  energies, see \citet{Bedaque:2002mn} for a review.

Although the details of the power counting scheme are not yet
  completely settled, see e.g., \citet{Valderrama:2014vra} and
  references therein, chiral nuclear forces work out well. In particular, before the advent of nuclear interactions from $\chi$EFTs, 
no consistent model of three-body and higher many-body nuclear forces 
existed. In addition to the incorporation of
symmetries from QCD, the advantage of $\chi$EFT approaches is the
possibility to extend the interactions in a consistent way to three-
and many-nucleon systems. Comprehensive reviews can be found in
\citet{Epelbaum:2008ga} and \citet{Machleidt_11}.  For recent high-quality
chiral potentials see \citet{Perez:2014bua} and
\citet{Piarulli:2014bda}, which includes $\Delta$ resonances.
The $\chi$EFT approach
has been extended to include strangeness and interactions of the full
baryon octet, see \citet{Polinder06,Haidenbauer07,Haidenbauer13}.  In
this case, the LECs cannot be determined by experiment due to the lack of
relevant data in the hyperonic sector. Instead they have been fixed by
flavor $SU(3)$-symmetry.

Another promising possibility to relate nuclear forces to QCD is
lattice QCD. In principle, it is a tool to calculate hadron properties
directly from the QCD Lagrangian with Monte Carlo methods on a
discretized Euclidian space-time. It is, however, extremely expensive
in the numerical application even with  sophisticated 
state-of-the-art algorithms on high performance computers. For the moment, 
simulations can only be carried out with large
quark masses and the extrapolation to physical
masses is difficult. In addition, the lattice spacing has to be fine
enough and the volume large enough to 
avoid computational artifacts. Recent substantial efforts, see,
e.g., \citet{Beane_11,Aoki_12} for reviews, give hope  for future 
high precision predictions. This
could be interesting in particular for channels where only few
experimental data are available, e.g., the hyperon-nucleon
interaction, see, e.g., \citet{Beane06,Inoue:2010hs,Beane:2012ey}.

\subsubsection{Renormalization group methods and evolved potentials}
\label{subsec:renorm}
The strongly
repulsive core of two-body baryonic interactions renders multi-baryon
systems non-perturbative.  Thus correlations become extremely
important but are difficult to treat with many-body methods. The
repulsive core, although a distinct feature of baryonic forces, is not
directly affecting low-energy observables.  With renormalization group
(RG) techniques, the high-momentum part of the interaction related to
the repulsive core can be ``integrated out'' via a continuous change
in resolution by applying suitable unitary transformations. In this
way, the high-momentum part decouples from the low-momentum part and
three- and many-body forces emerge automatically from a pure two-body
force. During the evolution, all generated interactions are
``phase-shift equivalent'' and low-momentum observables are
preserved. Thus the description of scattering data remains as
precise as for the original interaction. The obtained RG-evolved
potentials are much more perturbative than non-evolved ones and
therefore simplify the baryonic many-body problem. In connection with
many-body perturbation theory (MBPT), i.e., a perturbative expansion
around the Hartree-Fock (HF) solution, RG-evolved interactions became
a great success for nuclear systems, see
\citet{Bogner09,Furnstahl:2013oba} for two recent reviews on the
subject.

Even though all high-precision ``bare'' nuclear forces are rather
different, almost unique RG-evolved potentials emerge at low momenta
\cite{Schwenk:2004hz}, often denoted as $V_{\mathrm{low-k}}$.
In \citet{Wagner_06,Schaefer_06} the same techniques have been applied
to nucleon-hyperon interactions. It turned out that the resulting
low-momentum interactions are different from each other because the
bare potentials are much less constrained.  Hence, it is not
surprising that there is a large spread in the results if they are
applied to dense matter with hyperons, see \citet{Djapo08,Djapo10}. 
This clearly shows the lack of relevant experimental data
concerning the hyperon-nucleon and hyperon-hyperon interactions.

\subsection{Many-body methods for homogeneous matter}
\label{sec:manybody}
The first step in studying strongly interacting matter is often the
investigation of homogeneous matter at vanishing temperature where
almost all methods discussed below can be applied.
Even if the basic few-body interactions were exactly known, the
theoretical modeling is not  a trivial task since any naive
perturbative expansion is likely to fail. 

The most simple method to treat the
many-body problem  beyond the perturbative level
is the HF approximation, see Sec.~\ref{sec:pheno}
and, e.g., the textbooks of
\citet{FetterWalecka,RingSchuck,GreinerMaruhn} for more details. The
idea is that each particle moves in a single-particle potential, the
``mean field'', generated by the average interaction with all other
particles. In practice, the many-body wave function is approximated as
an antisymmetrized product of single-particle wave functions, which are
determined self-consistently. Although generally very successful in atomic
physics and in chemistry, in a nuclear system a HF calculation
starting from conventional two-body interactions fails to reproduce
known properties of nuclear matter. In addition, the results are very
sensitive to the modeling of the short-range repulsive core of the
two-body interaction which is not fixed uniquely by scattering
data. This can be understood since the HF approximation neglects any
short-range correlations between the particles which arise from their
mutual interaction.

Two ways out of this problem are currently applied: either
correlations are explicitly included within the many-body approach, or,
instead of realistic few-body interactions, an effective, 
usually medium dependent, interaction is used within a HF approach. 

In Sec.~\ref{sec:abinitio}, we discuss different theoretical
{\em ab-initio}
frameworks to include correlations in a strongly interacting many-body
system. Interesting attempts to compare in a quantitative way
different {\em ab-initio} many-body methods 
can be found, e.g., in \citet{Baldo:2003yp,Bombaci:2004iu,Baldo_12}.
Sec.~\ref{sec:pheno} is devoted to models with different types of
phenomenological effective interactions.  Apart from the
textbooks, there exist many excellent reviews on the different
standard many-body methods, 
see, e.g., \citet{Muther_00,Baldo_11,Carlson:2014vla}.  Therefore we
do not aim to give a comprehensive and complete overview, but only present the general ideas.

\subsubsection{Ab-initio methods}
\label{sec:abinitio}

\paragraph{Self-consistent Green's function}
\label{sec:scgf}
The idea of the self-consistent Green's function (SCGF) method is that
the system's energy can be calculated conveniently from the
single-particle Green's function $\mathcal{G}$. 
It describes the propagation of a single-particle state $\psi$ from
time $t$ and position $\vec{x}$ to $t^{\prime}$ and $\vec{x}^{\prime}$ as
\begin{equation}
 \psi(t^{\prime},\vec{x}^{\prime}) = \int d^{3}x \:
 \mathcal{G}(t^{\prime},\vec{x}^{\prime},t,\vec{x}) \: \psi(t,\vec{x}) \: .
\end{equation} 
The level of approximation in the SCGF method is
controlled by the approximations made in order to determine the
single-particle Green's function. A thorough definition can be found
in any textbook on quantum theory at finite density and temperature,
see, e.g., \citet{FetterWalecka}. For a non-interacting
homogeneous system at zero temperature 
the Green's function can be written in momentum space 
as\footnote{For simplicity we assume non-relativistic kinematics. 
  A relativistic treatment does not change the general
  reasoning. We further consider the zero temperature limit. Finite
  temperature is easily included in the formalism, see,
  e.g., \citet{FetterWalecka,Frickthesis}.}
\begin{equation}
{\mathcal G}^0 (\vec{k},\omega) 
= \frac{\theta(|\vec{k}| - k_F)}{\omega - E^0(\vec{k}) + i \eta} + 
\frac{\theta(k_F - |\vec{k}|)}{\omega - E^0(\vec{k}) - i \eta} \: ,
\label{eq:g0}
\end{equation}
where $k_F$ denotes the Fermi momentum and $E^0(\vec{k}) =
\vec{k}^2/(2m)$ the non-interacting single-particle energy of
a particle with mass $m$. Any indices related to further quantum numbers of the
particle are suppressed for clarity.
The first term on the r.h.s.\ of Eq.~(\ref{eq:g0}) 
describes the propagation of a state outside the Fermi sea, a
{\em particle}, and the second term a state inside the Fermi sea.  Since
per definition all states in the Fermi sea are filled, it can
propagate inside the Fermi sea only as a {\em hole}, i.e., a
particle removed from the Fermi sea.

The energy density 
of the system can be
straightforwardly calculated from the trace of the single-particle
Green's function. For the non-interacting system the well-known
expression for an ideal Fermi gas is obtained. Of course, a dense baryonic system 
cannot be described as a Fermi gas of non-interacting particles. 
Thus the full interacting single-particle Green's function ${\mathcal G}$ 
has to be determined.
A calculation from a perturbative series in the interaction potential
is not viable in view of the strong baryonic interaction. 
At this point self-consistency is introduced in the form of Dyson's equation,
which schematically can be written as
\begin{equation}
{\mathcal G} = {\mathcal G}^0 + {\mathcal G}^0 \Sigma {\mathcal G}
\label{eq:dyson}
\end{equation}
with the one-particle irreducible\footnote{One-particle irreducible means that the diagrams
  cannot be disconnected by cutting a fermion line. It is obvious that
  the self-consistent resummation via Dyson's equation automatically
  generates all reducible terms.} 
self-energy $\Sigma$ which itself is determined by the interaction, 
${\mathcal G}^0$ and ${\mathcal G}$.

\begin{figure}[ht!]
\centering
\includegraphics[width=1.0\columnwidth,bb= 42 584 590 720]{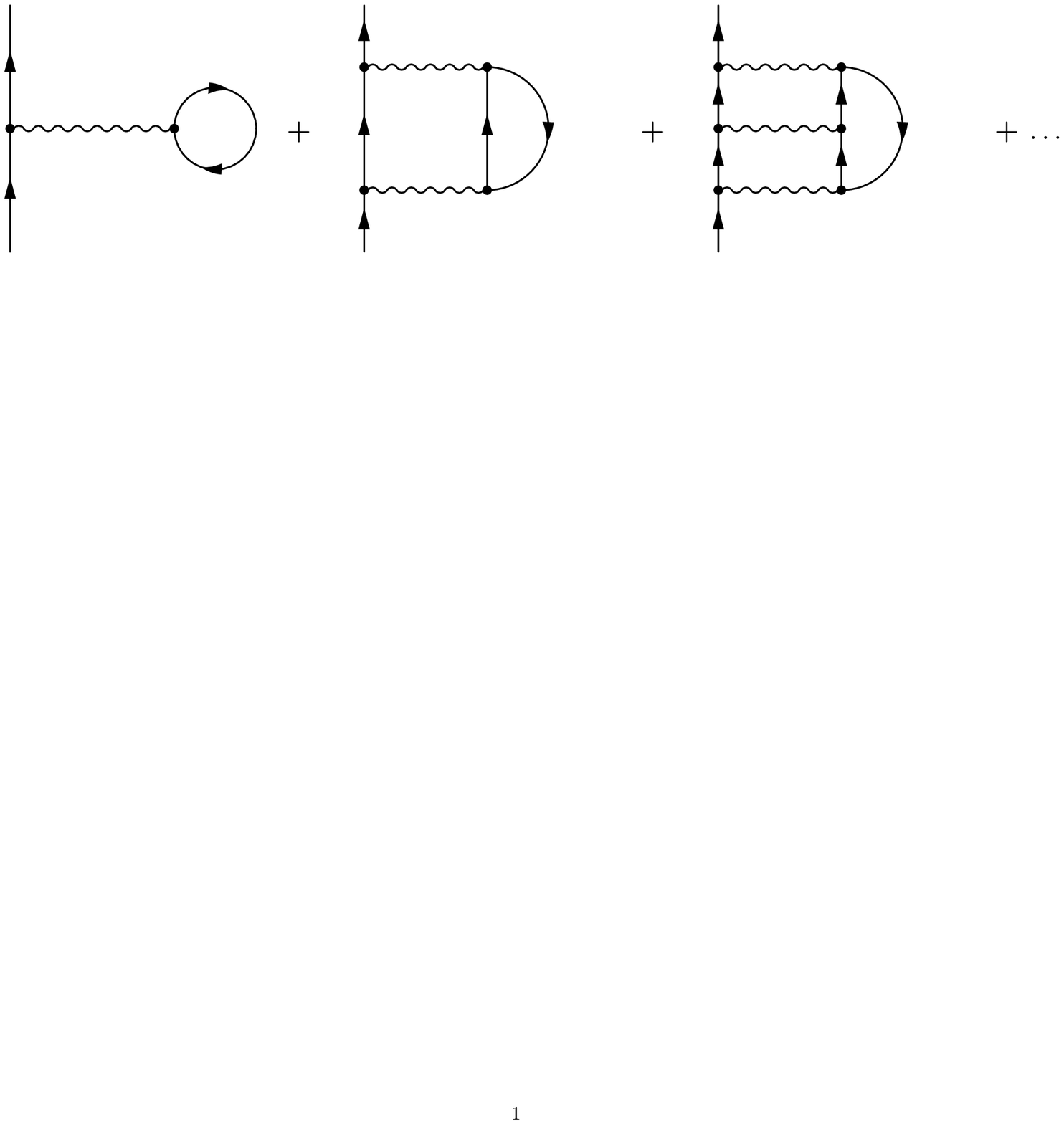}
\caption{Feynman diagrams illustrating the lowest-order contributions to 
  the self-energy in Dyson's equation, Eq.~(\ref{eq:dyson}), for the
  single-particle Green's function in the ladder approximation 
  (without exchange contributions). Plain
  lines represent fermion propagators and wavy lines an interaction.
\label{fig:ladder}}
\end{figure}

Retaining the lowest-order diagrams in the interaction, the
Hartree-Fock (HF) approximation can be derived with the SCGF formalism. 
Formally this means that in HF approximation the
$N$-particle Green's functions are (antisymmetrized) products of
single-particle Green's functions.
When going beyond the HF approximation, the dominant effect
should be multiple scattering processes with two participating
baryons. Under this assumption the so-called ``ladder approximation''
is obtained. The name originates from a diagrammatic
representation of this approximation for the single-particle Green's
function depicted in Fig.~\ref{fig:ladder}. In practice, the complete
two-particle $T$-matrix is introduced, describing an effective
two-particle interaction upon summing up all ``ladders'', see
Fig.~\ref{fig:ladder}. Note that the ladder approximation includes the
two contributions leading to the HF approximation.  The
equation for the $T$-matrix can be written as
\begin{equation}
\langle 1 2 | T| 3 4\rangle = \langle 1 2 | V| 3 4\rangle + \sum_{n n'} \langle 1 2|V |n n'\rangle {\mathcal G}_n {\mathcal G}_{n'} \langle n n'| T | 3 4\rangle~, 
\label{eq:tmatrix}
\end{equation}  
where $V$ represents the bare two-body interaction and a summation over
intermediate states has to be performed.

In the intermediate states of the $T$ matrix there can be no propagation of a
{\em particle} and a {\em hole} state. 
In ladder approximation the self-consistent
Green's function method thus sums up {\em particle-particle} and
{\em hole-hole} ladders to all orders. Physically
the ladder diagrams take into account multiple scattering of
{\em particles} and {\em holes} and can therefore, in
contrast to the HF approximation, describe the effect that
the strong short-range repulsion disfavors states when two particles
come very close to each other.
Reviews on the applications of the method to nuclear problems can be
found in \citet{Muther_00,Dickhoff_04}. The results for the EoS improve 
substantially in comparison to the HF approximation. 

The choice of the contributions included in the summation 
(ladder, ring, parquet)
has been more or less intuitive and was justified by the results.
It is not clear how the method can be
systematically extended in order to improve 
the results achieving self-consistency and avoiding
double counting. So far three-nucleon forces were not included in the ladder
approximation although their importance in nuclear systems is well known.
A method to include them in the SCGF formalism has
been developed recently in~\citet{Carbone_13}.

Another point is that low-temperature nuclear (and baryonic) matter is
unstable with respect to the formation of a superfluid or
superconducting state. This is the well-known Cooper instability
\cite{Cooper:1956zz}: a
fermionic many-body system with an attractive interaction tends to
form pairs at the Fermi surface. 
The instability shows up as a pole in the $T$-matrix
when the temperature falls below the critical temperature $T_c$ 
for the transition to the
superfluid/superconducting state, see, e.g., 
\citet{Thouless_60,Schmidt_90,Alm:1993zz,Stein_95}. 
Formally, the approach can be extended to
include the possibility of superfluidity and superconductivity by
introducing anomalous Green's functions describing pair
formation. However, practical SCGF calculations in the ladder
approximation are numerically already very demanding since the full
energy-dependence of the intermediate states has to be accounted
for. Therefore actual calculations are often performed  at
temperatures above $T_c$ and  extrapolated to zero temperature, see,
e.g., \citet{Frickthesis}. 

\paragraph{Brueckner-Hartree-Fock}
The Brueckner--Hartree--Fock (BHF) approximation is a widely used
microscopic many-body method developed by Brueckner, Bethe and others in the
1950's. Numerically it is less involved than the SCGF discussed above.
In a general framework it can be derived from
the Brueckner-Bethe-Goldstone hole-line expansion, truncated at the
two-hole-line level for the evaluation of the ground state energy. A
``hole-line'' has to be considered as the propagation of a
{\em hole}. This series can be roughly understood as an expansion in
density, see, e.g., \citet{FetterWalecka}. At low densities the
contributions with an increasing number of hole-lines should be
suppressed, thus ensuring good convergence.
It is generally assumed that the $n$-hole line contributions contain
the dominant part of the $n$-body correlations. Detailed and
pedagogical introductions can be found, e.g., in
\citet{Day_67,FetterWalecka,Baldo_01,Baldo_11}.

The BHF can be obtained from the SCGF approach in the ladder approximation
after some simplifications. The first one is to neglect the {\em
  hole-hole} contributions. The second one is to approximate the full
self-energy of the intermediate states by a quasi-particle
approximation.  The equation for the $T$-matrix,
Eq.~(\ref{eq:tmatrix}), then becomes an equation for the Brueckner
$G$-matrix by replacing $T$ with $G$,
except that the product of the two single-particle Green's functions in the
intermediate states is approximated for $T=0$ by 
\begin{equation}
{\mathcal G}_n {\mathcal G}_{n'} \to \frac{\mathcal{P}(n,n')}{\omega - E_n 
- E_{n'} + i\eta}
\end{equation}
where the Pauli operator $\mathcal{P}$ is nonzero only if both states lie outside the Fermi sea, 
i.e., they correspond to
two {\em particle} states. Note that the denominator 
does not
contain the full self-energy. In the quasi-particle approximation it has
the form of the non-interacting system.  

The single-particle energies $E$ are determined self-consistently from
the $G$-matrix in the following way (written in momentum space):
\begin{eqnarray}
E(\vec{k}) &=& \frac{\vec{k}^2}{2m} + U(\vec{k}) \: ,
 \\
U(\vec{k}) &=& \sum_{|\vec{k'}| < k_F} \langle k k'| G (E(\vec{k})
  + E(\vec{k'})) | k k'\rangle_A \: ,
\end{eqnarray}
where the subscript $A$ indicates that the matrix element has to be
antisymmetrized. 

The BHF method is not fully self-consistent. There remains some
freedom in determining the single-particle energy $E(\vec{k})$, see
\citet{Baldo_01} for a discussion. It has been shown by
\citet{Baldo:2001mv} that with a proper choice the three-{\em hole}
line corrections to the energy are small, indicating good convergence
of the series. The main problem is that the BHF method violates the
Hugenholtz-van Hove theorem~\cite{Hugenholtz_58} and, hence, is
thermodynamically inconsistent.
This theorem states that the
single-particle energy at the Fermi surface should equal the chemical
potential. Numerically the differences are of the order 10-20~MeV at
saturation density~\cite{Bozek_01}.

In BHF approximation, the description of nuclear matter
is improved substantially as compared with HF calculations. 
However, the saturation properties are not
satisfactorily reproduced with only two-body forces. It is generally 
believed that three-body forces are needed. In non-relativistic BHF
calculations they can be explicitly included, see,
e.g., \citet{Lejeune:2001bg,Zuo_02,Zuo:2002sg}. 
So far, there is no general consensus on how to improve BHF calculations 
systematically such that uncertainties are under control.

It is possible to extend the non-relativistic BHF formalism in order
to treat baryons in a special relativistic way. This Dirac-Brueckner
Hartree-Fock (DBHF) approach
\cite{Brockmann:1984qg,Horowitz:1986fr,Haar:1986ii,Brockmann:1990cn,Sammarruca_10} 
is computationally more involved than non-relativistic BHF calculations
and some ambiguities exist concerning the representation of the
in-medium $G$-matrix in terms of Lorentz invariants, see, e.g., the
discussion by \citet{GrossBoelting:1998jg}. However, the main advantage
is that an additional repulsion at high densities is obtained
since part of the three-body interaction is automatically generated
\cite{Brown:1985gt}.  Relativistic DBHF approaches also avoid a
problem of non-relativistic BHF calculations which can result in a superluminal
speed of sound at the high central densities of massive NSs.

\paragraph{Methods derived from the variational principle}
The Ritz-Raleigh variational principle is the basis for
variational approaches to the many-body problem.
It ensures that the trial ground
state energy %
\begin{equation}
E_{\rm trial} = \frac{\langle\Psi_{\rm trial}|H|\Psi_{\rm trial}\rangle}{\langle 
\Psi_{\rm trial}|\Psi_{\rm trial}\rangle} \: ,
\end{equation}
calculated from the system's Hamiltonian $H$
with a trial many-body wave function $\Psi_{\rm trial}$,
gives an upper bound for the true ground state energy of the
system. Correlations can be
embodied in the trial wave function, given for the $A$-baryon
system by 
\begin{equation}
\Psi_{\rm trial}(1, \cdots,A) = F(1,\cdots,A) \Phi_{\rm MF} (1,\cdots, A) \: .
\end{equation}
The operator $F$ is intended to transform the uncorrelated
wave function $\Phi_{\rm MF} (1,\cdots,A)$ to the correlated one.
$\Phi_{\rm MF}$ is an
antisymmetrized product (a Slater determinant) of single-particle wave
functions.  In practice, an ansatz is chosen for the trial wave function
and its parameters are varied in order to minimize
$E_{\rm trial}$. 
Once the trial wave function is determined, expectation values of other operators can be evaluated. 

The idea of the variational method is very simple and appealing. The
difficulty resides, however, in the details, namely the numerical
evaluation of the different expectation values. The first point is the
choice of the interaction Hamiltonian. As it stands, the method is
conceived for treating a local non-relativistic potential\footnote{See
  \citet{Walhout_96} for an attempt to generalize this method to
  relativistic systems within the path integral formalism.}. Thus,
some of the realistic potentials discussed in
Sec.~\ref{sec:fewbodyinteractions}, in particular those involving
energy-dependent meson exchange, cannot be used within this
approach. There are, however, potentials, which have been designed
specifically for variational methods. The most prominent example is
the series of Argonne $NN$-forces, see Sec.~\ref{sec:phenoforces}.
Most variational calculations include a three-body force in the
nuclear Hamiltonian.  Together with the Argonne forces, the Urbana
three-body forces are usually applied.  Further,
\citet{Gezerlis_13,Gezerlis_14} developed a local version of nuclear
interactions for Quantum Monte Carlo calculations from $\chi$EFT which
in principle is applicable.

The central task is to find a suitable ansatz for the correlation operator
$F$. At sufficiently low densities, two-body correlations
should be dominant in nuclear systems. This assumption
represents, for instance, the basis for the ladder approximation
discussed in Sec.~\ref{sec:scgf}. Within the variational methods
this assumption leads to an ansatz for the
two-body correlation operator $F_{2}$ which contains essentially the same
operational structure as the two-body interaction, see,
e.g., \citet{Fantoni_98,Muther_00,Carlson:2014vla}. Thus, it is written as a
sum of two-body operators, incorporating the nuclear spin-isospin
dependence, multiplied by radial correlation functions $f^{(m)}(r)$:
\begin{equation}
F_2(i,j) = \sum_m f^{(m)}(r_{ij}) \mathcal{O}^{(m)}(i,j) \: .
\label{eq:pair}
\end{equation}
The $f^{(m)}(r)$ are then determined by minimizing
$E_{\rm trial}$.  In practice, this is done employing different
techniques. 

In the nuclear context, Fermi-Hyper-Netted-Chain
(FHNC)~\cite{Fantoni_75,Pandharipande_79} calculations have 
proven to be efficient.  We note that up to now it is impossible to include
the complete operator structure of the most sophisticated Argonne
potentials in the correlation operators $F$. The spin-orbit
correlation, in particular, cannot be treated on the same footing,
since it cannot be chained. 

In coupled-cluster theory, which is based on ideas of \citet{coester_1958,coester_1960}, 
the correlation operator $F$ is represented in an exponential form
$ F = \exp(\hat{T}) $
with the cluster operator
\begin{equation}
 \hat{T} = \sum_{m=1}^{A} \hat{T}_{m}
\end{equation}
that is a sum of $m$-particle $m$-hole excitation operators $\hat{T}_{m}$.
The method, which is non-variational in practice, is utilized with great success in
quantum chemistry \cite{Bartlett:2007zza} and nuclear structure
calculations, see, e.g., \cite{Hagen:2007ew,Hagen:2010gd,Hagen:2012fb}. 
Early applications to nuclear matter can be found in \citet{Kummel:1978zz,day_1981}.
Nowadays, optimized chiral nucleon-nucleon interactions are implemented
\cite{Baardsen:2013vwa,Hagen_13}.

The variational ground state energy represents only an upper bound on
the exact ground state energy, and the deviation depends on the choice
of the trial wave function. The method of correlated basis functions
(CBF), see, e.g., \citet{Fantoni_98}, allows to improve on the variational
ground state. The idea is to add up 
second order perturbative
corrections to the ground state energy calculated with correlated
basis functions. The latter are determined by the variational
calculation from model basis functions. Taking only
$\Phi_{\mathrm{MF}}$ as model basis function, the usual variational
calculation would be recovered. Within CBF, other basis functions are
added, containing already some particle-hole excitations on the
initial wave function $\Phi_{\mathrm{MF}}$.

Another widely used method in nuclear physics to evaluate 
expectation values is the variational Monte-Carlo (VMC) approach,
see \citet{Guardiola_98} and \citet{Carlson:2014vla} for two reviews. Very
sophisticated calculations with this technique have been performed for
light nuclei, including two-body correlations, such as given in
Eq.~(\ref{eq:pair}), and triplet correlations, see,
e.g., \citet{Wiringa_14}. Since the computational effort
increases very rapidly with the number of nucleons, the EoS of
homogeneous nuclear matter, however, is extremely difficult to obtain.

\paragraph{Quantum Monte-Carlo}

Advancing computer technology has allowed for a rapid progress in
the application of
Monte-Carlo methods to nuclear systems in recent years. In
addition to the variational Monte-Carlo approach discussed in the previous section,
the ground state wave function and energy can be determined 
by evolving the many-body Schr\"{o}dinger equation in imaginary time.  
Monte Carlo sampling is thereby used to evaluate
the paths. In nuclear physics these methods suffer, however, from the
fermion sign problem and different approximations are
employed, for comprehensive reviews see
\citet{Guardiola_98,Carlson_12,Carlson:2014vla}.

The Green's function Monte Carlo (GFMC) method \cite{Carlson_87,Carlson_88}
gives very accurate results for light nuclei, but due to the nuclear
spin and isospin degrees of freedom, computing time increases exponentially with
the number of particles. Up to now, the largest systems treated are 
$^{12}\mathrm{C}$ and a system of 16 neutrons. The Auxiliary Field Quantum Monte
Carlo (AFQMC) approach \cite{Schmidt_99} introduces 
auxiliary fields by Hubbard-Stratonovich transformations
to sample the spin-isospin states.  This efficient
sampling allows for treating larger systems, with more than 100
nucleons. Finite size effects are expected to be small and AFQMC
calculations have been applied in the last years to homogeneous
matter, in particular neutron matter, but also to symmetric matter and
nuclear matter with hyperons, see Sec.~\ref{sec:theosmnm}.  

In spite of the recent progress, it is still not possible to perform
GFMC and AFQMC calculations with the full Argonne $v_{18}$ 
potential since some of the terms, again related to the spin-orbit
structure, induce very large statistical errors. Simplified potentials
have been developed~\cite{Pudliner_97,Wiringa_02} containing less
operators with readjusted parameters. 
Besides these two-body interactions, the Urbana three-body potentials are used. 
Recently, a local chiral potential has been
developed~\cite{Gezerlis_13} which is well suited for quantum Monte
Carlo techniques.

\paragraph{Chiral Effective Field Theory}
$\chi$EFT is very successful in describing nuclear
forces and has allowed to establish a link between the underlying
theory of QCD and nuclear physics, see
Sec.~\ref{sec:chiralBBforces}. Except for very light nuclei, where
direct numerical solutions of the Schr\"{o}dinger equation are possible,
these chiral forces are generally employed within standard
many-body techniques to  address  heavier nuclei
or homogeneous nuclear matter.  In recent years some effort has been
devoted to an alternative approach for homogeneous matter, namely
extending the idea of chiral perturbation theory directly to nuclear
matter calculations, i.e., developing an effective field theory (EFT)
for nuclear matter. Similar to nuclear forces in vacuum, pions and
nucleons are treated as explicit degrees of freedom and short-range
dynamics is comprised in local contact terms. The advantage of such an
EFT is that it establishes a power counting which allows to select at
a given order the relevant ones among the infinite number of
contributions and that one can determine an associated uncertainty, 
which is not possible in many other methods. 
The main difficulty resides in defining a well
adapted power counting scheme. 

For nuclear matter, the nuclear Fermi momentum $k_F$ enters as an
additional scale. It is considered as small, of the same order as the
pion mass. 
At saturation density it is approximately given as $k_F \approx 263$ MeV/c, which is 
indeed smaller than a typical hadronic scale.
Based on this assumption, different power counting schemes
have been developed. On the one hand, in the works of
\citet{Kaiser_02,Kaiser_03,Fritsch_05}, the vacuum chiral power counting has 
been applied directly to nuclear matter. 
On the other hand, \citet{Meissner_01,Oller_09,Lacour_09} argue
that a propagating nucleon in the medium cannot always be counted in
the standard way as $1/|\vec{k}|$ with $|\vec{k}|$ being a typical
nucleon three-momentum, but that there are ``non-standard'' situations
where it is to be counted as an inverse nucleon kinetic energy. In
practice, within this ``non-standard'' counting, certain classes of
two-nucleon diagrams have to be resummed. Another difference is that
these works include local nucleon-nucleon interactions
fixed by free nucleon-nucleon scattering in addition to nucleon
interactions mediated by pion exchange \cite{Oller_09,Lacour_09}.

\paragraph{Lattice methods}
The {\em  ab initio} approach to
solve QCD numerically becomes extremely complicated at finite densities.
The Fermion determinant in the medium turns complex
valued due to the appearance of the chemical potential. 
Consequently the integrals, which are evaluated in LQCD
with Monte Carlo methods, have no longer
positive weights.  Different approaches to avoid this
fermion sign problem have been suggested, e.g., reweighting
techniques \cite{Fodor:2001au}, the introduction of complex chemical
potentials \cite{deForcrand:2002ci}, or Taylor expansion
schemes \cite{Allton:2002zi}.  A crucial parameter in all of these
approaches is $\mu/T$,
which for these analyses is typically of magnitude one
or less.  However, this value is
significantly larger in the applications we are interested in,
in particular for NSs. Accordingly, no consistent cold matter 
or supernova EoS has been provided by LQCD so far.

In recent years lattice methods have been applied directly to nuclear
systems, see \citet{Lee:2008fa} for a review. In the so-called
``Nuclear Lattice Effective Field Theory'' (NLEFT), nucleons are
treated as point-like particles residing on the lattice sites. For the
interactions, EFT nuclear forces are employed consisting of nucleon
contact terms and potentially pion exchanges. These are represented on
the lattice as insertions on the nucleon world lines. Due to the
approximate $SU(4)$ spin-isospin symmetry of nuclear forces, NLEFT
suffers much less from the sign problem than lattice QCD. 
These methods
have been applied successfully to light and medium mass nuclei, see for
example~\citet{Epelbaum:2013paa} and \citet{Lahde:2013uqa},
and to dilute neutron matter up to roughly one tenth of
nuclear matter saturation density, see \citet{Lee:2008fa} and
references therein. For the moment, however, there are no computations
of denser systems. Lattice methods are used in the context of Quantum
Monte Carlo simulations, see~\citet{Wlazlowski_14} for a recent
example.

\paragraph{Perturbative QCD}
QCD is asymptotically free, viz. the coupling constant decreases logarithmically
with the energy~\cite{Gross:1973id,Politzer:1974fr}
and therefore, it is addressable by perturbative methods if 
the coupling constant turns small enough.
This has been exploited to describe the thermodynamics of dense deconfined
quark matter at finite 
temperature \cite{Freedman:1976xs,Freedman:1976dm,Freedman:1976ub}.
Recent efforts aimed to account for all second-order effects in an expansion
of the thermodynamic pressure of deconfined QCD~\cite{Kurkela:2009gj}.
This procedure gives valuable insights into the high density limit
of QCD and therefore provides an important constraint 
for the asymptotic behavior of the EoS at baryonic chemical potentials of several GeV~\cite{Kurkela:2014vha}.
However, results from matching the EoS obtained within this approach directly to
a nuclear matter EoS as suggested by \citet{Kurkela:2009gj} do not provide
a conclusive answer as in this domain non-perturbative
features cannot be neglected.

\paragraph{Dyson-Schwinger approach}
\label{sec_ds}
The Dyson-Schwinger (DS) formalism is a non-perturbative approach to
analyze QCD.  It starts from a generating functional (the partition
function of QCD). From there, coupled integral equations, the DS
equations, are derived for the $n$-point Schwinger functions of the
theory.  Formally, this approach is similar to that of self consistent
Green's functions we discussed in Sec.\ref{sec:scgf}, but now based on
the QCD Lagrangian.  A further successful proving ground for this
approach is QED.  Like in any many-body theory, every QCD Schwinger
function couples to further Schwinger functions of higher order.  This
implies an infinite hierarchy of DS equations which can be solved practically
only by introducing truncation schemes.  There is no strict
prescription how to truncate without erasing inherent properties
(e.g., symmetries) of the original theory. The truncation scheme
defines a specific model which then can be compared to experimental
data and subsequently used to predict observables.  The theoretical
framework for vacuum and in-medium studies and numerous applications
are reviewed in detail in, e.g.,
\citet{Alkofer:2000wg,Roberts:2000aa,Roberts:2012sv}.  Despite the
number of successful vacuum studies at zero and finite temperature it
has been used only rather recently to compute EoSs of dense homogeneous
quark matter in the deconfined phase 
\cite{Chen:2008zr,Klahn:2009mb,Chen:2011my,Chen:2015mda}.  
Prominent topics are superconducting
phases, see, e.g., \citet{Nickel:2006vf,Alford:2007xm}, and the role
of strange quarks \cite{Nickel:2006kc,Muller:2013pya}.  Further DS
studies investigate the critical line in the QCD phase diagram, see, e.g.,
\citet{Qin:2010nq,Fischer:2011mz,Bashir:2012fs,Gutierrez:2013sta}.
Although the DS approach promises insights from a QCD based framework,
no EoS has been obtained to date that covers the whole parameter space
required to perform, e.g., CCSN simulations. 
However, \cite{Klahn:2015mfa} have shown that both the NJL model
and the thermodynamic
bag model, see \ref{sec:qm}, can be understood as solutions of in-medium DS equations in
rainbow approximation assuming a contact interaction for the gluon propagator.

\subsubsection{Empirical/phenomenological approaches}
\label{sec:pheno}

Phenomenological approaches to describe dense matter are characterized
by the use of effective inter-particle interactions instead of
realistic forces.  They usually have a rather simple functional form
in order to allow them to be used in several applications, i.e., not
only uniform matter but in many cases also finite nuclei. However,
their structure can be guided by symmetry principles, power counting
arguments or insights from {\em ab-initio} 
approaches.
Effective Hamiltonians can be derived from more fundamental forces
using the formalism of density-matrix expansions
\cite{Negele71b,Dobaczewski:2010qp,Stoitsov:2010ha}.  In general, the
actual parameters of the interactions are not directly calculated from
underlying fundamental theories but they are determined empirically by
fitting to \mbox{(pseudo-)} observables that are calculated in certain
approximations of many-body theory. The interaction and the considered
model space are not independent. Systematic extensions of the
empirical approaches are not straightforward and usually require a
refit of the model parameters.

Most empirical descriptions of dense matter rely on the mean-field (MF)
approximation or use the closely related language of energy density
functionals. Originally, the mean-field approximation corresponds to
the Hartree approximation of the many-body state, i.e., a simple
product of single-particle wave functions, but the term is often used
to denote the HF approximation, too. The constituents are
considered as quasi-particles with modified properties,
e.g., effective masses that are different from their rest masses in
the vacuum. Quantities such as the energy density or the pressure of
the system can be expressed as functionals of the single-particle
densities and an EDF is derived. It can be
used as a starting point for the development of more refined EDFs that
take into account features such as exchange and correlation effects
going beyond the mean-field approximation.
It is well known from the basic theorems in density functional theory
\cite{Hohenberg:1964zz,Kohn:1965zzb,Kohn:1999zza,DrGr1990,FiNoMa2003}
that an EDF exists which yields the exact energy of the system's
ground state but its explicit form is not known. With suitable
extensions of mean-field EDFs guided by empirical information one can
try to come close to the exact EDF of the system, 
even if the interaction itself is not completely known.

A further set of phenomenological EoSs can be characterized
as purely parametric models which by itself are not based on
a description of the interaction of particles. Instead, a parameterized functional,
typically of the energy density, is either assumed or fitted to microscopically motivated
EoS. These models are useful to analyze astrophysical data. 
Examples are single and piecewise polytrope fits for nuclear matter \cite{Read_08}
and a linear fit for quark matter~\cite{Zdunik:2000xx,alford2013}.

\paragraph{Hadronic matter}
\label{sec:mf_models}

Self-consistent mean-field (MF) models 
are well developed. They are successful  in describing the properties of systems composed 
of nucleons and were first used for bulk nuclear matter. Nowadays
they are mainly applied in the description
of finite nuclei, see, e.g., the review by \citet{Bender:2003jk}.
The approaches can be divided into two main classes:
non-relativistic and relativistic models. 
The main distinction between them are 
the specific form of the interaction and the resulting
dispersion relation of the quasi-particles.

Non-relativistic approaches generally start from a  Hamiltonian 
\begin{equation}
 \hat{H} = \hat{T} + \hat{V}
\end{equation}
for the many-body system that contains the usual kinetic contribution 
\begin{equation}
 \hat{T} 
 = \sum_{i} \frac{\hat{p}_{i}^{2}}{2m_{i}}
\end{equation}
and a potential term $\hat{V}$ that varies from model to model. Usually it is given
as a sum
\begin{equation}
\hat{V} = \sum_{i < j} \hat{V}_{ij} + \sum_{i<j<k} \hat{V}_{ijk}
\end{equation}
of two-body ($\hat{V}_{ij}$) and three-body ($\hat{V}_{ijk}$) interactions. 
The latter are required in
order to reproduce the empirical saturation properties of nuclear matter.
The energy of the system is calculated 
under certain assumptions for the form of the many-body wave-function, usually within the HF
approximation. Pairing effects can be considered in the Hartree-Fock-Bogoliubov 
approximation. 
However, the original model interaction $\hat{V}$ cannot always be used in the 
pairing channel and a suitable
pairing interaction has to be specified separately.
Non-relativistic approaches are in danger to fail in the description of
dense matter at high densities, e.g., the EoS can become superluminal.

Relativistic models are commonly formulated in a field-theoretical language by 
defining a Lagrangian density $\mathcal{L}$ that serves as the starting point 
in order to derive the
field equations of the interacting particles. They constitute a set of 
coupled equations that
has to be solved self-consistently. 
Expressions for the energy density and pressure
are obtained from the energy-momentum tensor.

The foremost application of MF models is the description
of finite atomic nuclei but nuclear matter properties are easily obtained 
once the parameters of the 
effective model interaction are determined.
Depending on the selection of observables and preferences
in the fitting of the effective interaction, 
different parameterizations are obtained. For the most common MF approaches, several
hundred parameter sets are available in the literature. In the following, 
the most-used MF models, distinguished by the choice of the interaction, are considered.

\begin{itemize}
\item Mean-field models with Skyrme-type interactions

An effective zero-range interaction for HF calculations was introduced
by \citet{Sk1956,Skyrme:1959zz}.
After the pioneering calculations of nuclei by
\citet{VaBr1970,Vautherin:1971aw,Beiner:1974gc,Brack:1974gy},
it became very popular and found widespread use, see, e.g., the review article by
\citet{Stone:2006fn}.
The basic form of the Skyrme interaction between two nucleons $1$ and $2$ can be written as 
\begin{eqnarray}
\label{eq:V_Skyrme}
  \lefteqn{\hat{V}^{\rm (Skyrme)}_{12} = t_{0} (1+x_{0}\hat{P}_{\sigma}) \delta({\bf r}_{12})}
 \\ \nonumber & & 
 + \frac{1}{2} (1+x_{1}\hat{P}_{\sigma}) \left[ \hat{\bf k}^{\dagger} \delta({\bf r}_{12})
 +  \delta({\bf r}_{12}) \hat{\bf k}^{2} \right]
 \\ \nonumber & & 
 + t_{2} (1+x_{2}\hat{P}_{\sigma}) \hat{\bf k}^{\dagger} \cdot  \delta({\bf r}_{12}) \hat{\bf k}
 \\ \nonumber & & 
 + \frac{1}{6} t_{3} (1+x_{3}\hat{P}_{\sigma}) \delta({\bf r}_{12}) \varrho^{\alpha} ({\bf R}_{12})
 \\ \nonumber & & 
 + i W_{0} \left( \hat{\bf \sigma}_{1} + \hat{\bf \sigma}_{2} \right) 
 \cdot \hat{\bf k}^{\dagger}\delta({\bf r}_{12}) \hat{\bf k}
\end{eqnarray}
with parameters $t_{i}$, $x_{i}$, $\alpha$ and $W_{0}$. In Eq.~(\ref{eq:V_Skyrme})
$P_{\sigma} = (1+\hat{\bf \sigma}_{1} \cdot \hat{\bf \sigma}_{2})/2$ denotes the spin-exchange
operator, $\hat{\bf k} = (\nabla_{1}-\nabla_{2})/(2i)$ is the relative momentum and $\varrho$ the
total nucleon density. The relative
coordinate and center-of-mass coordinate are defined by ${\bf r}_{12}={\bf r}_{1} - {\bf r}_{2}$
and ${\bf R}_{12} = ({\bf r}_{1} + {\bf r}_{2})/2$, respectively.
The contribution with factor $t_{3}$ is a generalization that originates from an explicit three-body
term
\begin{equation}
\label{eq:v123}
 \hat{V}_{123} = t_{3} \delta({\bf r}_{1}-{\bf r}_{2}) \delta({\bf r}_{2}-{\bf r}_{3}) 
\end{equation}
in the original Skyrme interaction and was converted to a density dependent two-body interaction.
The parameter $\alpha$ controls the strength of the repulsion. 
The original three-body interaction (\ref{eq:v123}) corresponds to $\alpha=2$.
The contribution with factor $W_{0}$
generates the spin-orbit interaction in systems that are not spin-saturated. 
For the description of nuclei, contributions from the Coulomb interaction 
have to be considered in addition.
The potential (\ref{eq:V_Skyrme}) 
can be seen as an expansion in powers of the relative momentum ${\bf k}$.
Since it stops at second order, the interaction cannot be applied reliably 
in cases where the momenta 
of the nucleons reach high values, e.g., in nuclear matter at densities 
substantially above saturation.

Evaluating the energy density from the Hamiltonian in HF approximation yields
an EDF that depends on a number of single-particle densities and their
spatial derivatives. 
Since the interaction is of zero-range, exchange contribution are easily obtained
and only local densities appear in the EDF. Besides the usual single-particle number densities $n_{i}$, 
the kinetic-energy densities $\tau_{i}$, the currents ${\bf j}_{i}$, 
the spin-orbit densities ${\bf J}_{i}$, the spin densities ${\bf \sigma}_{i}$ etc.\ are relevant, see, e.g.,
\citet{Bender:2003jk} for details. 
In applications to nuclear matter, currents, spin densities and
spatial derivative of all single-particle densities vanish. The energy density becomes a more-or-less
simple functional in fractional powers of the number densities
\cite{Dutra:2012mb}. Obviously, an extrapolation to high 
densities can lead to divergences.
Several extensions of the standard Skyrme functional have been proposed, see, e.g.,
\citet{Margueron:2012qx,Margueron:2010zz,Margueron:2009rn,Margueron:2009jf,Lesinski:2007ys,Bender:2009ty,Dutra:2012mb,Hellemans:2011aa,Chamel:2011aa,Davesne2015}.
We note that not every Skyrme-type EDF can be derived from simple two- and three-body potentials in a MF approximation.
A number of well-calibrated parameterizations were proposed recently
\cite{Chabanat:1997qh,Chabanat:1997un,Agrawal:2005ix,Stone:2005mz,Goriely_09,Goriely:2009he,Kortelainen:2010hv,Kortelainen:2011ft,Washiyama:2012yc,Goriely:2013nxa,Goriely:2013xba,Kortelainen:2013faa}.
They predict many properties of nuclei
close to experimental values with rather small deviations. 
The performance of 240 Skyrme parameterizations under nuclear matter 
constraints was studied by \citet{RikovskaStone:2003bi,Dutra:2012mb}. 
Only a few satisfy all criteria that were selected by the authors. Some Skyrme forces show instabilities 
\cite{Chamel:2010wr,Kortelainen:2010wb,Hellemans:2013bza,Navarro:2013bda,Pastore:2014yua}
under particular conditions that may be cured with appropriate modifications 
of the EDF.

\item Mean-field models with Gogny interaction

Instead of a 
zero-range force as in the Skyrme case, the use 
of finite-range interactions
is an established approach in MF models. A sum of two Gaussians was suggested by
\citet{Brink:1967zz} for HF calculations. A review of phenomenological interactions in early
HF models is given by \citet{Quentin:1978ia}. In order to obtain quantitatively reasonable results, 
a density-dependent effective two-body interaction was added by Gogny which lead to the
present form
\begin{eqnarray}
\label{eq:V_Gogny}
  \lefteqn{\hat{V}^{\rm (Gogny)}_{12} =\sum_{j=1,2} \exp \left(-\frac{{\bf r}_{12}^{2}}{\mu_{j}^{2}} \right)}
 \\ \nonumber & & 
  \times
 \left( W_{j} + B_{j} \hat{P}_{\sigma} - H_{j} \hat{P}_{\tau} - M_{j} \hat{P}_{\sigma} \hat{P}_{\tau}\right)
 \\ \nonumber & & 
 + t_{3} (1+x_{0}\hat{P}_{\sigma}) \delta({\bf r}_{12}) \varrho^{\alpha} ({\bf R}_{12})
 \\ \nonumber & & 
 + i W_{ls} \left( \hat{\bf \sigma}_{1} + \hat{\bf \sigma}_{2} \right) 
 \cdot \hat{\bf k}^{\dagger}\delta({\bf r}_{12}) \hat{\bf k}
\end{eqnarray}
with parameters $\mu_{j}$, $W_{j}$, $B_{j}$, $H_{j}$, $M_{j}$, $t_{3}$, $\alpha$ and $W_{ls}$ and
the isospin-exchange operator $P_{\tau} = (1+\hat{\bf \tau}_{1} \cdot \hat{\bf \tau}_{2})/2$
\cite{Decharge79}.
The density-dependent and the spin-orbit contribution have the form of the 
corresponding terms in the Skyrme 
interaction, although with a different notation of the parameters. Due to the finite-range part 
in the Gogny interaction, it is technically more involved to consider the exchange contributions to the
energy density. On the other hand, divergences of a zero-range interaction are avoided in calculations
involving pairing channels.
Due to the more involved numerical calculations, there are only few parameterizations 
of the Gogny interaction that are used in practice \cite{goriely_08}. 
A collection of these parameter sets with a
comparison to predictions of nuclear matter properties can be found in 
the work of \citet{Sellahewa:2014nia}.

\item Relativistic mean-field and Hartree-Fock models

In relativistic approaches to nuclear matter and finite nuclei, a
field-theoretical formalism is employed where nucleons are represented
by Dirac four spinors $\psi_{i}$ and the nucleon-nucleon interaction
is modeled by an exchange of mesons. This description, called quantum
hadro dynamics (QHD), was originally seen as a fully field theoretical
approach
\cite{FetterWalecka,Walecka:1974qa,Chin:1974sa,Serot:1984ey,Serot:1992ti}
and treated with the respective formalism. Later, the view of an
effective description to be applied in rather simple approximations
prevailed since nucleons as composite objects cannot be considered as
fundamental degrees of freedom.  The common starting point in QHD
models is a Lagrangian
\begin{equation}
\label{eq:L}
 \mathcal{L} = \mathcal{L}_{\rm nuc} + \mathcal{L}_{\rm mes} + \mathcal{L}_{\rm int}
\end{equation}
that contains contributions of nucleons $i$
\begin{equation}
 \mathcal{L}_{\rm nuc} = \sum_{i=n,p} \bar{\psi}_{i} 
 \left( \gamma_{\mu} i \partial^{\mu} - m_{i}\right) \psi_{i}
\end{equation}
with rest mass $m_{i}$, of free mesons $\mathcal{L}_{\rm mes}$ 
and an interaction term $\mathcal{L}_{\rm int}$. 

In early versions of the model only isoscalar mesons such as the
(Lorentz-) scalar $\sigma$-meson and the (Lorentz-) vector
$\omega$-meson were considered in order to model the long-range
attraction and short-range repulsion of the nuclear interaction,
respectively, in symmetric nuclear matter. Isovector mesons were
added for the description of neutron-proton asymmetric systems.
In most models, the  vector isovector $\rho$-meson 
is considered, but an
isospin-dependent splitting of the neutron and proton Dirac
effective masses is only obtained when a  scalar isovector $\delta$-meson is
included. In contrast, the Landau effective masses
are different in asymmetric matter even without a $\delta$-meson.
Pseudoscalar mesons, such as the pion, or pseudovector
mesons are relevant in models that treat exchange effects explicitly
\cite{Lalazissis:2009nc}.  The mesons in the QHD approach share the
same quantum numbers with their counterparts observed in experiments,
however, they have to be seen as effective fields in the model that
serve to capture the essential features of the strong interaction in
the medium.  With the standard choice of mesons, the contribution of
free mesons to the Lagrangian (\ref{eq:L}) reads
\begin{eqnarray}
 \mathcal{L}_{\rm mes} & = & \frac{1}{2} \left( \partial_{\mu} \sigma \partial^{\mu} \sigma 
 - m_{\sigma}^{2} \sigma^{2}  \right) 
 \\ \nonumber & & 
 + \frac{1}{2} \left( \partial_{\mu} \vec{\delta} \cdot \partial^{\mu} \vec{\delta}
 - m_{\delta}^{2} \vec{\delta}\cdot\vec{\delta}  \right)
 \\ \nonumber & & 
 - \frac{1}{4} G_{\mu\nu}G^{\mu\nu} + \frac{1}{2} m_{\omega}^{2} \omega_{\mu} \omega^{\mu}
 \\ \nonumber & &
 - \frac{1}{4} \vec{H}_{\mu\nu}\cdot \vec{H}^{\mu\nu}
 + \frac{1}{2} m_{\rho}^{2} \vec{\rho}_{\mu} \cdot \vec{\rho}^{\mu}
\end{eqnarray}
with the usual field tensors of the vector mesons 
$G_{\mu\nu}  =  \partial_{\mu} \omega_{\nu} - \partial_{\nu} \omega_{\mu}$
and 
$\vec{H}_{\mu\nu}  =  \partial_{\mu} \vec{\rho}_{\nu} - \partial_{\nu} \vec{\rho}_{\mu}$.
In most approaches, it is assumed that mesons couple minimally to nucleons leading to
an interaction contribution of the form
\begin{eqnarray}
\label{eq:L_int}
 \mathcal{L}_{\rm int} & = & - \sum_{i=n,p} \bar{\psi}_{i} 
 \left[ \gamma_{\mu} \left( g_{\omega} \omega^{\mu} + \vec{\tau} \cdot g_{\rho} \rho^{\mu} \right)
 \right. \\ \nonumber & & \left.
 + g_{\sigma} \sigma + g_{\delta} \vec{\tau} \cdot \vec{\delta} \right]
 \psi_{i}
\end{eqnarray}
where $g_{i}$ ($i=\omega$, $\sigma$, $\rho$, $\delta$) denote the empirical coupling constants.
Their values are obtained by fitting to properties of nuclear matter or finite nuclei.
The coupling to scalar mesons modifies the Dirac effective mass 
$m_{i}^{\ast}$ of the nucleons.  
It is essential in order to obtain a realistic spin-orbit splitting in nuclei.
From the Lagrangian (\ref{eq:L}) the field equations for nucleons and mesons are derived. 
They have to be solved self-consistently, usually in the 
relativistic mean-field (RMF) approximation where meson fields
are treated as classical fields and negative-energy states of 
the nucleons are neglected (no-sea approximation). Scalar and vector densities of nucleons appear as source terms for the mesons.
Finally, a covariant energy density functional (CEDF) is obtained.

The basic version of QHD as discussed above can qualitatively describe the feature of
saturation in nuclear matter. It results from a competition of attractive scalar 
and repulsive vector self-energies,
$S_{i}$ and $V_{i}$, in the relativistic dispersion relation
\begin{equation}
E_{i} = \sqrt{p^{2}+(m_{i}^{\ast})^{2}} + V_{i}
\end{equation}
for a nucleon $i$ with momentum $p$
and Dirac effective mass $m_{i}^{\ast} = m_{i}-S_{i}$. With increasing density of the
medium, the scalar potential $S_{i}$ rises more slowly than the vector
potential $V_{i}$.  The EoS becomes very stiff, nevertheless, the
speed of sound does not exceed the speed of light.  A quantitative
description of nuclear matter and nuclei requires the extension of the
simple Lagrangian density (\ref{eq:L}) in order to simulate a
medium-dependent effective interaction. Several options have been
explored in the literature.

In early extensions of the model, non-linear (NL) self-interactions of the mesons were 
considered  by adding a contribution of the form
\begin{equation}
 \mathcal{L}_{\rm nl}  =  - \frac{A}{3} \sigma^{3} - \frac{B}{4} \sigma^{4} + \frac{C}{4}
 \left(\omega_{\mu} \omega^{\mu}\right)^{2}
\end{equation}
to Eq.~(\ref{eq:L}). Cubic and quartic terms of the
$\sigma$-meson were introduced by \citet{Boguta:1977xi}
and very satisfactory results were obtained, see, e.g.,
\citet{Reinhard:1986qq,Rufa:1988zz}. The addition of the quartic 
$\omega$ term by \citet{Sugahara_94} 
was motivated by comparison 
of the scalar and vector potentials with DBHF.
Later, self-couplings of isoscalar and isovector mesons were used to
modify the isospin dependence of the EoS 
\cite{Mueller:1996pm,Furnstahl:1996wv,ToddRutel:2005zz}.

Instead of adding explicit new terms to the Lagrangian density
(\ref{eq:L}), the coupling constants $g_{i}$ in (\ref{eq:L_int}) are
replaced by functionals $\Gamma_{i}$ of the nucleon fields in RMF
models that were inspired by results of DBHF calculations of nuclear
matter. Here it is found that effective density-dependent (DD)
nucleon-meson couplings can be extracted from the medium-dependent
nucleon self-energies. Usually, a dependence of the couplings
$\Gamma_{i}$ on the vector density\footnote{Note that despite the name,
  the vector density is a Lorentz scalar.} 
$n_{v}=\sqrt{j_{\mu}j^{\mu}}$ is assumed, which is defined in
a covariant way with the nucleon current $j_{\mu}$.  In the rest frame
  of a nucleus or nuclear matter, $n_{v}$ is
  identical to the baryon number density $n_{B}$. The
  density dependence of the couplings leads to so-called
``rearrangement'' contributions to the vector self-energies
\cite{Fuchs:1995as}.  This is essential in order to obtain a
thermodynamic consistent model.  The functional form for the
density-dependence of the couplings with rational and exponential
functions was suggested by a comparison to DBHF results 
in  an early parameterization of the DD-RMF model that was fitted to
binding energies of nuclei \cite{Typel:1999yq}. 
This approach has been followed in several publications 
\cite{Niksic:2002yp,Long:2003dn,Niksic:2004az,Typel:2005ba,Typel:2009sy,RocaMaza:2011qe}.
Alternative functions have been considered,  e.g., by
\citet{Gogelein:2007qa}. The density dependence
of the meson-nucleon couplings were also directly derived from the
nucleon self-energies \cite{Hofmann00,Hofmann:2000vz,Gogelein:2007qa}
where the momentum dependence of the self-energies was mapped to an
effective density dependence.

Details on applications of the NL and DD-RMF models can be found in a
number of reviews, see, e.g.,
\citet{Reinhard:1986qq,Rufa:1988zz,Reinhard:1989zi,Ring:1996qi,Serot:1997xg,Bender:2003jk,Furnstahl:2003cd,Vretenar:2005zz,Niksic:2011sg}.
The models are employed without taking 
the antisymmetrization of the many-body wave function  explicitly into account.
 The finite range of the effective interaction mediated by
the meson exchange requires extra computational efforts in order to
handle fully antisymmetrized many-body states similar as in
non-relativistic Gogny HF calculations. 
In addition, an entirely new
parameter set for the couplings 
has to be determined. Nevertheless,
relativistic Hartree-Fock (RHF) or Hartree-Fock-Bogoliubov models were
implemented \cite{Meng:2005jv,Long:2005ne,Long:2007dw,Long:2008fm}.

Treating the basic scalar-vector models as relativistic local quantum field theories, it was found that two-loop corrections lead to huge contributions.
It was concluded that the loop expansion does not converge and that the composite nature of the nucleons has to be respected, e.g., by introducing
non-local interactions as represented by form factors, see, e.g., \citet{Prakash:1991fx} and references therein.

Some generalizations of the RMF model extend the form of the
nucleon-meson interaction from minimal couplings to couplings of the
meson fields to derivatives of the nucleon fields.  An early version
is the model by \citet{Zimanyi:1990np} where scalar derivative
couplings were introduced that could be transformed to particular
nonlinear couplings of nucleons to the $\sigma$ meson.  General
first-oder derivative couplings were considered by
\citet{Typel:2002ck,Typel:2005ba}.  The approach was extended to
derivatives of arbitrary high order by
\citet{Gaitanos:2009nt,Gaitanos:2011yb,Gaitanos:2012hg} in the nonlinear
derivative (NLD) coupling model, and studied in various versions
\cite{Chen:2012rn,Chen:2014xha,Antic:2015tga,Gaitanos:2015xpa}. The
main feature of the NLD approach is the dependence of the nucleon
self-energies not only on the density but also on the energy or
momentum as in DBHF calculations. 
As a consequence, the momentum
dependence of the nucleon
optical potential in nuclear matter, which is extracted from the fitting of
proton-nucleus scattering data in Dirac phenomenology
\cite{Hama:1990vr,Cooper:1993nx},
can be reproduced up to nucleon energies of about 1~GeV.

The explicit appearance of meson fields in the Lagrangian density of
RMF approaches is suppressed in so-called relativistic point-coupling
(PC) models
\cite{Nikolaus:1992zz,Rusnak:1997dj,Burvenich:2001rh,Zhao:2010hi}.
Here, four-nucleon contact terms, including powers and derivatives
thereof, appear with free prefactors that need to be determined. They
can be seen as the result of expressing solutions of the meson field
equations in the QHD approach as functions of the relativistic source
densities and their spatial derivatives.  This resembles the
non-relativistic Skyrme HF approach.  A systematic expansion of
$\mathcal{L}$ in various densities, currents and their derivatives is
possible by using power counting arguments from principles of
effective field theory \cite{Furnstahl:2001un}. The form and
parameters of the PC approach can also be constrained by in-medium
$\chi$EFT \cite{Finelli:2002na,Finelli:2003fk,Finelli:2005ni}.

Nuclear matter characteristics of 263 RMF parameterizations  
were compared recently in \cite{Dutra:2014qga}. Similar as in the
case of Skyrme HF models, only a very small number is consistent 
with all nuclear matter constraints considered in that publication.

\item Quark-meson coupling model

An approach closely related to the previously discussed
relativistic descriptions of matter and nuclei is the quark-meson
coupling (QMC) model
\cite{Downum:2006re,RikovskaStone:2006ta,Thomas:2013sea,Whittenbury:2013wma}.
It explicitly considers nucleons as bound states of quarks which
couple to mesons in the surrounding medium. This leads to a polarization of the nucleon and the resulting
mass shift is calculated self-consistently. It can be expressed as a
polynomial in the $\sigma$ meson field similar as in NL-RMF
models. Properties of matter are calculated in the HF
approximation, including pions in addition to the standard 
scalar and vector mesons.  With a small number of parameters, 
results of similar quality as in (non)relativistic 
mean-field models or EDFs are obtained.

\item Other approaches to the nuclear energy density functional 

Besides the models discussed above that dominate the applications to the EoS, a number of independent alternative
approaches were developed in the past. Instead of non-relativistic
effective interactions of the Skyrme or Gogny type, other forms were
investigated, e.g., a density-dependent separable monopole (SMO)
interaction \cite{RikovskaStone:2002hh} or three-range Yukawa (M3Y)
type interactions \cite{Nakada:2003fw}.  The phenomenologically inspired approach
of \citet{Fa1998,Fayans:1999ts,Fayans:2001fe} exploits the
quasi-particle concept of Migdal's theory of finite Fermi systems.
The Barcelona-Catania-Paris (-Madrid) (BCP or BCPM) EDFs
\cite{Baldo:2004dx,BaScVi2008,Baldo:2012hw} are constructed by
interpolating between parameterizations of BHF results, obtained with
realistic nucleon-nucleon potentials, for the EoS of symmetric nuclear
matter and neutron matter.  Adding appropriate surface and spin-orbit
contributions, an excellent description of finite nuclei is obtained
with only a small number of parameters.

\end{itemize}

\paragraph{Quark Matter}
\label{sec:qm}

A proper QCD-based description of strongly interacting matter, in
particular in vicinity of the predicted deconfinement phase
transition, is desirable but currently not available since the theory
is challenging to solve at finite chemical potentials.  With a few
exceptions, the prevailing approach for the hadron-quark transition
region is to describe both phases separately and to interpolate in
between in terms of a phase transition construction, see
Sec.~\ref{sec:PT}.  Therefore, quark matter (QM) in the following has
to be understood as {\em  deconfined quark matter}.  In this phase one
can think of quarks as actual particles or quasi-particles with no
particularly complicated behavior or confinement properties.  Then it
is not surprising that typical approaches to describe quark matter
show many similarities to RMF models for nuclear matter.  To model
quark matter correctly it is important though to understand and to
account for the confinement mechanism in order to eventually
understand the phase diagram of strongly interacting matter in the
language of QCD. Therefore we briefly review corresponding
developments as far as they concern the EoS of dense matter.

\begin{itemize}
\item{Thermodynamic Bag Model}

The simplest, but still widely applied model for QM is a limiting case
of the MIT bag model~\cite{Chodos74}, which has been originally
developed to describe hadrons as quark bound states of finite size.
Confinement in this model is accomplished by endowing the finite
region with a constant energy per unit volume, $B$.  A special case is
a highly excited hadron in which quarks would then behave as an ideal
gas.  The latter idea has been followed to describe a system of
homogeneous, deconfined quark matter~\cite{Farhi:1984qu}.  The EoS is
that of an ideal Fermi gas of three quark flavors, where the bag
constant $B$ is added to the total energy density and subtracted from
the pressure in order to maintain thermodynamic consistency.  $B$ can
be understood as the pressure difference of confined and deconfined
quarks in vacuum.  The value of $B$ can be determined from more
sophisticated models~\cite{Cahill:1985mh}.  
The bag constant arises not solely due to
deconfinement but rather from the breaking of chiral symmetry.
Consequently it is density dependent~\cite{Buballa:1998pr} as well as
flavor dependent~\cite{Buballa:2005}.  Already in perturbative QCD the
thermodynamic bag models treatment of quarks as free non-interacting
fermions does not hold and requires
corrections~\cite{Freedman:1976ub,Fraga:2001id}, which can be
generalized to a simple power series expansion of the pressure in the
quark chemical potential~\cite{Alford:2004pf}.  Similarly, a
phenomenologically motivated expansion considers diquark
contributions to the pressure~\cite{Alford:2002rj}.  Recently, an
extension of the thermodynamic bag model has been suggested which
accounts for the breaking of chiral symmetry and the 
influence of vector interactions~\cite{Klahn:2015mfa}.

\item{Nambu--Jona-Lasinio type models and extensions}

One of the prominent features of QCD is the dynamical breaking of chiral
symmetry as the mechanism which generates most of the
hadrons masses.
The idea that the nucleon mass can be understood as the self-energy
of a fermion in analogy to the energy gap of a superconductor has
been developed at a time where the notion of quarks 
did not even exist~\cite{Nambu_61a,Nambu_61b}.
The NJL model is based on a Lagrangian for
a fermion field with a quartic, chirally symmetric local 
interaction of the form,
\begin{eqnarray}
 \mathcal{L} & =  &
  \bar{\psi}
 \left( \gamma_{\mu} i \partial^{\mu} - m\right) \psi
 \\ \nonumber & & 
 + G\left\{
 (\bar{\psi}\psi)^2+(\bar{\psi}i\gamma_5\vec\tau\psi)^2
 \right\} \: ,
\label{eq:njllagrangian}
\end{eqnarray}
where  $\psi$ is understood as 
a quark fermion spinor, $m$ is the bare quark mass,
$G$ a coupling constant and $\vec{\tau}$ the isospin matrices.
The similarity to RMF models is evident.
A Fierz transformation of the interaction gives access to
all possible quark-antiquark interaction channels. 
These techniques are explained in detail by
\citet{Klevansky:1992qe,Buballa:2005} for vacuum and in-medium applications.
The coupling $G$ can be understood as a particular choice of a generalized form factor
for nonlocal current-current interactions
\cite{Schmidt:1994di,Bowler:1994ir}.
An important feature of the approach is that the 
quartic terms --- as hallmark of NJL type models --- can 
be shown to bosonize into baryon, meson and diquark contributions
to the partition function, and
hence to the thermodynamic potential 
\cite{Kleinert:1976xz,Roberts:1987xc,Cahill:1988bh,Hatsuda:1994pi}. 
This bosonization property is exploited to develop
an understanding of hadrons as quark bound states in 
the medium~\cite{Bentz:2001vc,Wang_11,Blaschke:2013zaa}.
The original NJL model's success rests on its ability to
describe the breaking of chiral symmetry.
It fails to describe the infrared behavior of QCD which
addresses confinement.
It has therefore been suggested to extend the model by
an imaginary chemical potential expressed in terms of the Polyakov loop
as a possible order parameter of deconfinement \cite{Fukushima:2003fw,Ratti:2005jh}.
A modification of the Polyakov loop potential in these PNJL models
due to a finite quark chemical potential $\mu$ has been introduced by \citet{Dexheimer:2009hi}.
Further extensions result from variations of the previous models, e.g.,
nonlocal PNJL models without \cite{Blaschke:2007np} 
and with \cite{Contrera:2012wj} $\mu$-dependent Polyakov loop potential.
The PNJL model is used to study the QCD phase diagram at finite temperatures
and densities \cite{Fukushima:2008wg}.
A further approach to account for confinement in 
NJL type models suggests to introduce an infrared cut-off
to remove unphysical quark-antiquark thresholds \cite{Ebert:1996vx}.
The similarity of the NJL model to the RMF approach for nuclear matter
suggests that in analogy to the non-linear Walecka model 
higher order coupling channels,
i.e., multiquark or quark-meson interactions,
will affect in particular the high density behavior 
of the EoS~\cite{Benic:2014jia,Zacchi:2015lwa}.
\end{itemize}

\subsection{Clustered and non-uniform matter}
\label{sec:clusteredmatter}
At subsaturation densities and not too high temperatures, 
nucleonic matter is no
longer uniform since it becomes instable with respect to variations in
the particle densities. There are various criteria to identify the
onset of instabilities, both in static and dynamic approaches, see
subsection \ref{sec:PT} for details.  Spatial structures can develop
on different length scales.  In stellar matter, nucleons can form
nuclei or clusters of different sizes and shapes due to the interplay
between the short-range nuclear interaction and the long-range
electromagnetic interaction.  If the size of the clusters is small as
compared to their mean free path, the matter can still be described as
a homogeneous system, however with 
cluster degrees of freedom in addition to 
nucleons and leptons.  At higher densities, e.g., in the
so-called `pasta phases' in the inner crust of NSs, the
density variations have to be treated explicitly. In low-density cold
matter, nuclei arrange themselves in
a lattice and a crystal structure develops,
e.g., in the outer crust of NSs, and a new length scale
emerges.  The appearance of cluster structures in matter can be
treated in various approximations that differ mainly with respect to the
choice of the basic degrees of freedom and to the description of
interactions.

\subsubsection{Nuclear Statistical Equilibrium}
\label{sec_nse}
The most basic approach to describe clustered matter 
is given by NSE models, which are
sometimes just called ``statistical models''. They are
characterized by assuming a statistical ensemble of different nuclear species
and nucleons in thermodynamic equilibrium. In particular, the chemical potentials
of nuclei are not independent. They are given by Eq.~(\ref{eq_nse}).
NSE models are not only used in simulations of CCSNe but 
also for nucleosynthesis calculations and in the context of 
SNIa, see, e.g., \citet{seitenzahl_09}. 

In its simplest form, the ideal NSE, a mixture of non-interacting ideal gases 
assuming Maxwell-Boltzmann statistics is utilized. In chemistry, this description 
is known as ``mass-action law''. In the ideal NSE 
approach the abundance ratio of nuclei is determined by a Saha equation, which
originally was used to describe the population of ionization states 
in atoms. 
Instead of classical Maxwell-Boltzmann statistics, 
the correct quantum statistics of the particles can also be implemented. 
Often 
the corresponding Fermi-Dirac distribution is adopted only for nucleons.

Standard NSE models do not take into account the effects of the strong 
interaction between the constituents explicitly, e.g., correlations in
nucleon-nucleon scattering states are neglected. However, in some models 
the interactions in the nucleonic component are 
incorporated by employing a mean-field description of homogeneous matter, see section \ref{sec:mf_models}.

A huge variety of NSE based models can be found in the
literature, employing various extensions and different levels of
sophistication. Here, we will only summarize the most important
aspects which are typically discussed in the context of these
models. For some selected models we will give further details in
Sec.~\ref{sec:generalEoS}.

\paragraph{Nuclear binding energies}
Realistic EoS in the NSE model description require nuclear binding
energies as  basic input.  Different approaches are used: On
the one hand, values from theoretical models are employed. These can
be simple mass formulas such as liquid-drop like parameterizations or
more detailed nuclear structure calculations see, e.g.,
\citet{myers_90,myers_94,moeller_95,lalazissis_99,koura_05,geng_05}.
On the other hand, experimentally measured binding energies, see,
e.g., \citet{audi_03,audi_12}, are used directly. However, an
extension of these mass tables from experiments is required with the
help of theoretical approaches since exotic nuclei that are not yet
studied experimentally can be encountered. 
Shell effects in the structure of
nuclei have a significant impact on the distribution at low
temperatures, as will be illustrated in Sec.~\ref{sec:nse_eos}. Binding energies of nuclei inside matter are modified
as compared with their vacuum values. These medium effects are often
introduced in NSE models in phenomenological approximations, see Sec.~\ref{sec_medium_mod_heavy} below.

\paragraph{Excited states}
\label{sec:nse_excited}
It is straightforward to include excited states of nuclei in an explicit way if 
their excitation energies are known experimentally. However, especially at high 
excitation energies and for very heavy or exotic nuclei, the experimental 
information on the levels and their properties is not complete. In this case,
theoretical level densities or internal partition functions 
can be used \cite{fai_82,engelbrecht_91,iljinov_92,Blinnikov:2009vm}.
Alternatively, a temperature 
dependence of the binding energies is introduced \cite{botvina08}. 
The works of \citet{rauscher_00,rauscher_01,rauscher_03} provide nuclear 
partition functions in tabular form, for temperatures up to 24~MeV and a wide 
range of nuclear masses. These calculations are based on both experimental data 
and a back-shifted Fermi-gas model. More recently, similar tables were provided 
by \citet{goriely_08}. 
Problems with divergences 
of the original Fermi-gas model \cite{Bethe:1936zza}
at low excitation energy can be solved \cite{GrFe:1985}. However, the general reliability 
of the employed densities of state formulas can be questioned.
For an investigation of effects of excited states on the 
supernova EoS and the stellar collapse, see, e.g., 
\citet{mazurek_79,nadyozhin_04,liu_07}.

\paragraph{Coulomb interaction}
In matter with the condition of electric charge neutrality, 
the Coulomb interaction among nucleons and nuclei is screened due to the
background of electrons and possibly muons. Some NSE models 
neglect this Coulomb screening. Others include it by using only the one-body 
Wigner-Seitz approximation, see, e.g., \citet{llpr1985}. However, there are 
very detailed calculations available that were obtained from 
the study of single- and even multi-component plasmas 
at different temperatures, 
see, e.g., \citet{chabrier_98,chugunov_09,potekhin_09a,potekhin_09b,Potekhin_13}
and section \ref{sec:Coul} for more details. 
Typically the results are provided in form of fitting formulae. 
The simulations have reached 
a high numerical precision and deviations from the linear mixing rule for binary plasmas
were found to be small \cite{DeWSlaCha:1996}. For a discussion of 
different approximations of Coulomb interactions in supernova EoS and 
the application of some of the aforementioned models, see, e.g., 
\citet{nadyozhin_05,Blinnikov:2009vm}.

\paragraph{Medium modifications of heavy nuclei}
\label{sec_medium_mod_heavy}
Some statistical models employ explicit medium corrections of the
binding energies of heavy nuclei. These can be due to temperature or
due to the presence of unbound nucleons. In both cases, the surface
and bulk properties of nuclei are modified as compared to the
vacuum at zero temperature. One aspect is a temperature dependence of
the symmetry energy \cite{dean_02,agrawal14} and of effective nucleon
masses \cite{donati_94,Fantina2012}. Obviously, such temperature
effects are related to excited states and internal partition functions
but the problem is approached from a different perspective. Effects of
the unbound nucleons on nuclei are often extracted from
nucleons-in-cell calculations, see Sec.~\ref{sec:nuc_in_cell}. 
For instance, \citet{Papakonstantinou_13} and \citet{Aymard_14}
calculated the binding energy
shifts for Skyrme interactions in the local-density approximation and
in the Extended Thomas-Fermi approximation, respectively. It
was pointed out that the definition of the binding energy shifts has
to be consistent with the definition of clusters where one has to
distinguish coordinate-space and energy-space clusters.

\paragraph{Cluster dissolution}
\label{sec:clusterdissolution}
The application of the standard NSE is limited to rather
low densities. This model cannot describe the dissolution of nuclei
with increasing densities, the Mott effect, which is mainly driven by
the Pauli principle \cite{RoMuSchu1982,RoSchmMuSchu1983}.  
When the nuclear saturation density is approached, a transition to uniform  
nucleonic matter
is often enforced with the help of the excluded-volume mechanism, which
represents a classical, phenomenological approach to describe the
dissolution of nuclei at high densities in a geometrical picture.
Different variants of excluded-volume effects can be found in the
literature.  In the simplest case, the total volume of the
thermodynamic system is replaced by the so-called free volume that is
the total volume reduced by the volume occupied by the particles of
finite size \cite{Rischke:1991ke}.  This approach is well known from the
EoS of a van-der-Waals gas.  General expressions for excluded-volume
effects can be found, e.g., in \citet{yudin_10,yudin_11,Typel:2016srf}. 
More detailed
models solve for the exact canonical partition function, taking into
account finite volumes of the particles and/or assuming a certain
geometry of the particles. An important example is the hard-sphere
model \cite{carnahan_69,mulero_08}. Note that the excluded-volume 
mechanism is also commonly used in the context of relativistic heavy-ion
collisions (HICs) to describe the freeze-out of particle and their yields in
a hadron resonance gas model \cite{gorenstein_81,andronic_12}.
A problem of the excluded-volume approach is the occurrence of a superluminal
speed of sound at high densities and hence the EoS becomes acausal
\cite{Rischke:1991ke,Venugopalan:1992hy}.

\subsubsection{Single nucleus approximation}
\label{sec_nse_sna}
Instead of considering the distribution of all nuclei, the chemical
composition is sometimes simplified by assuming a representative
single heavy nucleus, unbound nucleons, and possibly $\alpha$ particles
and other light nuclei in the description.  
\citet{burrows_84} have shown that this so-called single nucleus approximation
(SNA) has only a small impact on thermodynamic quantities. However,
there can be significant differences between the average mass and
charge number of heavy nuclei in a full NSE model and the
corresponding values of the representative nucleus employing the SNA
\cite{souza_09}, see also Sec.~\ref{sec:nse_eos}. 
Furthermore, the conclusions of \citet{burrows_84}
are not applicable if the composition is dominated by light nuclei. In
this case it is not possible to consider a ``representative light
nucleus'' due to the small number of nuclei involved and the large
variability of their binding energies, see, e.g.,
\citet{Hempel_11a,Hempel:2015yma}. We want to point out that considering a
statistical ensemble of all nuclei, i.e., going beyond the SNA, is
particularly relevant for the determination of electron-capture rates
during core-collapse, see Sec.~\ref{sec:ccdynamics}. 

\subsubsection{Virial Expansion} 
\label{sec:veos}

Correlations between the constituents of a low-density gas of particles at finite 
temperature can be considered in the virial equation of state (VEoS). It provides corrections 
to the NSE approach to clustered matter. The original formulation goes back to
\citet{BeUh1936,Beth:1937zz} and uses a description based on 
the grand canonical ensemble. The VEoS relies on a 
series expansion of the grand canonical potential
\begin{eqnarray}
 \lefteqn{\Omega(T,V,\{ \mu_{i} \}) = -TV} 
\\ \nonumber & & 
 \times \left( \sum_{i} \frac{b_{i}}{\lambda_{i}^{3}} z_{i} 
 + \sum_{ij} \frac{b_{ij}}{\lambda_{i}^{3/2}\lambda_{j}^{3/2}} z_{i}z_{j} 
 + \dots \right)
\end{eqnarray}
in powers of the particle fugacities
$z_{i}=\exp\left[ \left(\mu_{i}-m_{i}\right)/T\right]$ 
where $\mu_{i}$ is the chemical potential of particle 
species $i$ including the rest mass $m_{i}$. 
The quantities $\lambda_{i}=\sqrt{2\pi/(m_{i}T)}$
denote the thermal wavelengths and $b_{i}$ are 
the degeneracy factors
of the single particles.
The virial coefficients $b_{ij}$, $b_{ijk}$, \dots\ are simple functions of the temperature.
They contain 
information on the two-, three-, \dots, many-body correlations 
in the system. In particular,  the second virial coefficients
\begin{equation}
\label{eq:b_ij}
 b_{ij}  =  \frac{\lambda_{i}^{3/2}\lambda_{j}^{3/2}}{2\lambda_{ij}^{3}}
 \int dE \: \exp\left( - \frac{E}{T}\right) D_{ij}(E) 
\end{equation}
with $\lambda_{ij}=\sqrt{2\pi/[(m_{i}+m_{j})T]}$
depend only on the energies $E_{l,k}^{(ij)}$ 
of the two-particle bound states 
and the two-body scattering phase shifts $\delta_{l}^{(ij)}$ in channels $l$ that
appear in the function
\begin{eqnarray}
  \lefteqn{D_{ij}(E)  = } \\ \nonumber & &
 \sum_{l} \left[ \sum_{k} g_{l,k}^{(ij)}\delta(E-E_{l,k}^{(ij)})
 +  \frac{g_{l}^{(ij)}}{\pi} \frac{d\delta_{l}^{(ij)}}{dE} \right]
\end{eqnarray}
with appropriate degeneracy factors $g_{l,k}^{(ij)}$ and
$g_{l}^{(ij)}$.  Note that the separation of bound state (first term) and
scattering contributions (second term) is not unique, since a partial
integration of the energy integral (\ref{eq:b_ij}) leads to
equivalent expressions with a different partitioning.  
Because experimental information can be used directly in the evaluation, a
model-independent approach is obtained. Quantum statistical effects
can be incorporated easily.  For contributions beyond second order
see, e.g., \citet{Pais:1959zzb,Dashen:1969ep,Bedaque:2002xy,Liu2009}.  
The VEoS is only applicable for small fugacities $z_{i} \ll 1$ or, equivalently
$n_{i}\lambda_{i}^{3} \ll 1$ with the particle number density $n_{i}$.
These thermodynamic conditions are found, e.g., in the neutrinosphere
of supernovae.  In present applications of the VEoS to stellar matter
only correlations on the level of the second virial coefficient 
due to the strong interaction are considered.

\subsubsection{Quantum statistical approach}
\label{sec:qs}
As mentioned above, the transition from inhomogeneous to uniform
matter cannot be easily described within the NSE approach nor within
the VEoS. For that purpose, NSE models are extended phenomenologically by
adding a classical excluded volume correction, see 
Sec.\ \ref{sec:clusterdissolution}.  A systematic
description of correlations and in particular the Mott effect
in an interacting many-body system is 
given by the quantum statistical (QS) approach
\cite{RoMuSchu1982,RoSchmMuSchu1983}.

In general, many-body methods can provide spectral functions that
contain all information on correlations. Prominent
peaks in a spectral function can be identified with the corresponding
quasiparticles, e.g., deuterons in nucleonic matter.  As an
approximation, quasiparticles with shifted energies can be introduced
in the practical calculation of an EoS.  These energies depend on the
nucleon densities, the temperature and the momentum of the
quasiparticle in the medium. They include effects of Pauli blocking 
and can be parameterized with more or less sophistication
\cite{Ropke:2008qk,Typel:2009sy,Ropke:2011tr}.
The energy shifts enter in the determination of the particle densities
and finally in the EoS.

Going beyond the simple quasiparticle approximation for the spectral function
in the QS approach leads to a form of the grand canonical potential that closely resembles
the corresponding expression of the VEoS. This generalized Beth-Uhlenbeck
approach has been
formulated and studied in detail with realistic two-nucleon potentials
\cite{Schmidt_90,Stein:1998mg}. In this case, the EoS can be
formulated with a modified
expression for the second virial coefficient. The quasi-particle
energies in the bound state contribution are affected by the in-medium
energy shifts and the phase shifts in the scattering contribution have
to be calculated from the in-medium T-matrix for two-body scattering.

\subsubsection{Generalized relativistic density functional}
\label{sec:grdf}
The modification of cluster binding energies in the medium, which is a
basic feature of the QS approach, can be implemented in quasiparticle
models.  The generalized relativistic density functional (gRDF)
approach \cite{Typel:2009sy,Typel:2013zna,Typel:2012esa} is such a
model. It is an extension of a RMF model with
density dependent nucleon-meson couplings that is constrained by fits
to properties of finite nuclei.  In the gRDF model, nucleons and
clusters are included as explicit degrees of freedom.  Two-nucleon
scattering correlations are described by effective resonances in the
continuum with parameters that are adjusted in order to reproduce the
model-independent virial EoS at low densities
\cite{Voskresenskaya:2012np}.  All particles couple to the meson
fields with appropriately scaled strengths resulting in density
dependent scalar and vector self-energies. In addition, mass shifts of
composite particles are implemented in a parameterized form, which are
derived from the QS approach, in order to account for Pauli blocking
from the nucleons in the medium. A further change of the cluster
binding energies is caused by the screening of the Coulomb field due
to the electronic environment in stellar matter. The density
dependence of the binding energy shifts and meson-nucleon couplings
leads to ``rearrangement'' contributions in the vector self-energy and
thermodynamic quantities, which guarantee a thermodynamically
consistent approach.  The model interpolates  between the
correct low-density limit given by the virial EoS and the
supra-saturation case of purely nucleonic matter.
Medium dependent mass shifts of light clusters were also incorporated in
other models that study pasta phases in compact star matter
\cite{Avancini:2012bj,Pais:2015xoa}.

\subsubsection{Nucleons-in-cell calculations} 
\label{sec:nuc_in_cell}

In all models that were described above, matter is treated as a
uniform system of interacting particles. They are
assumed to be point-like or to have a finite volume. The formation of
inhomogeneous structures in subsaturation matter at low temperatures
can be studied in calculations where a non-uniform distribution of
nucleons inside a cell of given shape and size is considered. Here,
quantal and classical methods can be distinguished. In stellar matter,
electrons are usually not treated explicitly but considered as a
uniform background gas. The methods are mostly applied to study the
formation of clusters or ``pasta'' phases in the crust of neutron
stars, the corresponding phase transitions 
and the effects of the Coulomb interaction, 
see section \ref{sec:PT}. Beyond static properties for an EoS, also the
dynamical response, e.g., in neutrino scattering, or hydrodynamic
quantities such as viscosities and conductivities can be studied.

In the classical molecular dynamics (CMD) approach
\cite{Horowitz:2011cn,Piekarewicz:2011qc,Dorso:2011gp,Dorso:2012db,Schneider:2013dwa,GimenezMolinelli:2014dha,Molinelli:2014uta},
a certain number of nucleons is placed inside a box of given volume to
reproduce a fixed density. The velocities of the particles are
chosen to represent a Maxwellian distribution at given
temperature. The particles interact via a two-body potential where the
Coulomb contribution is essential. The time evolution of the system
is followed by solving a set of classical coupled equations of
motion, which is possible even for a rather large number of particles.
In quantum molecular dynamics (QMD) models
\cite{Maruyama:1997rp,KiMaNiCh2000,Watanabe:2001ix,Watanabe:2002sf,Watanabe:2002ty,Watanabe:2003hz,Matsuzaki:2005sw,Watanabe:2006qf,Sonoda:2007sx,Watanabe:2009vi,Maruyama:2012bi},
particles are not treated as point-like objects as in CMD calculations
but as wavepackets, e.g., of Gaussian
shape. 
The motion of the centroid of the wavepackage is
followed classically. 
Quantum-statistical effects, such as
antisymmetrization or shell effects, are not accounted for in
molecular dynamics (MD) simulations. The Pauli exclusion principle can
be incorporated approximately with appropriately designed
contributions to the potentials.  Central questions of MD calculations
are the fragment recognition \cite{Dorso:1992ch,Strachan:1997zz}, the
chemical composition
\cite{Horowitz:2007hx,Horowitz:2009af,Dorso:2012pt,Caplan:2014gaa},
the appearance of different, sometimes complicated, shapes in the
density distribution and their topological characterization
\cite{Watanabe:2003xu,Dorso:2012pc,Alcain:2013bza,Schneider:2014lia,Horowitz:2014xca}
and phase transitions \cite{Watanabe:2004tr,Alcain:2014fma}.
Structure functions and quantities related to the dynamical response
can be extracted as well
\cite{Horowitz:2004pv,Horowitz:2004yf,Horowitz:2005zb,Horowitz:2008vf,Caballero:2008tx,Horowitz:2009af}.

A description of matter inside a cell based on particle density
distributions instead of localized classical particles is a widely
adopted approach for inhomogeneous systems.  A number of different
methods is available. They differ in the physical input, the 
approximations and  the
numerical complexity. Many approaches that
model the formation of large nuclei surrounded by a gas of
nucleons employ the Wigner-Seitz approximation where the size of the
cell is determined by the neutrality condition, i.e., the total charge
of baryonic matter inside the volume considered is compensated by the
electronic charge. At low densities, a spherical cell, which is
centered around individual nuclei, is usually assumed. When other
geometries, mostly rectangular shapes, are considered, 
it is common to presume a periodic density distribution in space.

The (compressible) liquid drop model (LDM)  belongs to the class of
microscopic-macroscopic approaches where the energy of matter inside
the cell is parameterized with several individual contributions such as
bulk, surface, etc.
\cite{Baym:1971ax,Lattimer:1981hn,Ravenhall:1983uh,LaLaPeRa1983,Lattimer:1985zf,Lattimer:1991nc,Lorenz:1992zz,Watanabe:2000rj,Douchin:2001sv,Oyamatsu:2006vd,Nakazato:2009ed,Nakazato:2010nf,furusawa11,furusawa13}.
They depend on the size of the heavy nucleus and the density of the
surrounding gas of nucleons.  The detailed structure of the
expressions can be guided by energy density functionals, often
non-relativistic Skyrme parameterizations.  Due to its convenience in
the numerical application, the LDM is one of the earliest approaches to
model inhomogeneous stellar matter.

Other widely employed methods are the Thomas-Fermi (TF) approximation
or extensions thereof.  Here, the density distribution of fermions in
the cell is determined using a given EDF for nucleonic matter.  In the
most simple form, a local density approximation of nuclear matter is
obtained, but corrections, e.g., related to surface effects and the
finite range of interactions, can be incorporated in the
calculation. The shape of the density distributions can be given by a
simple functional form with few parameters or can be determined fully
self-consistently. Technically, the free energy of the cell is
minimized variationally and the chemical potentials of the nucleons
are obtained for fixed particle numbers. Both, non-relativistic
and relativistic
\cite{Sumiyoshi:1995np,Cheng:1997zza,shen_98,Avancini:2008zz,Avancini:2008kg,Avancini:2010xb,Avancini:2010ch,Okamoto:2011tc,Shen:2011qu,Zhang:2014uma}
models are available. The main disadvantage of the TF approach is
that shell effects are not included.  They can be incorporated with
the Strutinski method
\cite{BrJeCh1976,Brack:1985vp,Onsi:2008pi,Pearson:2012hz,Pearson:2015hla}
but this correction to the TF approach is not widely used for
astrophysical EoS.  The formation of inhomogeneous matter inside cells
of various shapes is also studied by using the DFT formalism in
relativistic approaches, see, e.g.,
\cite{Maruyama:2005vb,Gogelein:2007pb}, where shell effects are
included but only in the mean-field approximation without explicit
particle exchange.

The next level of complexity is reached in HF calculations, which take
the antisymmetrization of the nuclear many-body wave function within
the cell fully into account.  This approach is most frequently
utilized with non-relativistic potentials such as the zero-range
Skyrme interaction, e.g., in the pioneering works of
\citet{Negele:1971vb,Ravenhall:1972zz,Bonche:1980ry,BoVa1982,Bonche:1985zz}.
Shell effects were found to disappear quickly with increasing
temperature at 2-3~MeV and nucleon pairing effects are relevant only
below $\approx$~1~MeV \cite{BrQu1974,Brack:1974gy}.  The evolution of
the density distribution inside the cells and consequences for the EoS
are investigated in several works, see, e.g.,
\citet{Magierski:2002jr,Sil:2002yk,Gogelein:2007pb,Newton:2009zz,Gusakov:2009kc,Pais:2012js,Pais:2014hoa,Papakonstantinou_13}. Also
time-dependent HF calculations were employed, both in the standard
formalism
\cite{Schuetrumpf:2012cj,Schuetrumpf:2012dia,Schuetrumpf:2014aea} and
in an extended approach using dynamical wavelets for single-particle
wave functions \cite{Sebille:2009zz,Sebille:2011zz}.  A correct
treatment of antisymmetrization beyond the individual cell volume was
developed recently
\cite{Vantournhout:2010pf,Vantournhout:2010wv,Vantournhout:2011zb}.

\subsection{Phase transitions}
\label{sec:PT}
Theoretical models for the EoS are mostly concerned with calculations
of single phases where a thermodynamic potential is extremized locally
for its set of natural state variables, see Sec.~\ref{sec:thermodynamics}.  
In the case of thermodynamic instabilities, the equilibrium state of the
system is obtained from a global minimization/maximization of
the appropriate thermodynamic potential allowing for a phase
transition with coexistence of different phases, i.e., macroscopic
regions in space with different values of the various densities but
identical intensive variables , see, e.g., \citet{LandauLifshitzV}.

Examples of phase transitions in the present context are the
liquid-gas phase transition in nuclear matter 
(see, e.g.,
\citet{Barranco:1980zz,Muller:1995ji,Gulminelli_03,Hempel_13}),
the hadron-quark transition, which is expected to occur at very large
baryon number densities and/or temperatures (see, e.g.,
\citet{Collins:1974ky,prakash_95,Steiner:2000bi,Mishustin:2002ys,Bhattacharyya:2009fg,Yasutake:2012dw,Hempel_13}),
and the gas/liquid-solid phase transition, which is relevant in the
formation of the crystalline crust during the cooling of proto-neutron
stars \cite{Chamel:2008ca}. A phase transition could also be caused by the appearance of
new particle species such as hyperons in dense hadronic matter
\cite{SchaffnerBielich:2000wj,SchaffnerBielich:2002ki,Gulminelli:2012iq,Gulminelli:2013qr}.

\subsubsection{Thermodynamic description of phase transitions}
In the following, $F(T,\{N_{i}\},V)$ is chosen as the thermodynamic potential 
for the discussion.
The free energy describes
the equilibrium thermodynamics of a system 
if it is a convex function of the extensive variables, i.e., $\{N_{i}\}$
and $V$, and a concave function of the temperature $T$.  
Then the conjugate intensive variables, the chemical
potentials $\mu_{i}=\left. \partial F/\partial
N_{i}\right|_{T,\{N_{j\neq i}\},V}$ and the pressure
$p=-\left. \partial F / \partial V \right|_{T,\{N_{i}\}}$ are constant
throughout the volume $V$.  
The free energy of a particular
theoretical model is locally convex 
in the subspace of extensive variables
if all eigenvalues of the
stability matrix $M$ with entries
\begin{equation}
 M_{ij} = \left. \frac{\partial^{2} F}{\partial Q_{i} \partial Q_{j}} \; ,
 \right|_{T,Q_{k,k\neq i, k \neq j}} 
\end{equation}
are positive. Here, $Q_{i}$, $Q_{j}$, $Q_{k}$ are variables from
the set of the conserved charges $\{N_{i}\}$ or the volume $V$.  
If at least one eigenvalue of $\boldmath{M}$ is zero or negative, this
point in the space of variables is metastable or unstable,
respectively.  All unstable points are enclosed by the so-called
spinodal in the space of conserved charge numbers $N_{B},N_{Q}$, etc.
Besides this thermodynamic criterion there are other dynamical
approaches to identify the instability region, e.g., collective
excitations with RPA calculations, the Vlasov equation formalism, 
or Fermi liquid theory
\cite{PeRa88,HePeRa93,Margueron:2004sr,Providencia:2006mm,Brito:2006se,Providencia:2006ux,Ducoin:2008xs,Ducoin:2008ic}. 

The spinodal is enclosed by the binodal that connects all
points with identical temperature, chemical potentials, and
pressure.  Points inside the binodal belong to the phase coexistence
region. Here the free energy for given conserved total charge numbers
can be lowered as compared to the locally calculated value by
considering the coexistence of two phases $p=I,II$ with different
volumes $V^{p}$ such that $V^{I}+V^{II}=V$. The particle numbers
$N_{i}^{p}$ in the individual phases are in general different but all
chemical potential are identical $\mu_{i}^{I} = \mu_{i}^{II}$ in the
two phases as well as the other intensive variables ($P^{I}=P^{II}$,
$T^{I}=T^{II}$). This  is called the Gibbs condition 
for thermodynamic equilibrium. Points in coexistence always
lie on a binodal.  Hence, it is sufficient to know the thermodynamic
potential on this line/surface to construct the system properties for
thermodynamic conditions inside the binodal.

If there is only a single
conserved charge, 
it is easily checked whether the free energy of a particular
theoretical model is a convex function of the only conserved charge
for given $T$ and $V$. If not, the binodal degenerates to two separate
points and the well-known Maxwell construction of phase transitions
with isotherms, e.g., in the pressure - density diagram is obtained.
Changing the temperature of the system, the two distinct points in
coexistence can collapse into a single point that defines the
critical point and the corresponding critical temperature and
critical pressure. If this topology applies, it is possible to move
from one phase to the other around the critical point without crossing
the binodal, e.g., in the temperature-density plane.
In some cases, e.g., for the
quark-hadron phase transition, there is often no consistent model for
the complete range of the thermodynamic variables
and two different models are used.
Then, a transition from one phase to the other without crossing
a phase separation line is not possible, see also \citet{Hempel_13}.

In case of more than one conserved charge the complete set of Gibbs
conditions applies and the topology of the binodals and
spinodals becomes more complex
\cite{Glendenning92,Muller:1995ji,Barranco:1980zz,iosilevskiy10,Hempel_13}. New
features appear: In the higher dimensional parameter space, the
critical point turns into a critical line or critical hypersurface and
several topological endpoints can be defined.  However, it is always
possible to map the Gibbs construction with several independent charge
numbers to a technically more simple Maxwell construction
\cite{Ducoin:2005aa,Typel:2013zna}. This is achieved by applying
Legendre transformations to the free energy $F$ that replace all
conserved charge numbers except one with the corresponding chemical
potentials. Thereby a modified thermodynamic potential is found that
depends only on a single particle number as in the standard Maxwell
case.  

\subsubsection{Coulomb effects}
\label{sec:Coul}

In models of the EoS for NSs and CCSNe the equilibrium
with respect to the strong as well as the electromagnetic interaction
between all constituents has to be considered.  As a consequence, the
phase structure of dense matter is substantially affected by the
interplay of these short and long-range forces in competition with
entropy.  In addition, the specific condition of charge neutrality
applies.  At not too high temperatures the appearance of clusters and
crystalline structures is expected, see also
Sec.\ \ref{sec:clusteredmatter} and below.  Closely connected to both
features is the possible occurrence of ``pasta'' phases.

If macroscopic phases coexist, different
assumptions  for the treatment of charge neutrality can be made. 
Either one requires local charge neutrality, i.e.,
each phase is charge neutral, or the
system is  charge neutral as a whole and  the phases are
allowed to carry a net charge.  
We note that the assumption of charged
phases but global charge neutrality contradicts the
assumption of the thermodynamic limit, if interpreted strictly
as the Coulomb energy would diverge for such a system.
Nevertheless, it is considered as a reasonable
simplification in many
situations, see \citet{Martin_15} for a discussion regarding the inner NS crust. 
Depending on the choice for realizing the charge neutrality condition,
the chemical equilibrium conditions for phase coexistence
\cite{Hempel:2009vp} lead to different qualitative properties of the
phase transition \cite{Glendenning92,iosilevskiy10,Hempel_13}, in
particular to the feature of Coulomb frustration
\cite{Gulminelli_03,Napolitani:2006au,Chomaz:2005xn,Hasnaoui:2012yr}.

In more advanced approaches, surface effects and the finite-range of
interactions are explicitly taken into account.  A crucial ingredient  
is the surface tension between phases. 
Typically, for high surface tensions, the phases tend to
approach a configuration which resembles the case of local charge
neutrality \cite{Heiselberg_93,Maruyama_08,Yasutake:2012dw}. For very
low surface tensions, however, the configuration can be similar to the
case of global charge neutrality without finite-size effects.

The liquid-gas phase transition in nuclear matter predicts the
coexistence of high-density and low-density phases of macroscopic size
below a critical temperature.  If Coulomb effects are included,
electrons have to be added to compensate the positive proton charge.
Phase transitions with macroscopic, charge-neutral phases in
coexistence would create large electric fields at the interfaces that
the system tries to avoid \cite{voskresensky03}. This can be seen
in the analysis of instabilities where density fluctuations of
certain wavelengths are preferred.  Clusters of finite size emerge
that are surrounded by a low-density gas of nucleons. The density of
the electrons throughout the system is nearly constant due to the
large incompressibility of such a high-density Fermi liquid.  Often a
constant electron density is assumed, with some exceptions, see, e.g.,
\citet{Maruyama:2005vb,endo2006,yasutake2011}.  The formation of
clusters in uniform nuclear matter is well studied in several models
that consider the distribution of nucleons and electrons in a cell of
given geometry, see section \ref{sec:nuc_in_cell}.  Usually, the
Wigner-Seitz approximation is applied. 
At low temperatures and rather low densities, a single cluster in a spherical
cell is the preferred geometry \cite{Lamb:1978zz,Douchin:2000kx}.
With increasing density it becomes more advantageous to develop
structures of cylindrical or planar geometry and a sequence of
``pasta'' phases is found, see, e.g.,
\citet{Watanabe:2003jz,Gupta:2013rda} and references given in sections
\ref{sec:nuc_in_cell} and \ref{sec:crust}. This was observed in early models for
cells with different symmetry with simple energy density functionals
\cite{HaSeYa1984,OyHaYa1984,WiKo1985,Oyamatsu:1993zz}.  
In principle, every change from one
geometry to another represents a phase transition of its own.  More
refined calculations with less restrictions to the spatial
distributions of the densities found that transitions from clustered
matter at low densities to uniform matter at high densities exhibit
weaker discontinuities, i.e., the phase transitions are much less
pronounced \cite{Newton:2009zz,Pais:2014hoa}.  
A nearly complete quenching of the traditional liquid-gas phase transition can
occur \cite{Gulminelli_03}.

When stellar matter is cooled down at a given density, the size of
clusters grows as observed, e.g., in spherical Wigner-Seitz cell
calculations. At very low densities, the average distance between the
clusters is large and the short-range strong interaction  
can practically be neglected. Coulomb and thermal
energies  drive the thermodynamic behavior of the system that
can be seen as a plasma  of ions and electrons.  
At very low
temperatures a phase transition from the gas phase to a crystalline
phase is expected \cite{Salpeter:1961zz,Baym:1971pw}.  The ratio of
the strengths of the Coulomb interaction and of the thermal energy is
measured by the parameter
\begin{equation}
 \Gamma_{i} = \frac{Z_{i}^{5/3} e^{2}}{a_{e}T}
\end{equation}
with the charge $Z_{i}$ of the cluster and the electronic length scale
\begin{equation}
 a_{e} = \left( \frac{3n_{e}}{4\pi}\right)^{1/3}
\end{equation}
depending on the electron density $n_{e}$. When the temperature
approaches zero, $\Gamma_{i}$ diverges.  Fully ionized electron-ion
plasmas have been studied in detail
\cite{BruSaTe1966,Ha1973,PoHa1973,chabrier_98,Farouki:1993zz,Potekhin:2000ye,Chabrier:2002hx,Da2006,Cooper:2008di,potekhin_09a,potekhin_09b,potekhin_10}.
From classical Monte Carlo simulations, see
Sec.\ \ref{sec:nuc_in_cell}, of a one-component plasma (OCP) it is
known that the phase transition from the gas to the crystal phase
occurs at $\Gamma_{i} \approx 175$.  The exact value
will depend on the details of the theoretical model. In particular, the
often employed Wigner-Seitz approximation is insufficient to capture
the correlations that are induced by the Coulomb interaction and
determine the location of the phase transition.  For a mixture of
different ionic species, corrections to the OCP result apply
\cite{Ogata:1993zz,chabrier_98,nadyozhin_05,chugunov_09}.  For
$\Gamma_{i} \to 0$ the Debye screening limit in a plasma is obtained.
For $\Gamma_{i} \to \infty$ a body-centered cubic (bcc) lattice of
ions immersed in a uniform sea of electrons is found as the ground
state \cite{Baym:1971ax,Baym:1971pw}.  At finite temperatures lattice
vibrations contribute to the thermodynamic potential of the system
\cite{BaPoYa2001}.  With increasing temperature these thermal
excitations will lead to the melting of the crystal.

An amorphous structure instead of a crystal as ground state at zero
temperature seems to be unlikely but the actual ion lattice type could
be very sensitive to detailed conditions
\cite{IcIyMi1983,Magierski:2001ud}.  However, \citet{jog1982}
  found that a phase composed of interpenetrating cubic lattices of
  different nuclei can be preferred in certain density regions.

\section{Constraints on the EoS}
\label{sec:constraints}

Models for the EoS can be constrained by different observables, 
which originate mainly from three different
sources: 1. laboratory measurements of nuclear properties and
reactions, 2. theoretical {\em ab-initio} calculations, and 3. observations
in astronomy.  These constraints can test different regions in the
space of thermodynamic variables.  They are, in the best case,
independent of each other and different aspects of an EoS model can be
checked, see,
e.g.,
\citet{Klahn_06,Lattimer:2006xb,Tsang:2012se,li_13,Lattimer:2012xj,
  Lattimer:2014sga,Stone:2014wza, Horowitz:2014bja} for elaborate
discussions.
Although we will discuss a fair amount of constraints keep in mind that these
are usually limited to very restricted domains in the phase diagram (e.g. saturation properties
are properties of symmetric matter, NSs are cold and do not explicitly probe the EoS at given density, etc.). 
Therefore, theoretical
models are required to interpolate between or even extrapolate away from these constrained regions.
While it is desirable that these models by themselves do not add further uncertainties one has to be 
cautiously aware that this is not necessarily the case.

Properties of nuclear matter are usually characterized by a number of parameters
that are related to the leading contributions in
an expansion of the energy per nucleon 
\begin{equation}
 E(n_{B},\delta) = E_{0}(n_{B}) + E_{\rm sym}(n_{B}) \delta^{2} + \mathcal{O}(\delta^{4})
\end{equation}
in the isospin asymmetry $\delta = (1-2Y_{q})$. 
Both the energy per nucleon of symmetric matter
\begin{equation}
\label{eq:E_0}
 E_{0}(n_{B}) = m_{\rm nuc} - B_{\rm sat} + \frac{1}{2} K x^{2} + \frac{1}{6} Q x^{3} + \dots
\end{equation}
and the symmetry energy
\begin{equation}
\label{eq:E_sym}
 E_{\rm sym}(n_{B}) = J + L x + \frac{1}{2} K_{\rm sym} x^{2} + \dots
\end{equation}
can be expanded close to nuclear saturation in the deviation 
$x = (n_{B}-n_{\rm sat})/(3 n_{\rm sat})$ 
of the baryon density $n_{B}$ from the saturation density $n_{\rm sat}$.
Equations (\ref{eq:E_0}) and (\ref{eq:E_sym}) define the binding energy at saturation
$B_{\rm sat}$, the incompressibility $K$, the skewness $Q$, 
the symmetry energy at saturation $J$, 
the symmetry energy slope parameter $L$ and the symmetry incompressibility $K_{\rm sym}$.
In general, nuclear matter parameters are
strongly correlated among each other as well as to properties of nuclei
and neutron stars, see, e.g.,
\citet{Klupfel:2008af,Kortelainen:2010hv,Lattimer:2012xj,Lattimer:2014sga}.
In recent years, many studies focused on obtaining constraints for the
symmetry energy $E_{\rm sym}$ and its density dependence, see, e.g.,
the contributions to the topical issue by \citet{LiRaVeVi20014}. 
The most important constraints on the nuclear matter parameters and the EoS 
will be discussed in more detail in the following subsections.

\subsection{Terrestrial Experiments}
\label{sec:exconstraints}
\subsubsection{Systematics from nuclear masses and excitations}
\label{sec_constraints_from_masses}

The most basic and least ambiguous constraints for the EoS come from
properties of nuclei, most notably nuclear masses~\cite{audi_03,audi_12} 
and density distributions~\cite{DeJager:1987qc,AnMa2013}.  
An extrapolation to infinite mass numbers yields the
corresponding nuclear matter parameters.  
Besides the saturation point
at saturation density of $n_{\rm sat}\approx 0.15 – 0.16$~fm$^{-3}$, and
the corresponding value of the binding energy of $B_{\rm sat}  \approx 16$~MeV,
accurate constraints on the symmetry energy $E_{\rm sym}$ and its
density dependence are obtained.  
There exists a linear correlation between
the symmetry energy at saturation $J$ and the slope parameter $L$
\cite{Lattimer:2012xj,Lattimer:2014sga}. 
This correlation is very robust and validated in
various theoretical approaches, see, e.g., 
\citet{Kortelainen:2010hv,Fattoyev:2011ns,Nazarewicz:2013gda}.

The aforementioned correlation is based on ground-state binding
energies. Instead of ground-state binding energies,  \citet{Danielewicz:2013upa} 
considered excitation
energies to isobaric analog states and charge invariance to derive
constraints for the symmetry energy. In a comprehensive analysis,
Skyrme HF calculations were used to derive an acceptable region for
$E_{\rm sym}$ at densities from 0.04 to 0.16~fm$^{-3}$.  They also
extracted a constraint for $J$ and $L$ 
which significantly overlaps with the constraint from nuclear
masses.  At baryon densities $n_B \sim 0.105$~fm$^{-3}$,
the constraint of \citet{Danielewicz:2013upa} is the tightest, with an
excellent accuracy of $\pm 1.2$~MeV. For higher densities the
constraint rapidly deteriorates, for lower densities it gets slightly
worse. This ``bottleneck'' region was previously noted by
\citet{Brown:2000pd,trippa_08,roca-maza_13b}. Nuclear energy density
functionals with different values of $J$ and $L$ that are
fitted to binding energies of nuclei often show a crossing of their
symmetry energies and/or their neutron matter EoS in this region. The
corresponding density can be interpreted as an average value of the
densities in finite nuclei.  \citet{Danielewicz:2013upa} combined
their analysis of isobaric analogue states with measurements of
skin thicknesses to arrive at tighter constraints of $J= (30.2 – 33.7)$ MeV and
$L = (35 – 70)$ MeV.

Obviously, constraints on the symmetry energy become tighter, if the
experimental knowledge about binding energies is extended to very asymmetric nuclei. 
Many of the current high-precision mass measurements
have been made possible by Penning-trap mass
spectrometers or mass spectrometry with storage rings
in combination with radioactive beams, see, e.g.,
\citet{Wolf_13}.  Binding energies of nuclei are also crucial for
nucleosynthesis calculations and 
the location of the drip lines \cite{erler_12}. They can
also be used directly in the EoS of the outer crust of cold NSs, see,
e.g., \citet{Baym:1971pw,Wolf_13,Kreim:2013rqa}, and
Sect.\ \ref{sec:crust}.

\subsubsection{Nuclear resonances}
\label{sec:nuclear_resonances}
Nuclear resonances in the form of collective excitations of finite nuclei 
contain important information about the isoscalar and isovector properties of 
the nucleon-nucleon interaction. For example, \citet{Paar:2014qda} performed a global 
statistical analysis of experimental results for different collective 
excitations with emphasis on correlations between different observables. 
In addition to nuclear masses and charge radii, 
they considered the anti-analog giant dipole resonance, the 
isovector giant quadrupole resonance, the dipole polarizability of $^{208}$Pb 
and the pygmy dipole resonance transition strength in $^{68}$Ni. Employing a 
certain class of relativistic nuclear EDFs, this lead to 
tight constraints for $J=(32.5\pm 0.5)$~MeV, 
$L=(49.9\pm4.7)$~MeV and the crust-core transition density in NSs. 
Interestingly, the former values are fully compatible with the final results of 
\citet{Lattimer:2012xj} with $J=(29.0-32.7)$~MeV and $L=(40.5-61.9)$~MeV
and of \citet{Lattimer:2014sga} with $L=(44-66)$~MeV.

\paragraph{Giant monopole resonance}
Constraints for the nuclear incompressibility $K$ can be deduced from fitting 
results of theoretical models to experimental data on
the isoscalar giant monopole 
resonance (ISGMR), also called the breathing mode. 
However, it is perceived in the literature that the 
extraction of $K$ from ISGMR data is not unambiguous as it 
relates to the density dependence of the symmetry energy 
in the models
\cite{piekarewicz_04,shlomo06,sharma09}. 
For  example, RMF models often obtain larger values for $K$ in the range of 250 to 270 MeV 
\cite{piekarewicz_04} than non-relativistic models.

Recently, \citet{Khan:2013tqa} reanalyzed the problem of model-dependencies.
They showed that the data actually 
constrain the density-dependent incompressibility around the 
crossing density of 0.1~fm$^{-3}$, by using both relativistic
and non-relativistic EDFs. Therefore constraints on $K$ depend also on the 
skewness parameter $Q$ of the functional used to analyze the data. 
The situation is similar for the extraction of the symmetry energy from nuclear 
masses, c.f., Sec.~\ref{sec_constraints_from_masses}.

A very comprehensive list of theoretical calculations of $K$ from the
literature was given by \citet{Stone:2014wza}. In the same article, a
reanalysis of the ISGMR was  performed, 
based on a liquid drop approach to the
description of the vibrating nucleus. 
Interestingly, it was found that $K$ lies in the range of 250 to 315~MeV, 
which is significantly higher than the generally accepted
values of $K=248 \pm 8$ MeV \cite{piekarewicz_04} or
$K = (240 \pm 20)$~MeV \cite{shlomo06}. 
The authors achieved consistency
with the latter values provided the ratio of the surface to volume contributions,
$K_{\rm surf}/K_{\rm vol}$ in a leptodermous expansion
is close to $-1$, as predicted by a majority of mean-field models. However, in their
analysis it seems that the experimental data favor a ratio
different from $-1$. The high values of $K$ are thus related to a
different surface contribution to the ISGMR compared with other works
employing mean-field models.  Note that \citet{Stone:2014wza} 
are able to explain the ISGMR of tin isotopes, which was found
to be a startling problem of nuclear structure by
\citet{piekarewicz2010}.

\paragraph{Giant dipole resonance}
The nuclear isovector giant dipole resonance (IVGDR) can be
used to constrain the symmetry energy. From measured centroid energies
for a liquid droplet model one obtains a correlation between the
volume and surface part of the symmetry energy of finite nuclei
\cite{lipparini_89,Lattimer:2012xj}, which can be transformed into a correlation
between $L$ and $J$. 
\citet{trippa_08} found that the IVGDR gives the
tightest constraints on $E_{\rm sym}$ around $n_B=0.1$~fm$^{-3}$ with
$23.3 < E_{\rm sym}(0.1~\rm{fm}^{-3}) < 24.9$ MeV, by analyzing the
IVGDR by a variety of Skyrme models. \citet{Lattimer:2012xj} used
different functional forms of $E_{\rm sym}$ to extract the correlation
between $L$ and $J$. 
Their results  show a significant overlap
with other constraints, see Fig.~1 in \citet{Lattimer:2014sga}.

The Pygmy dipole resonance (PDR)  at 
excitation energies much below the IVGDR is
also sensitive to the symmetry energy 
\cite{Klimkiewicz:2007zz,carbone_10}. 
\citet{reinhard_10,daoutidis_11} 
argued that it is not possible to extract constraints on $J$ and $L$ from PDR 
strengths because their correlation with the symmetry energy is too weak. 
This was studied in more detail by \citet{reinhard_13} who found that the
correlation between the accumulated low-energy strength and the symmetry energy
``dramatically depends on the energy cutoff'' used. Furthermore, the authors 
came to the conclusion that the low-energy dipole excitations cannot be 
interpreted in terms of a collective PDR mode. 

\paragraph{Electric dipole polarizability}
\citet{tamii_11} reported a precise measurement of the electric dipole 
response of $^{208}$Pb from proton inelastic scattering. 
The extracted electric 
dipole polarizability $\alpha_{D}$ is  
correlated to the neutron skin thickness 
\cite{lipparini_89,reinhard_10,piekarewicz_12}, which  in turn is
correlated  with $L$, see Sec.~\ref{sec:skins}. 
Using this two-step process, 
\citet{Lattimer:2012xj} obtained an anti-correlation between $L$ and $J$, with 
significant overlap with other constraints.

Recently \citet{roca-maza_13} found that the product $\alpha_D J$ is
much better correlated with the neutron skin thickness of $^{208}$Pb
and $L$ than the polarizability $\alpha_D$ itself. 
After reanalyzing the experimental
results of \citet{tamii_11}, \citet{roca-maza_13} obtained a linear
correlation between $J$ and $L$. Adopting a value of $J=31\pm2$ MeV,
this resulted in $L = 43 \pm (6)_{\rm expt} \pm (8)_{\rm theor} \pm
(12)_{\rm est}$, where ``expt'' denotes experimental, and
``theor'' theoretical uncertainties, while ``est'' originates from the
uncertainty in $J$. \citet{Tamii:2013cna} obtained a linear correlation
between $J$ and $L$. 
They pointed out that the difference to the
anti-correlation found by \citet{Lattimer:2012xj} results from
how they analyzed the data. 
\citet{Lattimer:2014sga} revised the results
of \citet{Lattimer:2012xj}, taking
the improved correlation of \citet{roca-maza_13} into account. 

\citet{zhang2015} analyzed the data from \citet{tamii_11} in yet another way.
Instead of constraining nuclear matter properties at normal nuclear density, 
they showed that $\alpha_D$ of $^{208}$Pb puts stringent constraints on the 
symmetry energy, or almost equivalently the pure neutron matter EoS, at 
subsaturation densities significantly below $n_{\rm sat}$. 
Their final results for the 
subsaturation EoS are consistent with the experimental constraints of 
\citet{Tsang:2008fd} and \citet{Danielewicz:2013upa}.
In addition, they obtain agreement
with various theoretical works for the neutron matter EoS, which are
included in Fig.~\ref{fig:neutronmatter}.
The recent study of \citet{Hashimoto:2015ema}
determined  the dipole polarizability of 
$^{120}$Sn, which is strongly correlated with that of ${}^{208}$Pb, 
experimentally from proton inelastic scattering. 

\subsubsection{Neutron skin thicknesses}
\label{sec:skins}

The density distributions of nucleons and their root-mean-square (rms) radii 
$\sqrt{\langle r_{i}^{2} \rangle}$
change rather smoothly for nuclei in the valley of stability when the mass number increases.
However, the proton and neutron radii
are not in general equal. 
Neutron-rich nuclei develop a neutron skin
with thickness 
$\Delta r_{np} = \sqrt{\langle r_{n}^{2} \rangle} - \sqrt{\langle r_{p}^{2} \rangle}$.
The charge distributions and charge radii of many nuclei are well-known experimentally, e.g.,
from elastic electron scattering or isotope shift measurements, see 
\citet{Angeli:2009zz,AnMa2013}
and references therein. In contrast, neutron radii of nuclei and thus neutron skin thicknesses
are much less precisely determined.

For the measurement of neutron radii of nuclei, experiments with
particles that probe the neutron distribution with the help of the
strong or weak interaction have to be utilized.  Typical examples are
proton scattering experiments
\cite{Ray:1979qv,Ray:1979zz,Klos:2007is,Terashima:2008zza,Zenihiro:2010zz},
isovector giant dipole excitations by inelastic $\alpha$-particle
scattering \cite{KrBaBoBrHaKaNyTiWo1994}, (${}^{3}$He,t) charge
exchange reactions \cite{Krasznahorkay:1999zz}, the excitation of
pygmy dipole resonances \cite{Klimkiewicz:2007zz} or the study of
antiprotonic atoms
\cite{Trzcinska:2001sy,Jastrzebski:2004yn,Brown:2007zzc}.  Parity
violation in elastic electron scattering is used in the lead radius
experiment PREX at Jefferson Lab
\cite{Horowitz:1999fk,Horowitz:2012tj,Abrahamyan:2012gp}. In this type
of approach the weak form factor of the nucleus is measured and it is
primarily determined by the neutron density
distribution. Unfortunately, the deduced neutron skin thickness of
$0.302 \pm 0.175 ({\rm exp.}) \pm 0.026 ({\rm model}) \pm 0.005 ({\rm
  strange})$~fm carries a large uncertainty.  It is expected to
diminish in future experimental runs.  A noticeably smaller value of
$\Delta r_{np} = 0.15 \pm 0.03 ({\rm stat.}) {}^{+0.01}_{-0.03} ({\rm
  sys.})$~fm was recently reported from experiments of coherent pion
photo-production at the MAMI electron facility \cite{Tarbert:2013jze}.

\citet{Brown:2000pd} found a strong correlation of the neutron skin
thickness of ${}^{208}$Pb with the derivative
$\left. d E(n_{B},\delta)/dn_{B}\right|_{n_{B} = n_{0},\delta = 1}$
of the neutron matter EoS at a
density $n_{0}=0.1$~fm$^{-3}$ in non-relativistic HF calculations with
18 different parameterizations of the Skyrme interaction.  These
observations triggered many theoretical and experimental studies to
explore the relation of isospin dependent properties of nuclei to the
EoS, in particular the density dependence of the nuclear symmetry
energy $E_{\rm sym}(n)$.  
An extension of the Skyrme HF calculations 
in similar studies of RMF models \cite{TyBr2001},
general density functionals in the context of EFT
\cite{Furnstahl:2001un} or the droplet model \cite{Warda:2009tc}
showed the same correlation, which can also be expressed as a
correlation between $\Delta r_{np}$ and the slope parameter $L$. More
recent representations of the $\Delta r_{np}$ - $L$ and similar
correlations of isospin dependent properties can be found in
\citet{Centelles:2008vu,Chen:2010qx,RocaMaza:2011pm,Tsang:2012se,Gaidarov:2012iv,Vinas:2014zla,Vinas:2013hua,Gaidarov:2014qla}.
The consequences for the properties of NSs, such as radius or proton
fraction, were studied, e.g., by
\citet{Horowitz:2000xj,Horowitz:2001ya,Horowitz:2002mb,ToddRutel:2005zz,steiner_05,Avancini:2007sd,Avancini:2007yz,Avancini:2007zz}. The
origin of the $\Delta r_{np}$ - $L$ correlation, its bulk and surface
contributions in nuclei and the relation to Landau-Migdal parameters
are discussed by
\citet{Dieperink:2003vs,Centelles:2010qh,Warda:2010qa}. 
 The neutron skin thickness provides a correlation between $L$ and $J$ as well and
shows a decreasing of $L$ with increasing $J$, in contrast to other
correlations of that type \cite{Lattimer:2012xj}. 
The present data of the $\Delta r_{np}$ -
$L$ correlation is derived only from mean-field calculations of
nuclear structure. 
However, the neutron skin
thickness could be modified by nucleon-nucleon correlations and clustering at the
surface of the nucleus \cite{Typel:2014tqa}.

\subsubsection{Heavy-ion collisions}
\label{sec:HIC}
The EoS of warm or hot, strongly interacting matter can be constrained
in laboratory experiments with HICs. Depending on the beam energy,
the impact parameter, the choice of observables, and the combination
of projectile and target nuclei, very different conditions can be
explored. In the early phase of almost central
collisions of about one GeV, high densities of up to four times the nuclear
saturation density and temperatures of about $40-50$~MeV 
can be reached for a very short time, see e.g.~\cite{Blaettel:1993uz,Fuchs:1997we}. 
In the later stages of a collision, more peripheral or less energetic reactions,
properties of dilute matter at temperatures below 
the critical temperature of the liquid-gas phase transition ($15-20$~MeV) 
and subsaturation densities can be studied.  
We will not consider here ultra-relativistic HICs probing matter 
at very low baryon density and high temperatures.
Since the physics of HICs is a huge field on
its own, we mention only the most important aspects relevant for this
review.

There are fundamental differences between matter in HICs and in
compact stars. Temperatures and densities can be similar as in CCSNe,
but matter in HICs is usually more isospin symmetric,
cf.\ Sec.~\ref{sec:range}. Furthermore, the fireball in a HIC has a
finite size with a fixed number of nucleons that are not necessarily
in thermal equilibrium. This limits for example the maximum mass
number of nuclear clusters formed under these conditions. Matter in
compact stars, which can be treated in the thermodynamic limit, has to
be charge neutral, whereas there is a net charge in HICs fixed by the
initial charge of the two colliding nuclei. In HICs Coulomb
interactions are typically neglected because of the high kinetic
energies. Characteristic timescales in HICs are of the order of a few
fm/c and do not allow for equilibrium with respect to weak
interactions. On the contrary, in catalyzed NSs full equilibrium is reached.  In
compact stars, weak equilibrium with respect to strangeness-changing
reactions is usually assumed, whereas the net strangeness in HICs is
zero.  These differences have to be taken into account when comparing
astrophysical EoSs with constraints from HICs.

The analysis of HICs requires the comparison of measured data to rather
complex theoretical simulations since a dynamical process has to be
followed.  These models are based on different approaches that aim to
solve the relevant transport equations. On the one hand, a set of
Boltzmann-type equations for the single quasi-particle distribution
functions is considered, 
see, e.g., \citet{Danielewicz:1982kk,Danielewicz:1982ca}.
They can be derived consistently as an
approximation of non-equilibrium Kadanoff-Baym theory including
collision and sometimes fluctuation terms
\cite{Bertsch:1988ik,Buss:2011mx}.  On the other hand, simulations
with molecular dynamics models in classical approximations, possibly
including antisymmetrization effects, are also employed
\cite{Aichelin:1991xy,Ono:1992uy,Hartnack:1997ez}. 
One major challenge is to
predict the distribution of observed particles and fragments reliably.
In these models the EoS does not enter directly but the interactions between all
particles which participate in the collision,  as well as in-medium cross
sections of the relevant reactions, which are usually parameterized in
a convenient form \cite{Li:2005jy}. One important aspect is the
momentum dependence of the interaction because particle momenta attain
much larger values in HICs than in nuclei~\cite{Chen:2013uua,Xu:2014cwa}.  
Hydrodynamic descriptions~\cite{Welke:1988zz,Gale:1989dm,Huovinen:2006jp,Gale:2013da}, 
which can make direct use of an
EoS, are more appropriate for studying the evolution of the
high-density phase of a collision, in particular in (ultra-)
relativistic HICs. However, one has to change to a different approach at
later times when the system expands, the density drops and fragments
are formed.

There are several observables in HICs that are sensitive to particular
features of  in-medium interactions that determine the EoS at
supra-saturation densities, see, e.g., \citet{Fuchs:2005yn}.  Not only
nucleons but also mesons, such as pions or kaons, as well as light
nuclei, e.g., ${}^{2}$H, ${}^{3}$H, ${}^{3}$He, and ${}^{4}$He, see,
e.g., \citet{Chajecki:2014vww}, are valuable messengers for the
properties of the medium at high and low densities, respectively.

The collective flow of nucleons exhibits a distinct azimuthal
distribution~\cite{Welke:1988zz}, which can be characterized with coefficients in a
Fourier analysis. The transverse flow in peripheral reactions seems to
be mainly sensitive to the momentum dependence of the mean-field. The
elliptic flow, in contrast, depends strongly on the maximum
compression that is reached and it is correlated with the stiffness of
the EoS. The analysis of laboratory experiments in comparison with
simulations indicates that the incompressibility of symmetric nuclear matter cannot
be too high~\cite{Welke:1988zz,Danielewicz_02,FOPI:2011aa,Fevre:2015fza},
see subsection \ref{sec:compatibility} and Fig.\ \ref{fig:Danielewiczeos+}.

The collision region with the highest densities is best studied with
particles that are produced only there and interact weakly with the
medium after their formation.  Although being a rare probe due to
their subthreshold production \cite{Hartnack:2011cn}, kaons seem to be
a good choice. Their  observation in HICs points  towards a rather low 
incompressibility $K$ with values below $250$~MeV~\cite{Fuchs:2000kp,Sturm:2000dm,Hartnack:2005tr}. 
Consequences of these constraints from HICs on compact star properties were explored,
e.g., by \citet{Sagert:2011kf}. For a discussion of the interplay
between HICs and astrophysical data see, e.g., \citet{aichelin08}.

In recent years, HIC experiments for constraining the EoS mainly focused
on the isospin dynamics~\cite{Li:2002yda,Baran:2004ih,Li:2008gp,DiToro:2008zm,Wolter:2008zw,Tsang:2012se,Cozma:2013sja,DeFilippo:2013ipa,Ademard:2013xzq,Kohley:2014qha}
in order to explore the properties of isospin asymmetric matter in more detail. 
Yield ratios of particle pairs with the same mass but different isospin, 
such as n/p, $\pi^{+}/\pi^{-}$
\cite{Xiao:2013awa},
or fragments ${}^{3}$H/${}^{3}$He \cite{Yong:2009te}
have been intensively investigated. The consideration of single or double ratios 
has the advantage that systematic experimental uncertainties are reduced
and the sensitivity is increased.
A possibility to amplify isospin-dependent effects is the comparison of
collisions with different combinations of projectiles and targets
with more or less neutron excess. 
For example, the isospin diffusion in the
neck region of peripheral and mid-central collisions of ${}^{112}$Sn/${}^{124}$Sn
nuclei is sensitive to the symmetry potential~\cite{Tsang:2008fd,Tsang:2012se}.

The density dependence of the symmetry energy at moderate to high densities
was studied with the help of n/p ratios and their elliptic flow difference
\cite{Cozma:2011nr,Russotto:2011hq,Russotto:2012jb,Russotto:2013fza} as well as
$\pi^{+}/\pi^{-}$ ratios \cite{Reisdorf:2006ie}. The analysis of the latter 
results within transport model simulations 
suggests a decrease of the symmetry energy at high densities, 
which is in conflict with most EoS models~\cite{Xiao:2009zza,Xie:2013np}.
Also, the puzzling results for the effective mass splittings of nucleons in intermediate
HICs \cite{Coupland:2014gya,Zhang:2014sva}
still need a satisfactory explanation \cite{Kong:2015rla}.
Only more accurate measurements will allow to set tighter bounds 
on the symmetry energy at high densities.

\begin{figure}
\includegraphics[width=1.0\columnwidth]{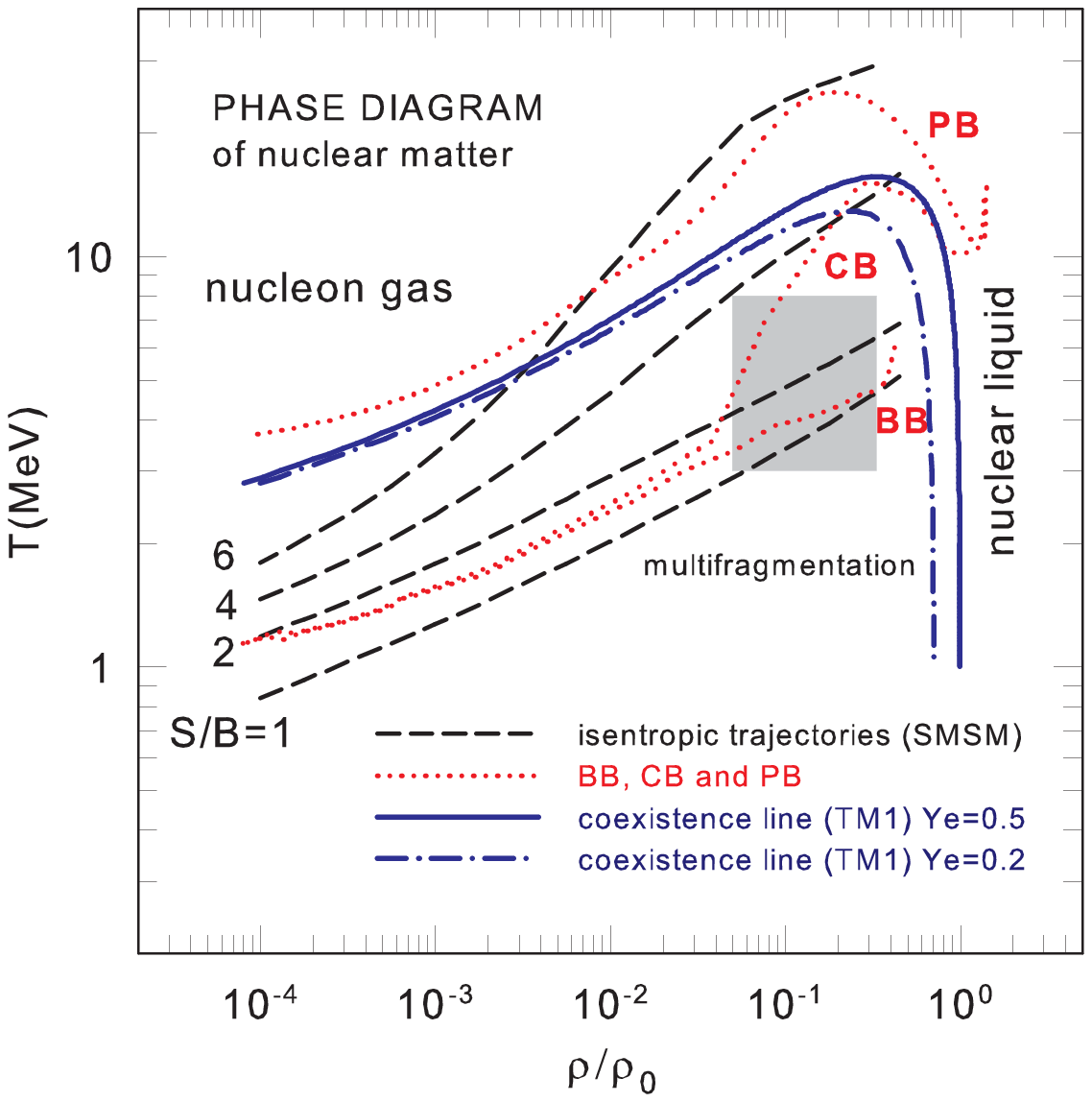}
\caption{(color online) 
\label{fig:multifrag}Nuclear phase diagram in the temperature–baryon density 
plane. Solid black and dashed-dotted purple lines indicate boundaries of the 
liquid–gas coexistence region for symmetric and asymmetric matter calculated 
with TM1 interactions \citep{Sugahara_94}. The shaded area corresponds to 
typical conditions for nuclear multifragmentation reactions \citep{botvina08}. 
The dashed black lines are isentropic trajectories characterized by constant 
entropy per baryon, $s$ = 1, 2, 4 and 6 calculated with the SMSM 
\citep{botvina08}. The dotted red lines show results of a CCSN simulation 
from \citet{sumiyoshi2005}
just before bounce (BB), at core bounce (CB) and post bounce (PB). 
Figure taken from \citet{Buyukcizmeci_13}. }
\end{figure}
Multifragmentation reactions probe conditions very similar to matter
in CCSNe, as is illustrated in Fig.~\ref{fig:multifrag}, where
typical conditions for CCSNe and multifragmentation reactions are
indicated, see the caption for details. In these reactions, a
thermalized system of nuclear matter is formed that is characterized
by subnuclear densities and temperatures of $(3–8)$~MeV. The
de-excitation of the system occurs via nuclear multifragmentation,
i.e., break-up into many excited fragments and nucleons.  For the
theoretical description of such reactions, for instance the Statistical
Multifragmentation Model (SMM) is used, that will be presented in more
detail in Sec.~\ref{sec:nse_eos}.  Statistical models describe
accurately many characteristics of the nuclear fragments observed in
the experiments: cluster multiplicities, charge and isotope
distributions, various correlations and other observables; see,
e.g.,~\citet{gross1990,bondorf95,botvina08}.

\begin{figure}
\includegraphics[width=1.0\columnwidth]{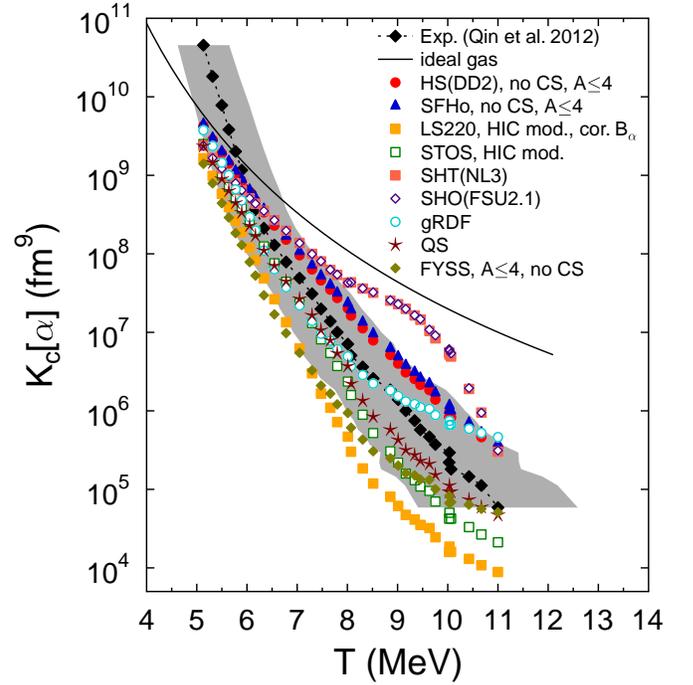}
\caption{(color online) Equilibrium constants of $\alpha$-particles
  extracted from HIC experiments (black diamonds) in
  comparison with those of various theoretical models, which are all
  adapted for the conditions in HICs, as far as possible. The grey band
  is the experimental uncertainty in the temperature
  determination. The black line shows the equilibrium constant of the
  ideal gas model. For details, see \citet{Hempel:2015yma}.}
\label{fig:kalpha}
\end{figure}

The observation of light nuclei, which are emitted in 
Fermi-energy HICs, allows to determine
the density and temperature of warm dilute matter
from experiments \cite{Kowalski:2006ju,Natowitz:2010ti,Wada:2011qm}. 
The derived symmetry energies of the clustered matter clearly indicate
an increase as compared to those obtained in model calculations of uniform matter
that is assumed to be composed solely of nucleons. 
The thermodynamic conditions are similar
to those in the neutrinosphere of  CCSNe
\cite{Horowitz:2014bja}. 
From the observed yields of nucleons and clusters, it was possible
to extract the in-medium binding energies and Mott points of light
clusters \cite{Hagel:2011ws} with the help of 
chemical equilibrium constants \cite{Qin:2011qp}.
\citet{Hempel:2015yma} refined the study of \citet{Qin:2011qp},
taking into account the differences between matter in HIC and CCSNe.
Results for the equilibrium constant 
of the $\alpha$-particle are presented in Fig.~\ref{fig:kalpha}.
A comparison of many EoSs for warm dilute matter clearly shows that
simple NSE descriptions are not sufficient to reproduce experimental data  \cite{Hempel:2015yma}.

\subsection{Neutron matter calculations}
\label{sec:theosmnm}

The simple isospin structure of pure neutron matter simplifies the
nuclear interaction Hamiltonian, such that {\em ab initio}
calculations can be carried out more easily than in the case of
general asymmetric nuclear matter.  Calculations using the different
many-body techniques introduced in Sec.~\ref{sec:manybody} with
well-calibrated interactions are available for a large range in
densities, see, e.g., the review by \citet{Gandolfi:2015jma} and
references therein.  They can serve as important constraints for the
EoS models which we discuss in Sec.~\ref{sec:generalEoS}, although
their results are not directly applicable to astrophysical objects.

At very low densities, neutron matter is dominated by $s$-wave
interactions with a large scattering length, $a = -18.5$~fm, indicating
that the two-neutron system is almost bound. Neutron matter at these
densities is close to the unitary limit, explored experimentally for
cold fermionic atoms \cite{Ho:2004zz,HoZa2004}.  

At intermediate densities, up to roughly nuclear matter saturation
density, and at higher densities, relevant for NSs, many
different calculations from {\em ab-initio} methods exist. 
We mention some calculations, without claiming completeness for the  list
below. Seminal results are the variational
calculations by \citet{Friedman_81} and \citet{Akmal98} using the
Urbana/Argonne nuclear two-and three-body forces. BHF calculations are
reported for instance by \citet{Baldo97,Zhou_04}, DBHF results by
\citet{vanDalen_04, Sammarruca_06, Sammarruca_12}. \citet{Rios_09}
compare SCGF calculations for thermodynamic properties of hot neutron
matter with the corresponding BHF calculations. In
\citet{Carbone_thesis} SCGF results including an effective
three-nucleon force are obtained for finite temperature and
extrapolated to vanishing temperature. 
\citet{Horowitz_05a,Horowitz:2006pj} apply the virial expansion 
to dilute neutron matter at nonzero temperature.
An early application of the virial expansion was the description of a
neutron gas in supernovae by \citet{BuCo1977} using the soft-core Reid potential.
QMC calculations for zero temperature
neutron matter using different versions of the Argonne/Urbana nuclear
potentials are presented, e.g., by 
\citet{Carlson_03,Gandolfi_09,Wlazlowski_09,Gandolfi_12}. 
Recent QMC~\cite{Gezerlis_13,Gezerlis_14, Roggero_14, Wlazlowski_14} and
coupled cluster~\cite{Baardsen:2013vwa,Hagen_13} calculations 
employ chiral potentials. Neutron matter is particularly
interesting for chiral forces, since only a few low-energy constants (LECs)
accompanying the
contact terms are involved up to next-to-next-to-next-to leading order
(N$^3$LO) including three- and four-nucleon forces. In addition, at
least up to roughly saturation density, the MBPT results with RG
evolved and unevolved chiral forces are in very
good agreement, showing that neutron matter in this range behaves
perturbatively to a very good approximation, see,
e.g., \citet{Kruger_13}. Comparison with QMC calculations corroborate
the perturbative nature of neutron matter at these
densities~\cite{Gezerlis_14}. Calculations of neutron matter with
chiral forces can be found
in~\citet{Hebeler:2009iv,Tews:2012fj,Kruger_13,Hebeler_14}. 
In-medium $\chi$EFT following different power-counting schemes is
applied to neutron matter at zero and nonzero temperature,
e.g., by~\citet{Lacour_09,Fiorilla_12,Drischler_13}.
\begin{figure}
\includegraphics[width=1.0\columnwidth]{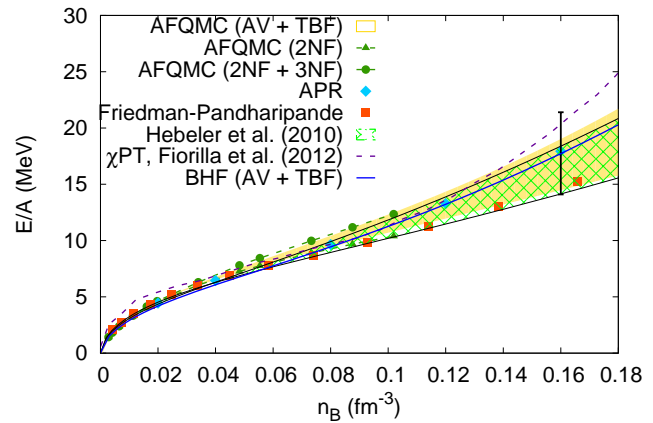}
\caption{(color online) Comparison of results for the energy per baryon of neutron 
matter at $T= 0$ from different {\em ab initio} approaches. The vertical bar represents the range given at saturation density in \citet{Kruger_13}. For details see text.
 \label{fig:neutronmatter}}
\end{figure}

In Fig.~\ref{fig:neutronmatter} we show the energy per baryon of pure
neutron matter as a function of baryon number density as obtained in a
few of the different approaches cited above.  The yellow region
corresponds to the AFQMC calculations of
\citet{Gandolfi_09,Gandolfi_12} with the Argonne two-body potential.
The width of the band indicates uncertainties related to the different
phenomenological three-body forces (TBF). The AFQMC results of
\citet{Wlazlowski_14} with chiral forces (green lines and symbols) including only two-nucleon
interactions (2NF) and those including a three-nucleon force (2NF +
3NF) are in very good agreement with the former ones.  The variational
results of \citet{Friedman_81} give lower values than \citet{Akmal98}
and lie at the lower boundary of the AFQMC calculations.  The width of
MBPT results by \citet{Hebeler:2013nza} shows mainly uncertainties in
the three-nucleon forces employed.  The constraint derived by
\citet{Kruger_13} at saturation density comparing different chiral
forces and cutoff schemes within a MBPT calculation is shown as a
vertical error bar.  Fig.\ \ref{fig:neutronmatter} also displays the
$\chi$EFT results by \citet{Fiorilla_12} and BHF calculations from
\citet{Vidana10a}.  The latter use Argonne plus phenomenological
three-body potentials.

We address two points concerning these different results. First,
three-nucleon forces (and potentially more, depending on the
resolution scale and density) are important in neutron matter and add
substantial repulsion at and above saturation density. This is seen
for example by comparing QMC calculations with and without
three-nucleon forces. Secondly, all results, with phenomenological or
chiral forces, applying different many-body techniques, are in
reasonable agreement up to saturation density.  This shows that the
{\em ab-initio} many-body calculations 
represent a reliable constraint on the EoS of neutron matter up to
nuclear densities, see subsection \ref{sec:compatibility} and
Fig.\ \ref{fig:Danielewiczeos+}.

The situation is different for calculations of symmetric nuclear matter.
They have been a cornerstone for many-body methods since decades. However,
the empirical saturation point is very
difficult to obtain.  In addition, symmetric matter is unstable with
respect to cluster formation at densities below saturation which is strongly
temperature dependent and leads to an increase of the binding
energy. Therefore theoretical many-body calculations of symmetric
matter are not as reliable as for neutron matter, and cannot serve as a
constraint on the EoS at present. Instead phenomenological models are 
adjusted to the empirical properties of symmetric matter.

\subsection{Astrophysical Observations}
\subsubsection{Neutron star masses and radii}
\label{sec:neutron_star_masses}
Presently, the main astrophysical constraint stems from the
measurements of two very massive NSs in NS-white dwarf systems which
have been reported with unprecedented high precision.  For the first
binary system, the determination is based on Shapiro delay; a general
relativistic effect~\cite{demorest_10}. It yields a mass of  $(1.928 \pm 0.017)$~M$_\odot$~\cite{Fonseca:2016tux}.
In the second case a well-known
structure model for the white dwarf is combined with the analysis of
orbital data to obtain a mass of $2.01 \pm 0.04 $~M$_\odot$ for the
NS~\cite{Antoniadis_13}. 
There are indications of even more
massive NSs, e.g., in black widow and redback
systems~\cite{vanKerkwijk:2010mt, Romani_12,Kaplan:2013hii}. In these
cases, the pulsar is accompanied by a low mass companion of a few
0.001~M$_\odot$ \ (black widows) or near 0.2~M$_\odot$ (redbacks),
which is bloated and strongly irradiated by the pulsar.  However, the
analysis of these systems is much more model dependent than for NS-white dwarf systems. 
In particular, the companion's light curve has to be modeled inducing large
uncertainties in the mass determination.  Although the most probable
mass for the NS indicates a very massive object, the results do not yet
reach the same reliability as the mass determinations of
\citet{Fonseca:2016tux} and \citet{Antoniadis_13}. 
This also holds for the NS in the eclipsing X-ray binary Vela X-1,
where a high mass of $2.12 \pm 0.16$~M$_\odot$ has been reported by 
\citet{falanga2015}.

Smaller NS masses have been measured in various binary systems;
see~\citet{Lattimer:2012nd} for a recent compilation. In some cases
masses have been derived very precisely from the orbital
parameters of the system without much model dependence in the
analysis. 
Particularly precise measurements have been performed for several binary
NS systems giving masses close to the canonical value of 1.4~M$_\odot$. 

At the other end, the lowest NS masses could be
interesting for constraining the EoS via their formation
history. Originally, \citet{Podsiadlowski:2005ig} suggested 
to consider pulsar B in the double pulsar system J0737–3039, with a very low
and precisely measured mass  of $(1.2489 \pm 0.0007)$~M$_\odot$. If it
originates from the collapse of a progenitor star with O-Ne-Mg 
core and the loss of matter during the formation of the NS is negligible,  
the baryon number, or equivalently the corresponding baryon mass $M_B$ for the
NS, is strongly constrained from the properties of the
white-dwarf progenitor. 
Its mass has been determined to be $1.366
M_\odot\le M_B\le 1.375$~M$_\odot$~\cite{Podsiadlowski:2005ig}, assuming
a stationary non-rotating object. \citet{Kitaura:2005bt} conclude on a
slightly smaller but similar mass of $M_B = (1.36 \pm 0.002)$~M$_\odot$
from simulations of an electron-capture supernova. 
A similar system, J1756-2251, has recently been observed
with a slightly lower gravitational mass of $(1.230 \pm 0.007)$~M$_\odot$
for the pulsar with the lower mass ~\cite{Ferdman:2014rna}. 
The constraint on the EoS arising
from the relation between gravitational and baryon mass of these low
mass NSs depends  strongly on assumptions. 
First, there is no complete consensus about the
formation history of these systems and the origin from a O-Ne-Mg
electron-capture supernova 
is not confirmed, see, e.g., \citet{Tauris_2013}.  
Secondly, already a possible baryon
loss of  1\% during the formation of the compact star broadens
the corresponding baryon mass region by increasing it by roughly a
factor of two. This effect is included in the constraints
derived by \citet{Kitaura:2005bt} but only for two particular EoSs.

\begin{figure}
\includegraphics[width=1.0\columnwidth,angle = -180,bb = 51 51 553 383]{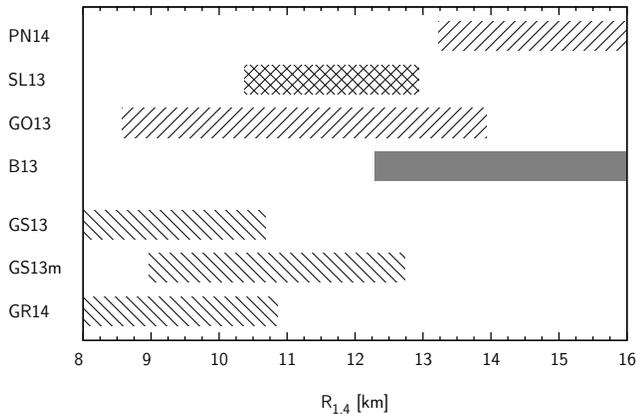}
\caption{(color online) Summary of recent NS radius
  estimations from observations for a star with the canonical mass
  of $1.4$~M$_\odot$. Shown are 2-$\sigma$ error bars. Figure courtesy of M. Fortin. 
  For details see table 2 of Ref.~\citet{Fortin_14}
  and text. \label{fig:nsradii}}
\end{figure}

The ultimate constraint on the EoS would be a determination of radius
and mass of the same object, see,
e.g., \citet{Read_08,Ozel_09a,Ozel_10,Steiner_13}. 
Recently, \citet{Sotani:2013dga}  discussed how
for low-mass NSs this could be translated into a constraint 
for a particular combination of $K$ and $L$.
Currently, radius observations are much more model dependent than mass
measurements, largely because radius measurements are much more 
indirect.  
Possible sources of systematic error 
include the composition of the atmosphere, the strength of the 
magnetic field, the distance to the source, interstellar
extinction, residual accretion in 
binaries, brightness variations over the surface, and the effects of 
rotation in sources with unknown spin frequencies, 
see \citet{Miller:2013tca, Potekhin:2014hja} for details.
The importance of uncertainties in
determining the radius depends on the type of the object
observed. Currently radii are extracted from four different types of
sources:
\begin{enumerate}
\item Isolated neutron stars (INSs). For INSs it is extremely
  difficult to determine the distance, the magnetic field and the
  composition of the atmosphere inducing altogether very large
  uncertainties on the radius determinations from these sources, see,
  e.g., the discussion by \citet{Potekhin:2014hja}.
\item Quiescent X-ray transients (QXTs) in low-mass X-ray
  binaries. The thermal emission from the surface of the NS
  can be observed in the quiescent phase, i.e., when the accretion of matter
  from the companion is absent or at least strongly reduced. They are
  promising sources for radii determinations, since the magnetic field
  of QXTs is low due to the accretion of matter. In
  addition, the atmosphere is likely to be composed of light elements
  (H or possible He) and if they are situated in globular clusters,
  the distance is well-known. Recent radius determinations from QXTs
  are shown in Fig.~\ref{fig:nsradii} as 
``SL13''~\cite{Steiner_13}, ``GS13'' and ``GS13m''~\cite{Guillot_13} and
 ``GR14'' \cite{Guillot_14}. 
    Although being promising sources, the results are still subject to
    many uncertainties. 
    For instance, there has been recent
  discussion about the NS's atmospheric composition in
  quiescent low-mass X-ray binaries (qLMXBs) in the globular cluster NG 6397. 
   \citet{Guillot_13,  Guillot_14} favor an unmagnetized hydrogen atmosphere and obtain a
  small radius of $R_{1.4} = 9.4\pm 1.2$ km (90\% confidence level)
  for a  $1.4$~M$_\odot$ NS~\cite{Guillot_14}. \citet{Heinke:2014xaa} argue that a
  Helium atmosphere is more probable which leads to approx.\ 2~km 
  larger radii. 
  ``GS13m''  therefore shows the result of \citet{Guillot_13} upon
  excluding the qLMXB in  NGC 6397.
\item Bursting NSs (BNSs). From these objects  very powerful photospheric
  radius expansion (PRE) bursts are observed. 
  Similar to QXTs, they have low magnetic fields and
  a light element atmosphere and, if situated in globular clusters, the
  distance can be well determined. The main uncertainties arise here
  from the modeling of the photospheric burst and no consensus has
  yet been reached, see, e.g.,
  \citet{Ozel_12,Galloway:2012me,Guver:2013xa,Steiner_13,Poutanen:2014xqa}. 
 Recent radius determinations from BNSs are shown in Fig.~\ref{fig:nsradii};
 ``PN14'' from \citet{Poutanen:2014xqa}, ``GO13'' from
  \citet{Guver:2013xa}, and  ``SL13''
  from \citet{Steiner_13}.
\item For rotation-powered millisecond pulsars (RP-MSPs) radii
  can be determined from the shape of the X-ray pulses.
  They are
  interesting, in particular if the mass is known from radio
  observations. The result of \citet{Bogdanov_13} and
  \citet{Verbiest_08} for J0437-4715 is shown  in
  Fig.~\ref{fig:nsradii} (``B13''). 
 Although with large uncertainties, the possible mass-radius region of neutron star XTE J1807-294 
 has been derived  by \citet{leahy11}.
   
\end{enumerate}
QXTs as well as BNSs are likely to rotate at
a frequency of few hundreds of Hz, inducing a non-negligible
rotational deformation  that complicates
the analysis of the X-ray
spectra. The latter effect is expected to affect the radii by roughly
10\%~\cite{Poutanen:2014xqa,Baubock:2014xha}. 
\citet{Ozel_15} include this rotational
correction in a combined analysis of observed QXTs and
BNSs. This common analysis of
12 sources statistically reduces the error on the final result for the radius obtained, $R_{1.5} = 10.1 - 11.1 $km.

In conclusion,  present radius determinations are subject to many
assumptions and uncertainties;  see also the discussion in
\citet{Potekhin:2014hja} and \citet{Fortin_14}.
Currently, they cannot provide as stringent constraints as some of the mass
measurements. 
However, much observational efforts are directed
to NS radius measurements. 
Future high-precision X-ray astronomy, such
as proposed, e.g., by the projects NICER, ATHENA+ or LOFT, 
and the gravitational wave signal of NS mergers expected for the near future (see Sec.~\ref{sec:binaries})
would help substantially to constrain radii and consequently the EoS of
NS matter.

Another interesting possibility to determine a relation between mass
and radius of a NS would be the observation of the
gravitational redshift at the NS surface. In
\citet{Cottam_02}, a value of $z=0.35$ has been deduced from
narrow absorption lines in the spectra of X-ray bursts from EXO
0748-676. This observation could, however, not be confirmed
later~\cite{Cottam_08}. In addition, the rotation frequency of the
source was measured to be of the order of $(400-500)$ Hz.
Therefore, one
expects not narrow but wider lines. 
As a consequence \citet{Lin_2010} concluded that these spectral lines
can actually not originate from the surface.
However, a more recent study by the same group suggests that 
line profiles from rotating NSs might actually be narrower than
initially predicted~\cite{Baubock:2012bj}.

\subsubsection{Neutron star cooling and rotation}
\label{sec:cooling}
While computing mass and radius of a NS requires only a known relation
between total pressure and total energy density, see Eq.\ (\ref{eqs:tov}), 
the cooling of NSs
depends on a detailed description of the interior composition which
determines the heat transport and amount of neutrino emission.  The
occurrence of superfluid states does barely influence the structure of
a NS but has great impact on cooling.  First, pair breaking and
formation are important neutrino emission channels. 
At a later stage, for temperatures below the corresponding critical temperature, the
related pairing gaps suppress the emission of neutrinos and reduce the
heat capacity and thermal
conductivity~\cite{Yakovlev:2004iq,Blaschke:2004vq,Page_06}. NS
cooling and the description of superfluid phases are reviewed, e.g.,
by \citet{Weberbook,Yakovlev:2004iq,potekhin15}. Although cooling
calculations face many difficulties due to a large number of not
precisely known quantities, the direct cooling observation of the
young, only about 330 years old NS in Cassiopeia A
\cite{Heinke:2010cr} over a period of 10 years promises to give direct
insight into its composition~\cite{Page:2010aw, Shternin:2010qi,
Blaschke:2011gc,Sedrakian:2013pva}. A recent analysis of 
{\em Chandra} observations suggests that the initially reported fast
cooling has to be considered with caution due to involved statistical
uncertainties~\cite{Posselt:2013xva} and possible instrumental
problems \cite{elshamouty13}. Nevertheless, the 
theoretical work has demonstrated the strong impact precise cooling
observations can have.

Neutron star rotation rates can be
determined precisely from pulsar observations. Theoretically, slowly rotating stars can be
described in the Hartle and Thorne
approximation~\cite{HartleThorne}, see, e.g., the textbook by
\citet{Weberbook}. Numerically precise
solutions~\cite{Nozawa_98} can be
obtained up to the mass shedding limit, the Kepler frequency, see
the textbook by \citet{bookfriedmanstergioulas}. The value of
the Kepler frequency depends on the EoS and an observed frequency
above the Kepler limit for a given EoS would clearly exclude the
underlying model. Currently observed rotation rates
\cite{Hessels_06,Kaaret_07} with a maximum of 716 Hz do not put
relevant constraints on the EoS~\cite{Haensel_09}, but this could change if
more rapidly rotating stars are observed in the future. Other
astrophysical observations can be used to derive EoS constraints,
e.g., quasi-periodic oscillations in soft-gamma ray repeaters, see,
e.g., \citet{Steiner_09}, \citet{sotani12}.  However, the modelling of
these events is complicated and often relies on additional model
assumptions.

\subsection{Summary of constraints on the symmetry energy}

\begin{figure}
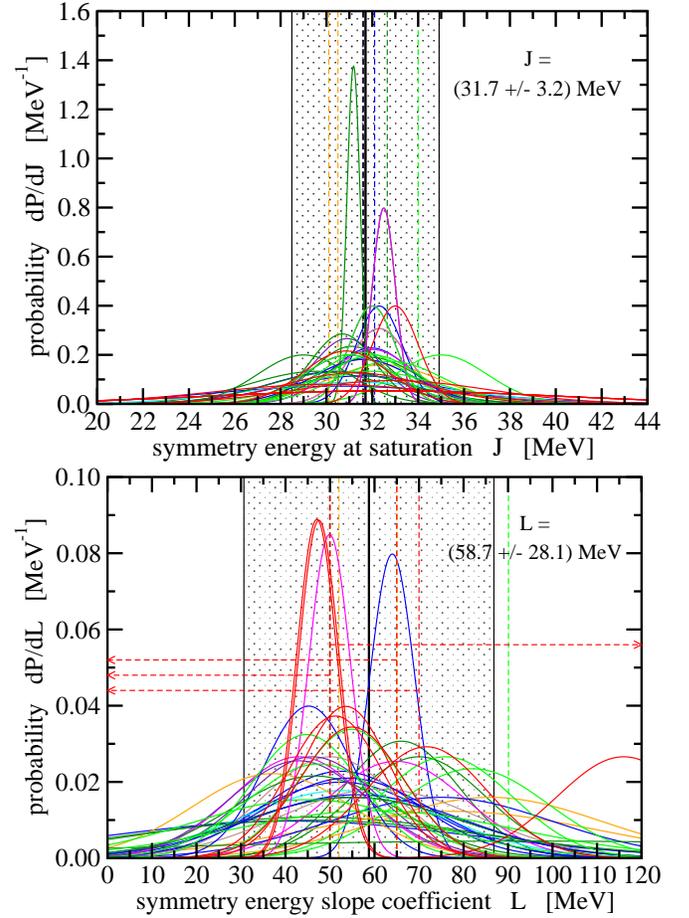

\includegraphics[width=1.0\columnwidth]{fig16a.eps}\hfill
\includegraphics[width=1.0\columnwidth]{fig16b.eps}\hfill
\caption{(color online) Probability
distribution of the symmetry energies at saturation $J$ (top panel)
and of the symmetry energy slope parameter $L$ (bottom panel) from various
studies. See text for details.}
\label{fig:JL-prob}
\end{figure}

Besides the constraints discussed above, further constraints 
on the symmetry energy at saturation and on the slope parameter
are collected in the literature, e.g., 
in \citet{Tsang:2012se,Lattimer:2012xj,Li:2013ola,Lattimer:2014sga}.
Extending these data collections,
the compilations in 
Figure \ref{fig:JL-prob}
depict the probability distributions of $J$ and $L$ values, 
respectively. For simplicity, the probability distributions
are assumed to be of Gaussian form with an area normalized to one. They are
centered at the obtained values for $J$ and $L$ with widths that are
given by the errors of the individual studies.  Dashed vertical lines
are used if no uncertainty is available. Allowed ranges with upper or
lower bounds are indicated by arrows.  
In general, nuclear matter parameters such as $J$ and $L$ and their errors
are correlated in models that are used in the analysis of experimental and observational
data. Since these correlations are rarely specified in the literature,
see, however, \citet{Lattimer:2012xj}, 
we treat $J$ and $L$ as independent quantities.
The origin of the constraints is encoded
in Figure \ref{fig:JL-prob}
by colors, see Table \ref{tab:JL_c}.
Averaging over this selection of results (excluding upper and lower
bounds) we find $J = (31.7 \pm 3.2)$~MeV and $L = (58.7 \pm 28.1)$~MeV
with an error for $L$ that is considerably larger than that for $J$. 

\begingroup
\squeezetable
\begin{table}
\caption{Sources of the data for the symmetry energies at saturation $J$ and slope parameters $L$ used in figure \ref{fig:JL-prob},
including the color code.
  \label{tab:JL_c}}
\begin{center}
\begin{tabular}{l| l}
\hline 
\hline 
Type of constraint 
& References \\ 
\hline
systematic of nuclear masses 
& \citet{Myers:1995wx} \\
(green lines) 
& \citet{Danielewicz:2003dd} \\
& \citet{Mukhopadhyay:2006ra} \\
& \citet{Klupfel:2008af} \\
& \citet{Kortelainen:2010hv} \\
& \citet{Liu:2010ne} \\
& \citet{Moller:2012} \\
& \citet{Lattimer:2012xj} \\
& \citet{Wang:2013yra} \\
& \citet{Vinas:2013hua} \\
\hline
neutron skin data and other 
& \citet{Centelles:2008vu}\\
nuclear structure information 
& \citet{Warda:2009tc} \\
(blue lines)
& \citet{Chen:2010qx} \\
& \citet{Chen:2011ek} \\
& \citet{Agrawal:2012pq} \\
& \citet{Dong:2012zza} \\
& \citet{Zhang:2013wna} \\
& \citet{Wang:2013zia} \\
& \citet{Danielewicz:2013upa} \\
& \citet{Vinas:2013hua} \\
\hline
nuclear resonances 
& \citet{Klimkiewicz:2007zz}\\
(magenta lines)
& \citet{Carbone:2010az} \\
& \citet{RocaMaza:2012mh} \\
& \citet{Colo:2013yta} \\
& \citet{Paar:2014qda} \\
\hline
dipole polarizability of nuclei 
& \citet{Roca-Maza:2013mla}\\
(indigo lines)
& \citet{Tamii:2013cna} \\
\hline
$\alpha$ and $\beta$ decay of nuclei 
& \citet{Dong:2013nma} \\
(brown lines)
& \citet{Dong:2012ah} \\
\hline
global nucleon optical potentials 
& \citet{Xu:2010fh} \\
(light blue lines) & \\
\hline
heavy-ion collisions 
& \citet{Tsang:2004zz,Tsang:2008fd} \\
(orange lines)
& \citet{Chen:2004si,Chen:2005ti} \\
& \citet{Li:2005jy} \\
& \citet{Shetty:2007zg} \\
& \citet{Sun:2010km} \\
& \citet{Kohley:2010zz} \\
\hline
theoretical calculations 
& \citet{Erler:2010zh} \\
(light green lines)
& \citet{Gandolfi:2011xu} \\
& \citet{Fiorilla:2011sr} \\
& \citet{Erler:2012qd} \\
& \citet{Hebeler:2013nza} \\
& \citet{Kruger:2013kua} \\
& \citet{Nazarewicz:2013gda} \\
\hline
properties of neutron stars 
& \citet{Newton:2009vz} \\
(red lines)
& \citet{Steiner:2010fz} \\
& \citet{Gearheart:2011qt} \\
& \citet{Steiner:2011ft} \\
& \citet{Wen:2011xz} \\
& \citet{Vidana:2012ex} \\
& \citet{Sotani:2012xd} \\
& \citet{Lattimer:2012xj} \\
& \citet{Steiner:2012xt} \\
& \citet{Sotani:2013jya} \\
\hline 
\hline
\end{tabular}
\end{center}
\end{table}
\endgroup

\section{Modelling the Equation of State}
\label{sec:generalEoS}

In this section, we describe models for the EoS, i.e., particular
realizations of the formal approaches that we introduced in
Sec.~\ref{sec:formal}. It is evident that the requirements on the EoS
are different depending on the astrophysical situation to which
they are applied, see Sec.~\ref{subsec:requirements}. We place
emphasis on the general purpose EoSs that cover the full
thermodynamic parameter range in $T$, $n_B$ and $Y_q$. 
An overview of the currently available
ones is presented in Sec.~\ref{sec:general_purpose}. The reason for
this is twofold.  First, there is a plethora of different EoSs
available in the literature applicable in a particular context,
especially for cold $\beta$-equilibrated NSs. To list here
all available models, including variations of free parameters
such as coupling constants in the phenomenological models, would be
fruitless. Secondly, excellent reviews already exist, see,
e.g., \citet{Lattimer:2006xb,Baldo_11,Lattimer:2012nd}. 
We therefore discuss only
some selected aspects of EoSs of cold $\beta$-equilibrated neutron
stars, see Sec.~\ref{sec:ns_eos}, and of EoSs describing homogeneous
matter at finite temperatures suitable for describing hydrostatic
proto-NS, see Sec.~\ref{sec:finite_t_eos}.  Sec.~\ref{sec:nse_eos}
will give a few representative examples of EoSs that describe
clusterization and nuclear statistical ensembles at finite
temperatures but that are restricted to sub-saturation densities.

Almost exclusively, phenomenological models have been used up to now
in the context of astrophysical applications due to the computational
complexity in the description of clustered matter. This concerns the
NS crust, see Sec.~\ref{sec:crust}, the NSE-type EoSs in
Sec.~\ref{sec:nse_eos}, and, in particular the general purpose EoSs
discussed in Sec.~\ref{sec:general_purpose}. The most advanced
descriptions of dense matter are found for particular conditions. In
fact, the {\em ab-initio} approaches 
discussed in Sec.~\ref{sec:formal}, if not restricted to pure neutron
matter, have only been applied to homogeneous nuclear matter at
various neutron to proton ratios with an interpolation to obtain the
$\beta$-equilibrated NS EoS. Some {\em ab-initio}
calculations exist at finite temperature as some of the methods, for
instance, SCGF, are easier to treat at nonzero temperature, but the
composition is fixed and matter is homogeneous, see
Sec.~\ref{sec:finite_t_eos} for some examples. It is desirable that in
the future reliable approaches will be developed to describe strongly
interacting matter for all relevant conditions needed in compact star
astrophysics. A first step is that the information obtained from
{\em ab-initio} neutron matter calculations,
experiments and NS observations is fully exploited to constrain the
general purpose models.

\subsection{Neutron star EoS}
\label{sec:ns_eos}

\begin{figure*}
\centering
\includegraphics[width=0.7\textwidth]{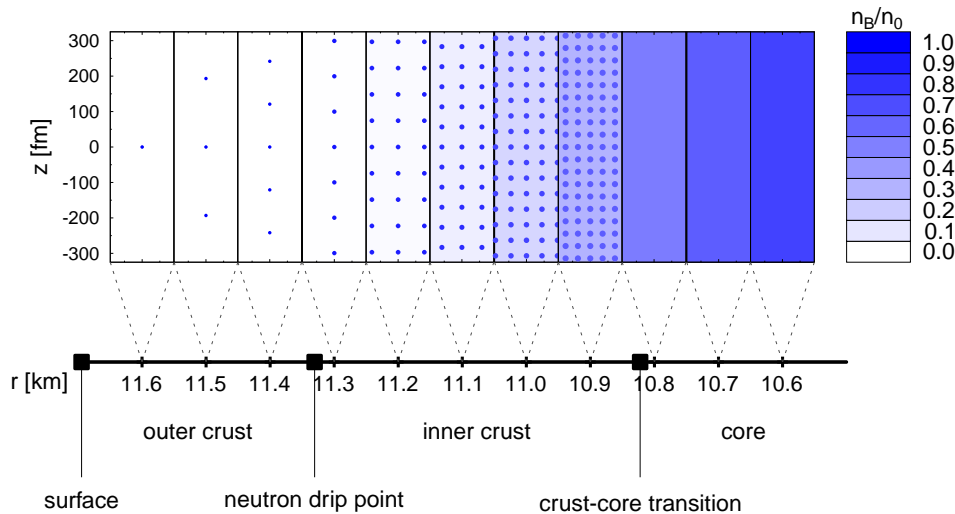}
\caption{Graphical representation of the structure and composition 
of the crust of a 1.44~M$_\odot$ NS. 
Each subpanel shows in color coding (see legend at right) the mean local 
density of the nucleons, 
for the position in the NS as indicated in the bottom 
part of the figure. The illustration uses the SLy4 EDF, 
the EoS of \citet{Ruester:2005fm} for the outer crust, 
and  the EoS of \citet{Douchin:2001sv} for the inner crust and the core.
(color online).}
\label{fig:ns_struc}
\end{figure*}

The physics of NSs has been discussed in detail in
several works, see, e.g.,
\cite{Baym:1975mf,Baym:1978jf,Glendenning:1997wn,Heiselberg:2000dn,Lattimer:2000nx,Lattimer:2004pg,Lattimer:2006xb,Sedrakian:2006mq,Chamel:2008ca,potekhin10,Lattimer:2010uk}.
 NS matter is charge neutral and can be considered as cold ($T=0$)
and in general in $\beta$-equilibrium. The EoS thus depends only on
one state variable, which can be conveniently chosen, e.g., to be the
baryon number density.  The EoS
entirely determines global properties of stationary NSs, such as
masses and radii. For nonrotating stars
with negligible magnetic field, they are found by solving the
Tolman-Oppenheimer-Volkoff (TOV) equations
\cite{Tolman:1939jz,Oppenheimer:1939ne}:
\begin{eqnarray}
\nonumber 
\frac{d P(r)}{d r}&=&-G 
\frac{\left[\epsilon(r)+P(r) \right]
\left[M(r)+4 \pi r^3 P(r)\right]}{r^{2}\left[ 1-\frac{2 G M(r)}{r}\right]} \: ,
\nonumber \\
\frac{d M(r)}{d r}&=&4 \pi \epsilon(r) r^2 \: ,
\label{eqs:tov}
\end{eqnarray}
relating the gravitational mass of the star, $M$, inside a radius $r$
to pressure $P$ and energy density $\epsilon$. The EoS in terms of $P$
and $\epsilon$ closes the system of equations. Despite the assumptions
of zero temperature and $\beta$-equilibrium, the calculation of the NS
EoS is not a trivial task, particularly if microscopic methods are
applied, see, e.g., \cite{Baldo97,Vidana10a,Schulze11,Baldo_11} and
references therein. 
The domain of validity of some {\em ab-initio} methods is restricted 
to rather low densities not exceeding nuclear saturation density substantially. 
The size of higher-order contributions in systematic expansions, 
such as chiral EFT approaches, increases 
with density, and the composition of matter at
suprasaturation densities is rather uncertain, see
Sec.~\ref{sec:hyppuzzle}. In order to cover the whole density range 
required to describe NSs, microscopic EoSs can be extended 
at high densities with generic parameterizations, such as piecewise polytropes. 
Thereby the uncertainty of the NS mass-radius relation 
and the dependence on the model parameters 
can be explored, see, e.g, \citet{Hebeler:2013nza}.

\subsubsection{Neutron star crust EoSs \& unified neutron star EoSs}
 \label{sec:crust} 
For mass densities below about $10^{4}$~g/cm${}^{3}$, an atmosphere of
partially ionized atoms and electrons forms the outer part of a NS
with an EoS given, e.g., by
\citet{Feynman:1949zz,Rotondo:2009cr,deCarvalho:2013rea}. At higher
densities, the spatial region that is made up of inhomogeneous
nucleonic matter and electrons not bound to nuclei in 
$\beta$-equilibrium is called the crust. 
It can be divided into an outer crust
with a plasma of nuclei and electrons as degrees of freedom and
an inner crust where also unbound neutrons exist.
Fig.~\ref{fig:ns_struc} gives a graphical representation of the state of matter
in the crust. The results shown employ the EoS of 
\citet{Ruester:2005fm} for the outer crust, where experimentally measured 
binding energies have been used in combination with nuclear 
structure calculations
with the SLy4 EDF. For the EoS of the inner crust and the core, the results of 
\citet{Douchin:2001sv} are taken, which are based on 
Thomas-Fermi calculations using
the same SLy4 EDF. To obtain the radial structure of the assumed NS with a 
mass of 1.44~M$_\odot$, the TOV equations (\ref{eqs:tov}) were solved.

The outer crust is composed of completely ionized nuclei in a sea of
electrons of almost constant density due to the large
incompressibility of the highly degenerate electron fluid. In the
standard picture, only a single nuclear species exists at a given
density. These nuclei form a body centered cubic (bcc) lattice of ions
as demonstrated, e.g., in classical one-component plasma simulations,
see Sec.~\ref{sec:nuc_in_cell}.  
The individual nuclei at their lattice sites can be identified as
the blue dots in Fig.~\ref{fig:ns_struc}.
At densities of about
$10^{7}$~g/cm$^{3}$ and below, a crystal of ${}^{56}$Fe nuclei is
expected to form.  With increasing density the
lattice constant decreases and the
electron chemical potential rises substantially. It becomes
energetically favorable to squeeze electrons into the nuclei,
converting protons to neutrons. A sequence of bcc lattices with
more and more neutron rich ions on the lattice sites appears the
deeper one penetrates into the NS. 
Each change from one to the next nuclear species is
connected to a phase transition with a jump in the density,
cf.\ section \ref{sec:Coul}. The series 
of nuclei in the outer crust is determined by their masses and
is influenced strongly by shell effects.  With
the recent progress to measure masses of very neutron-rich nuclei
experimentally with high precision, the order of ions in the outer
crust from ${}^{56}$Fe via ${}^{62}$Ni, ${}^{64}$Ni, ${}^{66}$Ni,
${}^{86}$Kr, ${}^{84}$Se, and ${}^{82}$Ge could be established
with increasing depth
\cite{Wolf_13,Kreim:2013rqa}. For higher densities, the nuclear masses
from theoretical models, e.g., liquid-drop or EDF type, have
been used to determine the chemical composition
of the outer crust. Since the early works of
\citet{Salpeter:1961zz,Baym:1971pw} the theoretical description of the
outer crust is well settled and the main changes result from
improvements in the theoretical description of exotic nuclei not
studied experimentally so far
\cite{HaZdDo1989,HaZd1990b,Haensel:1993zw,Ruester:2005fm,RocaMaza:2011pk,Pearson:2011zz,Wolf_13}.
  
At a mass density of approximately  $10^{11}$g/cm${}^{3}$ the neutron
drip density is reached, i.e., the neutron chemical potential becomes
too high for nuclei at the lattice sites to bind additional
neutrons. In Fig.~\ref{fig:ns_struc}, the contribution of 
these unbound neutrons, indicated by the blue background color,
becomes only visible at sufficiently high densities.
These unbound neutrons can propagate more or less freely
through the lattice, though their interaction with the lattice could
modify the crystalline structure~\cite{Kobyakov_13}. A proper
treatment of the periodic crystal structure and its effect on neutron
and electron properties requires a description using band structure
models as in solid state physics \cite{Pethick:1996yj,Chamel:2004in}.
Since the temperature is very low, effects of neutron pairing could be
important. This is less relevant for the basic thermodynamic
properties of the crust matter itself but it has to be considered
for dynamic processes and thermal properties, in particular neutron
star cooling. In view of neutron superfluidity, the so-called
entrainment effect has to be taken into account in hydrodynamic
descriptions of the NS's inner crust and core. In this case
the momentum of one fluid is not aligned with its particle current,
but depends on the particle currents of all other fluids
\cite{Chamel:2004in,Gusakov:2005jx,Carter:2004zr,Carter:2004pp,Carter:2004he,Chamel:2006rc,Gusakov:2009mb}.

The exact location of the neutron drip density is sensitive to details
of the theoretical model, in particular to the isospin dependence of
the effective interaction due to the very large neutron excess
encountered in the crust, see, e.g.,
\citet{Douchin:2000kx,Douchin:2001sv,Steiner:2007rr,Ducoin:2011fy}.
The properties of nuclei also change inside matter at high densities,
mostly in the inner crust of the NS when they are surrounded by a gas
of neutrons and the electric field is screened by the electrons.

Several studies are devoted to the description of nuclei inside matter
and the effects on the EoS, see, e.g.,
\citet{Baym:1971ax,BaBuIn1972,Ravenhall:1972zz,Negele:1971vb,Lamb:1978zz,Cheng:1997zza,Baiko:1999iw,Douchin:2001sv,Matsuzaki:2005sw,Ducoin:2008xs,Papakonstantinou_13,Raduta:2013fna,Aymard_14}.
In most cases, the Wigner-Seitz approximation in spherical cells
surrounding a single nucleus is employed, see section
\ref{sec:nuc_in_cell}.  With increasing depth inside the crust, 
nuclei approach each other and the action of the short-range nuclear
interaction beyond the size of an individual nucleus has to be
considered.  First, a strong deformation of nuclei and a change in the
shell structure is observed in model calculations, see, e.g.,
\cite{Oyamatsu:1993zz,OyYa1994,DoHaMe2000}, such that they finally
touch and 
the sequence of classical pasta phases is found, 
see, e.g.,
\cite{Buchler:1971zz,Ravenhall:1983uh,Watanabe:2003xu,Maruyama:2005vb,Avancini:2008kg,Watanabe:2009vi,Newton:2009zz}. 
The picture of the classical pasta phases with their specific
geometries and phase transitions changes if more general shapes are
allowed in full three-dimensional calculations with less restrictions
on the symmetries
\cite{Watanabe:2004tr,Nakazato:2009ed,Okamoto:2011tc,Schneider:2014lia,Schuetrumpf:2014aea}.
The extension of the inner crust, the types of pasta phases and the
transition density to uniform matter depends crucially on the density
dependence of the symmetry energy
\cite{Pethick:1994ge,Oyamatsu:2006vd,RocaMaza:2008ja,Porebska:2009gk,Grill:2012tp,Grill:2014aea}.
The pasta phase could be relevant for the neutrino transport in
the PNS and the subsequent cooling of the NS. For example
\citet{Horowitz:2014xca} showed that the pasta phase can reduce the
electrical and the thermal conductivities. The
reduced electrical conductivity 
might be related to the observed upper limit of
X-ray pulsar spin periods \citep{pons13}.

The crust has a subdominant effect on global properties of NSs such as mass or
radius. Therefore,  a crust EoS
is matched often to an EoS of uniform matter from an independent model
calculation.  Considerable effort is 
required in developing an EoS which describes matter from the
surface to the center of the NS in a unified manner, i.e., on the
basis of the same interaction model, including a description of
inhomogeneous matter in the crust. This is important for detailed 
predictions of NS radii and for
dynamical properties.
Only few such unified NS EoSs exist, see, e.g.,
\citet{Douchin:2001sv,Fantina_13,Miyatsu:2013hea,Baldo:2013ska,Gulminelli:2015csa}.
Unified NS EoSs can also be obtained from the general purpose EoS, which will
be discussed in Sec.~\ref{sec:general_purpose}, by applying zero 
(or negligibly small) temperature and $\beta$-equilibrium conditions. 
However, the aforementioned dedicated unified NS EoS
models often give a more detailed description of non-uniform NS matter. A
comparison of these classes of models is useful to investigate limitations
of general purpose EoSs regarding their description of nuclei in dense and cold 
matter. 

In contrast to a conventional star in a hadronic model, the
EoS is very different for strange stars, see, e.g.,
\citet{Haensel_86, Alcock_86}, and the structure of the crust is still a 
matter of debate, see, e.g., \citet{Alford:2006bx,Jaikumar:2005ne,Oertel_08}.

\subsubsection{Composition of the neutron star core}
\label{sec:hyppuzzle}

The composition of matter at suprasaturation densities reached in
the NS core is very uncertain and, in particular other
particles than nucleons and electrons are expected to appear. In the
literature muons, pions, kaons and their condensates, hyperons,
nuclear resonances and quarks have been considered, see,
e.g., \citet{Glendenning:1997wn}. There is even the possibility of
absolutely stable strange quark matter \citep{witten84,Farhi:1984qu} 
and pure strange stars \citep{Haensel_86, Alcock_86}. 
In this context, the recent discovery of two NSs with masses of about 2
M$_\odot$~\cite{demorest_10,Antoniadis_13,Fonseca:2016tux} 
has triggered intensive
discussions, since without an interaction, any additional degree of
freedom softens the EoS simply by lowering the Fermi energies of the
particles present. As a consequence, a lower maximum mass is obtained
and many older models containing additional particles are in
contradiction with the NS mass constraint.

Phenomenological quark models can easily be supplemented with the
necessary repulsion at high densities. 
As an example, for the NJL model Lagrangian of Eq.~(\ref{eq:njllagrangian}) 
this can be achieved by adding a vector interaction term of the form
\begin{equation}
 \mathcal{L}_{V} = 
 G_V (\bar{\psi}\gamma^\mu\psi) (\bar{\psi}\gamma_\mu\psi) \: .
\end{equation}
Maximum NS masses
above 2 M$_\odot$ can then be obtained, see, e.g.,
\citet{Klahn:2006iw,Alford07, Weissenborn11, Zdunik:2012dj}, and
\citet{Buballa:2014jta} for a recent review. \citet{masuda13} proposed
that the transition from hadronic to quark matter might be a
cross-over potentially leading to an increase of the maximum
mass. However, this scenario requires an ad-hoc interpolation scheme
to connect the hadronic and the quark phase, see also
\citet{Kojo:2014rca}.
The same scenario was recently applied in the
context of PNSs by \citet{masuda15}.

Hyperonic degrees of freedom are more difficult to
reconcile with a 2 M$_\odot$ NS. Most models that include hyperons
predict that they appear at $n_B \sim 2-3 n_{\rm sat}$ but lead at the
same time to maximum NS masses of $\sim 1.4$~M$_\odot$, well below the
highest observed ones. 
Sometimes this is called the ``hyperon puzzle''
in the literature \cite{lonardoni2014}. 
A similar effect is observed with nuclear
resonances \cite{drago14} and meson condensates. 
It is thus obvious that additional
repulsion is needed to stiffen the high-density EoS.

Different solutions have been proposed to overcome this problem. The
first one is that a transition to quark matter appears at sufficiently
low densities such that hyperons or other additional hadronic particles have not yet softened the EoS too much. A two family
scenario with low mass compact hadronic stars and high mass quark
stars has been recently discussed by \citet{Drago_13}.

Another possibility is to modify the interactions at high densities.
Hyperonic interactions have been  extensively studied in this
respect. Since experimental data are scarce and furnish
only weak constraints on the interactions at
subsaturation densities, even
less is known about the hyperon-nucleon ($YN$) and
hyperon-hyperon ($YY$) interactions at the relevant densities in the
core of NSs, see Sec.~\ref{sec:exdatafewbody}. Presently several phenomenological EoS models exist 
that contain hyperons and predict maximum NS masses
in agreement with observations, see, e.g.,
\citet{Hofmann00,RikovskaStone:2006ta,Bednarek:2011gd,
  Weissenborn11b,Weissenborn11c,Bonanno11, Colucci2013,Lopes2014,
  Banik:2014qja, vanDalen2014, Gomes2014, Oertel:2014qza}.
The crucial point is that the
interaction is adjusted to provide the necessary repulsion.

In microscopic models the missing repulsion for hyperons is more
difficult to obtain. Naturally, one would expect it to arise from
three-body forces. But, using a microscopic model based on the
BHF approach, \citet{Vidana10b}
found that even adding a phenomenological three-body force was not
enough to allow for the existence of stars that are massive enough to
be compatible with observations.  Recent relativistic DBHF
calculations~\cite{Katayama2014}, including automatically part of the
three-body forces, reproduce hyperonic NSs with two solar masses, but
with a nuclear EoS that is either too stiff or does not give enough
binding in contradiction with known properties of symmetric nuclear
matter at saturation. On the other hand, in recent calculations using
an auxiliary field diffusion Monte Carlo method
(AFQMC)~\cite{lonardoni2013a, lonardoni2013b,lonardoni2014}, the
authors have found that a sufficiently strong repulsive three-body
force, constrained by the systematics of separation energies in a
series of hypernuclei, can produce an EoS stiff enough to satisfy the
2 M$_\odot$ constraint, even if a strong model dependence due to the
phenomenological nature of the hyperonic two- and three-body forces is
apparent.  In conclusion, there are still many open questions
regarding the role of hyperons and other additional non-nucleonic degrees of 
freedom in NSs.

\subsection{EoS of uniform matter at finite temperature}
\label{sec:finite_t_eos}

The thermal properties of nuclear matter are an important subject on
its own and many studies are actually performed without an
astrophysical application. Examples are the {\em ab-initio}
calculations of neutron matter at finite
temperatures, some of which have been mentioned in
Sec.~\ref{sec:theosmnm}, or studies of the nuclear liquid-gas phase
transition which occurs at subsaturation densities if Coulomb and
finite size effects are neglected.  The liquid-gas phase transition is
an important aspect of the low-density nuclear matter EoS and has been
studied extensively in the literature, see, e.g.,
\citet{Barranco:1980zz,Muller:1995ji}.  An example is shown in
Fig.~\ref{fig:multifrag}.  In addition to BHF~\cite{baldo_99,
  Baldo_04} and DBHF~\cite{Haar:1986ii,Huber_98} calculations extended
to finite temperature, consistent SCGF calculations are reported in~\citet{Rios_08} and \citet{Fiorilla_12,Wellenhofer:2014hya,Wellenhofer:2015qba} 
study the phase diagram of nuclear matter applying
in-medium $\chi$EFT.

Here, we are mainly interested in EoS relevant for astrophysical
applications. Finite temperature effects are of particular relevance
for PNSs, CCSNe, and NS mergers,
and can be studied in HICs too. Although more scarce
than NS EoSs, there is a variety of works considering EoSs of
homogeneous matter at fixed entropies or
temperatures and hadronic charge fractions. 
Many of these EoS have
actually been developed for studying PNSs by considering
characteristic hydrostatic configurations that represent different
evolutionary stages. The history
concerning additional particles, such as hyperons, mesons or quarks,
is almost as long as for cold neutron
stars, reaching from the discussion of quark matter formation or meson
condensates to hyperons, see \citet{Prakash:1996xs} 
for an early review, and \citet{Pons:2000iy,Pons:2000xf,Bombaci:2011mx,
Menezes:2007hp,Dexheimer:2008ax,yasutake09,Beisitzer:2014kea,masuda15} 
for a few examples.

Often, PNSs are  approximated as isentropic,
i.e., having a constant entropy per baryon $s$, 
and/or a fixed lepton
or electron fraction ($Y_{L^{(e)}}$ or
$Y_{e}$, respectively) with or without trapped neutrinos.
Sometimes isothermal PNS are considered, too. If the crust and
envelope of the PNS are neglected, the situation is still similar to
that for cold NS: Matter is uniform and for each parameter combination
of $s$ or $T$ and $Y_{L^{(e)}}$ or $Y_e$ the EoS is still
one-dimensional, i.e., it depends only on one state variable, e.g.,
baryon number density. 

Most {\em ab initio} calculations of the EoS of nuclear matter at finite
temperature concern pure neutron matter, 
see, e.g., Sec.~\ref{sec:theosmnm}, or symmetric nuclear matter, 
since general asymmetric nuclear
matter asks for more involved computations. Some exceptions exist,
e.g., the EoS of \citet{togashi_13} uses the variational method,
starting from a nuclear Hamiltonian that is composed of the Argonne
v18 and Urbana IX potentials. The authors aim at providing a full
general purpose EoS that can be applied in astrophysical simulations
(see Sec.~\ref{sec:general_purpose}). However, in their first work,
they considered only uniform nuclear matter, but for various
temperatures and asymmetries. To simplify computations, the
frozen-correlation approximation is employed: the self-energies and
correlation matrix elements are evaluated at zero temperature.  This
approximation is motivated by the results of
\citet{baldo_99}. \citet{togashi_13} validate it by comparing with
results for fully minimized calculations. 

The frozen correlation approximation is standard in finite-temperature
BHF calculations aimed to model the PNS EoS, too,
see, e.g., \citet{Nicotra_05, nicotra_06, Burgio11, chen_12}. Effects
of hyperons and/or quarks were considered, too, within these works
relying on different models for the interactions. For
instance, in the study of \citet{chen_12}, for the quark phase a model
based on the Dyson-Schwinger equations of QCD was used,
cf.,~\ref{sec_ds}, whereas \citet{nicotra_06} apply the MIT bag
model. As known for BHF calculations, see Sec.~\ref{sec:hyppuzzle}, the
maximum mass of a cold NS including additional degrees of freedom such
as hyperons or quarks lies in general well below the canonical value
of 2~M$_\odot$. Exceptions are
the hybrid NS and PNS models of \citet{chen_12} where
much higher maximum masses in the vicinity of 2~M$_{\odot}$ were
found.

Applying phenomenological interactions to finite temperature matter,
an obvious question is whether the effective couplings, determined via
zero temperature properties of strongly interacting (mainly nuclear)
systems, depend on temperature. This was addressed by
\citet{fedoseew_15}, where the thermal properties of asymmetric
nuclear matter have been analyzed within a relativistic approach. The
parameters of a density-dependent relativistic hadron (DDRH) nuclear
field theory, similar to a density-dependent RMF model, have been
adjusted to reproduce DBHF results for in-medium self energies. In
particular, it was shown that the temperature modifications of the
nucleon-meson couplings is almost negligible. A comparison of the free
energy of nuclear matter using the Brussels Skyrme interaction with
the results of \citet{Fiorilla_12} leads to the same
conclusion~\cite{Fantina_private}.
In \citet{moustakidis_09}, where a
momentum-dependent finite-range term is added by hand to a Skyrme type
interaction, temperature-dependent couplings are obtained but this
dependence is weak up to temperatures of 30~MeV. 

\citet{Constantinou:2014hha} investigate thoroughly the finite
temperature properties of the bulk EoS. They employ the
potential model of \citet{Akmal98}, which is fitted to results from
variational calculations, and compare it with the typical Skyrme
EDF 
SKa from \citet{koehler_76}.  The latter
parameterization is also applied in the H\&W EoS, see
Sec.~\ref{sec:h_and_w}.  Analytical formulae are derived for all
thermodynamic state variables and their derivatives at finite
temperatures, simplifying the use in astrophysical applications. 
A similar study of the thermal properties of the EoS but for finite 
range interactions was recently published by \citet{constantinou15}.

\subsection{EoS of clustered matter at finite temperatures}
\label{sec:nse_eos}

Complementary to investigations of bulk properties of warm and
dense uniform matter, there are many works 
studying inhomogeneous warm matter
within NSE based models, see Sec.~\ref{sec_nse}. 
Most of them do not cover the full parameter space relevant 
for simulations of CCSNe or NS
mergers, partly since they are designed for a particular
application, e.g., multifragmentation experiments or nucleosynthesis
aspects. A typical problem is the omission of interactions and/or
medium modifications of nuclei. As a result,
the EoS does not provide a realistic description at high densities. 
Nevertheless, these models allow to investigate 
important aspects of the EoS of clustered matter, e.g.,
the chemical composition and the role of excited states.
The impact of the different model ingredients depends on the thermodynamic 
conditions.  

The chemical composition of matter at baryon densities 
roughly below $10^{-3}$~fm$^{-3}$ and a few MeV temperature 
is mainly driven by the nuclear binding energies and the 
treatment of thermal excitations.
Simple mass formulas provide binding energies for the widest possible range 
of nuclei that are considered in statistical models. 
A liquid-drop type mass formula
is used in the Statistical Model for Supernova Matter (SMSM)
\citep{botvina04,botvina08,Buyukcizmeci_14}. It is based on the Statistical
Multifragmentation Model (SMM) \citep{bondorf95,sagun_14} which has proven
successful in the analysis of fragment yields in low-energy HICs.
The parameters of the used liquid-drop mass formula, including temperature effects, 
have been calibrated by the analysis of experimental multifragmentation
data.
A liquid-drop parameterization is also used in 
the statistical model of \citet{Raduta_09,Raduta:2010ym}.
The NSE model of \citet{Blinnikov:2009vm} considers up to 20,000
nuclei, whose binding energies are taken from the theoretical mass
formula of \citet{koura_05}. Binding energies from \citet{myers_90,myers_94} 
are adopted in the work of \citet{ishizuka03} 
incorporating about 9000 different nuclei. 
Results of the microscopic-macroscopic FRDM model
\cite{moeller_95} or 
the Duflo-Zuker model \cite{Duflo:1995ep}
are used, e.g., in \citet{Gulminelli:2015csa}.
Tables with theoretical masses from
fully microscopic models, mainly EDFs, 
usually cover a smaller range of
nuclei. Experimental binding energies \cite{audi_03,audi_12} 
are available for an even smaller number of nuclei rather
close to the valley of stability. They are used as far as 
available in some NSE based models, 
but have to be supplemented by theoretical masses for more exotic nuclei.
The choice of different sources can result in artificial 
jumps of the isotopic abundances at the boundaries, 
see, e.g., \citet{Buyukcizmeci_13}.

\begin{figure}
\centering
\includegraphics[width=0.92\columnwidth]{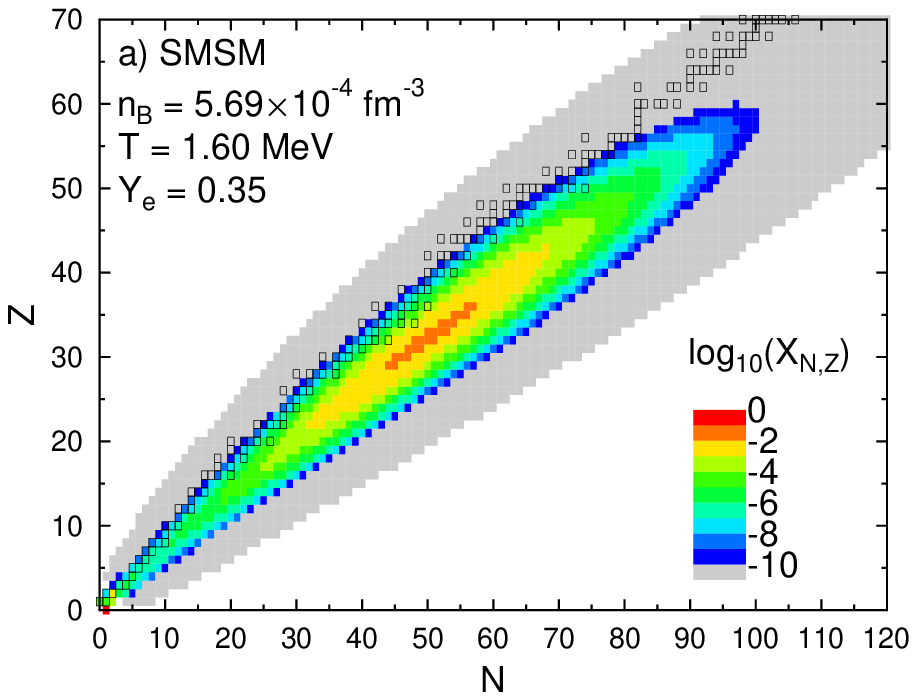}
\includegraphics[width=0.92\columnwidth]{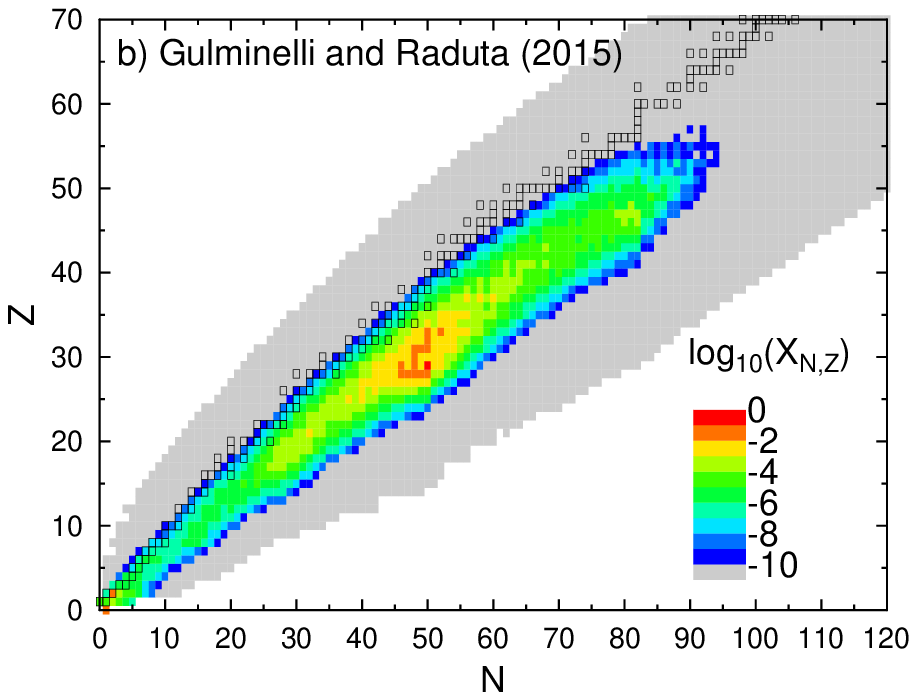}
\includegraphics[width=0.92\columnwidth]{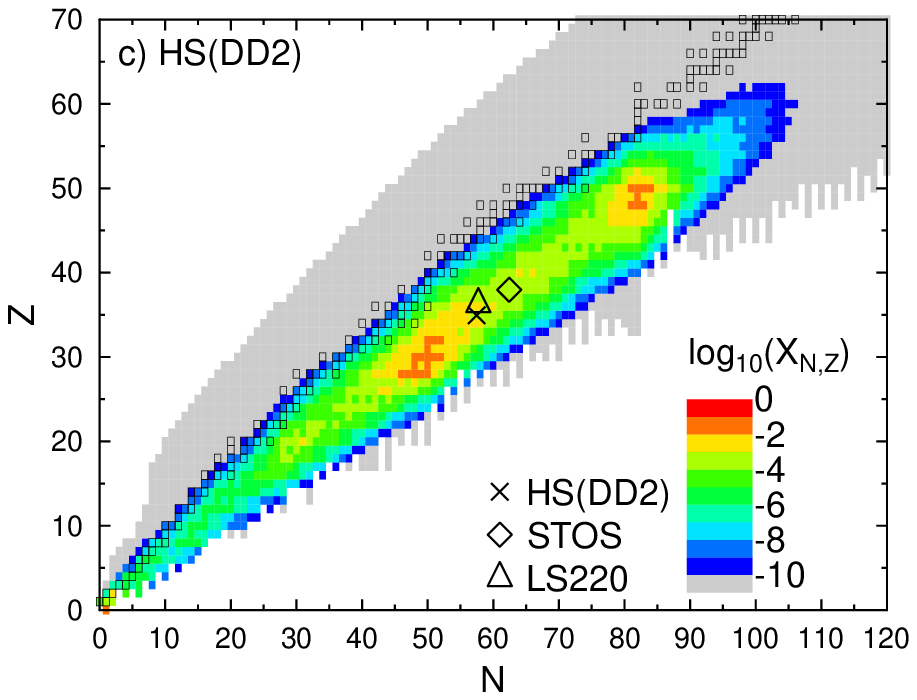}
\caption{Composition of matter in the center of a CCSN 
6~ms before bounce at thermodynamic conditions taken from a simulation
of \citet{Perego_15}. The color map shows the distribution of nuclei 
(mass fractions) in the SMSM \cite{Buyukcizmeci_14} (a), 
the EoS of \citet{Gulminelli:2015csa} (b) and 
the HS(DD2) model (c).  
The black cross indicates in panel (c) the average heavy nucleus. 
The black diamond and triangle show the representative heavy nucleus
of STOS and LS220, respectively, calculated within the SNA. (color online)}
\label{fig:nse_dist_new}
\end{figure}

 As an illustration, we compare in Fig.~\ref{fig:nse_dist_new} nuclear abundances
for typical conditions in the collapse phase of a CCSN for three 
statistical models.
The different predictions mainly reflect the extrapolation of 
binding energies to neutron rich nuclei. This depends on the 
choice of the mass model since the most abundant nuclei are situated outside the 
region where masses are experimentally known. A mono- or bimodal structure is obtained, 
peaked around magic numbers in panels b) and c), demonstrating the importance 
of shell effects, which are missing in models based on 
simple mass formulas, e.g., \citet{Raduta:2010ym,Buyukcizmeci_14}, as in panel a) or models that 
use the Thomas-Fermi approximation for the description of nuclei, e.g., in \citet{Aymard_14}. 
Broad and even bimodal distributions cannot be represented within the SNA, see section \ref{sec_nse_sna}, 
which is employed, e.g., in the general purpose models STOS or LS220, discussed in section
\ref{sec:general_purpose}. Nevertheless, the predictions for the average heavy nucleus show
only a moderate deviation compared to that of the statistical HS(DD2) model and
global thermodynamic quantities are only slightly modified \cite{burrows_84}. 
However, these differences in the composition
are relevant for electron capture reactions during the
collapse phase and thus can influence the dynamics of a CCSN, as will be
discussed in Sec.~\ref{sec:ccdynamics}.
For very neutron-rich conditions, the range of nuclei considered in the table, 
which is indicated by the grey region in Fig.~\ref{fig:nse_dist_new}, also matters. 
Different rules for determining the boundary are employed, e.g., 
vanishing neutron separation energies or binding energies.

With increasing temperature, excited states of nuclei are
populated. These are considered explicitly in models that use
temperature dependent degeneracy factors. Usually, level densities of
Fermi-gas type are employed, e.g., those of \citet{iljinov_92} in
\citet{Raduta_09,Raduta:2010ym} or of \citet{fai_82} in
\citet{ishizuka03}, see Sec.~\ref{sec:nse_excited}. 
An alternative approach is to incorporate
temperature effects directly in the mass model, e.g., in the SMSM, by
introducing a temperature dependence in the coefficients of the mass
formula, in particular in bulk and surface contributions to the
energy. Part of the differences in Fig.~\ref{fig:nse_dist_new}
  also result from the treatment of excited states. The actual
treatment in the model also influences the dissolution of heavy
clusters with increasing temperature and the change of the abundance
distributions. They are dominated more strongly by light clusters at
high $T$, which are usually handled independently of the heavier
nuclei using experimental binding energies and sometimes correct quantum
statistics.  At very low densities, the finite-temperature EoS is given model
independently by the virial EoS, see Section \ref{sec:veos}.  The
VEoS with light species (p, n, d, t, ${^{3}}$He, $\alpha$) was
discussed by \citet{PrSiUs1987} for conditions in HICs. In
\citet{Horowitz:2005nd} neutrons, protons and $\alpha$-particles
were considered as basic constituents and experimental information
on binding energies and phase shifts was used in the calculation of
the second virial coefficients.  \citet{O'Connor:2007eb} added
${}^{3}$H and ${}^{3}$He nuclei, and even heavier species were
included in \citet{Mallik:2008zt}. A relativistic VEoS was given
already by \citet{Venugopalan:1992hy} for an interacting gas of
nucleons, pions and kaons.  A general finding, present in all
models, is that deuterons, tritons and helions appear abundantly for
typical conditions of supernova matter in addition to
$\alpha$-particles
\cite{Sumiyoshi08a,Heckel09,Typel:2009sy,Hempel:2015yma,Pais:2015xoa}.
The presence of light clusters can modify weak interaction rates and therefore
the dynamics of astrophysical processes, see Sec.~\ref{sec_application_in_astro}.

With increasing baryon density, above approximately
$10^{-3}$~fm$^{-3}$, interaction effects start to play a role and
medium effects on the binding energies have to be incorporated in the
statistical models. They modify the chemical composition in comparison
to results of pure NSE models that use vacuum binding energies.  In
essentially all models, the screening of the Coulomb potential due to
the electron component is taken into account in the Wigner-Seitz
approximation, see \ref{sec:Coul}.  The repulsion of nuclei is often
modeled by an excluded-volume mechanism, see
\ref{sec:clusterdissolution}, e.g., in \citet{sagun_14} and
\citet{Raduta_09}. In \citet{Raduta:2010ym} the cluster partition sums
are solved by using Metropolis Monte Carlo techniques which allow to
consider configuration dependent excluded-volume corrections. A
comparison of predictions from the geometric excluded-volume approach
with those of the more microscopically inspired approach using mass
shifts, see \ref{sec:qs}, is presented by \citet{Hempel:2011kh}
concentrating on light nuclei.  The dissolution of clusters with
increasing density cannot be described properly in basic statistical
models unless an excluded-volume mechanism or mass shifts are
considered.  The treatment of the homogeneous nucleonic matter
contribution is relatively similar in most NSE based EoS models,
employing, if at all, phenomenological mean-field approaches with
various interactions, see Sec.~\ref{sec:compatibility} for a
discussion of the compatibility with present day constraints.
Interactions of unbound nucleons are not included in the SMSM,
therefore it is only applicable to dilute matter. A comparison of
different NSE-type models, can be found in \citet{ishizuka03} and
\citet{Buyukcizmeci_13} with detailed and comprehensive analyses of
the nuclear composition.

\subsection{General purpose equations of state}
\label{sec:general_purpose}

In this section we describe EoSs which cover the full thermodynamic
parameter range necessary for astrophysical simulations of CCSNe or NS
mergers. Such EoS not only have to be available for finite
temperatures and different charge fractions, but should include a
description of non-uniform matter at subsaturation densities where
nuclei appear and a description of homogeneous matter at high
densities and/or temperatures.

There exist only a few such EoSs. To give an overview, 
we summarize their
particle content, disregarding leptonic degrees of freedom here
and in the following, and some key properties for cold NSs of the
different models in Table~\ref{tab:eos3d}. We indicate if the EoS
is publicly available in tabulated form or as a computer code
(see appendix~\ref{app:databases} for a list of different online
resources). Key nuclear matter properties of the nuclear
interaction models, which are used in the EoS models of
Table~\ref{tab:eos3d}, are given in Table~\ref{tab:eos_para}. 

We remark that  all of the presently
available general purpose EoSs are included in the discussion,
even though many of them are in strong disagreement with some
astrophysical, experimental, or theoretical
constraints. However, several of them are still used for
reference applications.  Any `benchmarking' of EoSs
depends on which constraints are chosen from the many available in
the literature, and there is not a single model that fulfills all of
them, and not even the very limited set of constraints that we will
consider below.  A general purpose EoS with
particular deficits can be interesting  because it depends very much on the
astrophysical context and the  specific application whether a
constraint is relevant or not. For example, cluster formation at
low densities seems to be more important than the neutron matter
EoS for the dynamics of CCSNe (see Sec.~\ref{sec:ccdynamics})
but in NS mergers probably the opposite is the case. Due to the very
limited number of general purpose EoSs and their importance for
astrophysical applications,  we first give a
complete overview. A critical discussion
follows  at the end in  Sec.~\ref{sec:compatibility}.

\begingroup
\squeezetable
\begin{table*}
\caption{Characteristic properties of the currently existing general purpose EoSs. Top part: EoSs containing nucleons and nuclei,
  bottom part: EoSs including additional hadronic or quark
  degrees of freedom. Listed are the
  nuclear interaction model used, the 
  included particle degrees of freedom, the maximum mass $M_{m\rm ax}$ of cold,
  spherical (non-rotating) NSs and their radii at a fiducial
  gravitational mass $M_G$ of 1.4~M$_{\odot}$. We have included in addition the
  compactness $\Xi = G M_G/ R$  
  of the maximum mass configuration.
  In nuclear interactions labeled with *, the nucleon masses have been changed 
  to experimental values without a refitting of the coupling constants. This 
  induces  a marginal change of the interaction.
  \label{tab:eos3d}}
\begin{center}
\begin{tabular}{c| c c c c c c l}
\hline 
\hline 
Model              & Nuclear &  Degrees & $M_{\rm max}$ & $R_{1.4 {\rm M}_{\odot}}$ & $\Xi$ & publ. & References \\ 
                   & Interaction &  of Freedom   & (M$_\odot$)        & (km)                &       &  avail.           & \\ 
\hline
H\&W               & SKa      & $n,p,\alpha,\{(A_i,Z_i)\}$                & 2.21\footnote{Values taken from \citet{marek09b}.}
 &   13.9 $^a$ &      & n & \citet{ElHi1980,hillebrandt1984} \\ 
LS180              & LS180    & $n,p,\alpha, (A,Z)$                       & 1.84 & 12.2 & 0.27 & y & \citet{Lattimer:1991nc} \\ 
LS220              & LS220    & $n,p,\alpha, (A,Z)$                       & 2.06 & 12.7 & 0.28 & y & \citet{Lattimer:1991nc} \\ 
LS375              & LS375    & $n,p,\alpha, (A,Z)$                       & 2.72 & 14.5 & 0.32 & y & \citet{Lattimer:1991nc} \\ 
STOS               & TM1      & $n,p,\alpha, (A,Z)$                       & 2.23 & 14.5 & 0.26 & y & \citet{shen_98,Shen:1998by,Shen:2011qu} \\ 
FYSS               & TM1      & $n,p,d,t,h,\alpha,\{(A_i,Z_i)\}$          & 2.22 & 14.4 & 0.26 & n & \citet{furusawa13} \\ 
HS(TM1)            & TM1*     & $n,p,d,t,h,\alpha,\{(A_i,Z_i)\}$          & 2.21 & 14.5 & 0.26 & y & \citet{Hempel09,Hempel_11a} \\ 
HS(TMA)            & TMA*     & $n,p,d,t,h,\alpha,\{(A_i,Z_i)\}$          & 2.02 & 13.9 & 0.25 & y & \citet{Hempel09} \\ 
HS(FSU)        & FSUgold* & $n,p,d,t,h,\alpha,\{(A_i,Z_i)\}$          & 1.74 & 12.6 & 0.23 & y & \citet{Hempel09,Hempel_11a} \\ 
HS(NL3)            & NL3*     & $n,p,d,t,h,\alpha,\{(A_i,Z_i)\}$          & 2.79 & 14.8 & 0.31 & y & \citet{Hempel09,Fischer_13} \\ 
HS(DD2)            & DD2      & $n,p,d,t,h,\alpha,\{(A_i,Z_i)\}$          & 2.42 & 13.2 & 0.30 & y & \citet{Hempel09,Fischer_13} \\ 
HS(IUFSU)          & IUFSU*   & $n,p,d,t,h,\alpha,\{(A_i,Z_i)\}$          & 1.95 & 12.7 & 0.25 & y & \citet{Hempel09,Fischer_13} \\ 
SFHo               & SFHo     & $n,p,d,t,h,\alpha,\{(A_i,Z_i)\}$          & 2.06 & 11.9 & 0.30 & y & \citet{Steiner_12} \\ 
SFHx               & SFHx     & $n,p,d,t,h,\alpha,\{(A_i,Z_i)\}$          & 2.13 & 12.0 & 0.29 & y & \citet{Steiner_12} \\ 
SHT(NL3)           & NL3      & $n,p,\alpha,\{(A_i,Z_i)\}$                &  2.78 &  14.9    & 0.31 & y & \citet{shen_11_nl3} \\ 
SHO(FSU)        & FSUgold  & $n,p,\alpha,\{(A_i,Z_i)\}$                &  1.75 &   12.8   & 0.23     & y & \citet{shen_11_fsu} \\ 
SHO(FSU2.1)        & FSUgold2.1&$n,p,\alpha,\{(A_i,Z_i)\}$                &  2.12 &  13.6    & 0.26     & y & \citet{shen_11_fsu} \\ 
\hline 
LS220$\Lambda$  & LS220   & $n,p,\alpha,(A,Z),\Lambda$                & 1.91 & 12.4 & 0.29 & y  & \citet{Oertel:2012qd,Gulminelli:2013qr} \\ 
LS220$\pi$      & LS220   & $n,p,\alpha,(A,Z),\pi$                    & 1.95 & 12.2 & 0.29 & n  & \citet{Oertel:2012qd,Peres_13} \\ 
BHB$\Lambda$       & DD2     & $n,p,d,t,h,\alpha,\{(A_i,Z_i)\}, \Lambda$ & 1.96 & 13.2 & 0.25 & y & \citet{Banik:2014qja} \\ 
BHB$\Lambda\phi$   & DD2     & $n,p,d,t,h,\alpha,\{(A_i,Z_i)\}, \Lambda$ & 2.11 & 13.2 & 0.27 & y & \citet{Banik:2014qja} \\ 
STOS$\Lambda$      & TM1     & $n,p,\alpha,(A,Z),\Lambda$                & 1.90 & 14.4 & 0.23 & y & \citet{Shen:2011qu} \\ 
STOSYA30           & TM1     & $n,p,\alpha,(A,Z),Y$                      & 1.59 & 14.6 & 0.17 & y & \citet{ishizuka_08} \\ 
STOSYA30$\pi$      & TM1     & $n,p,\alpha,(A,Z),Y,\pi$                  & 1.62 & 13.7 & 0.19 & y & \citet{ishizuka_08} \\ 
STOSY0             & TM1     & $n,p,\alpha,(A,Z),Y$                      & 1.64 & 14.6 & 0.18 & y & \citet{ishizuka_08} \\ 
STOSY0$\pi$        & TM1     & $n,p,\alpha,(A,Z),Y,\pi$                  & 1.67 & 13.7 & 0.19 & y & \citet{ishizuka_08} \\ 
STOSY30            & TM1     & $n,p,\alpha,(A,Z),Y$                      & 1.65 & 14.6 & 0.18 & y & \citet{ishizuka_08} \\ 
STOSY30$\pi$       & TM1     & $n,p,\alpha,(A,Z),Y,\pi$                  & 1.67 & 13.7 & 0.19 & y & \citet{ishizuka_08} \\ 
STOSY90            & TM1     & $n,p,\alpha,(A,Z),Y$                      & 1.65 & 14.6 & 0.18 & y & \citet{ishizuka_08} \\ 
STOSY90$\pi$       & TM1     & $n,p,\alpha,(A,Z),Y,\pi$                  & 1.67 & 13.7 & 0.19 & y & \citet{ishizuka_08} \\ 
STOS$\pi$          & TM1     & $n,p,\alpha,(A,Z),\pi$                    & 2.06 & 13.6 & 0.26 & n & \citet{nakazato08} \\ 
STOSQ209n$\pi$     & TM1     & $n,p,\alpha,(A,Z),\pi,q$                  & 1.85 & 13.6 & 0.21 & n & \citet{nakazato08} \\ 
STOSQ162n          & TM1     & $n,p,\alpha,(A,Z),q$                      & 1.54 &      &      & n & \citet{nakazato13} \\ 
STOSQ184n          & TM1     & $n,p,\alpha,(A,Z),q$                      & 1.36 & ---\footnote{$M_{\rm max}$ below 1.4~${\rm M}_{\odot}$.}  &      & n & \citet{nakazato13} \\ 
STOSQ209n          & TM1     & $n,p,\alpha,(A,Z),q$                      & 1.81 & 14.4 & 0.20 & n & \citet{nakazato08,nakazato13} \\ 
STOSQ139s          & TM1     & $n,p,\alpha,(A,Z),q$                      & 2.08 & 12.6 & 0.26 & y & \citet{sagert_12,fischer_14} \\ 
STOSQ145s          & TM1     & $n,p,\alpha,(A,Z),q$                      & 2.01 & 13.0 & 0.25 & y & \citet{sagert_12} \\ 
STOSQ155s          & TM1     & $n,p,\alpha,(A,Z),q$                      & 1.70 & 9.93 & 0.25 & y & \citet{fischer_11} \\ 
STOSQ162s          & TM1     & $n,p,\alpha,(A,Z),q$                      & 1.57 & 8.94 & 0.26 & y & \citet{sagert_09} \\ 
STOSQ165s          & TM1     & $n,p,\alpha,(A,Z),q$                      & 1.51 & 8.86 & 0.25 & y & \citet{sagert_09} \\ 
\hline 
\hline
\end{tabular}
\end{center}
\end{table*}
\endgroup

\subsubsection{Nucleons and nuclei as degrees of freedom}
\label{sec:sneos_nucleonic}
\paragraph{H\&W}
\label{sec:h_and_w}
The EoS of \citet{hillebrandt1984,hillebrandt1985} is one of the first
EoSs that was suitable for CCSNe simulations and which is
still in use today \cite{janka_12}. At low densities, a NSE model
based on the work of \citet{ElHi1980} is applied including 470
different nuclei: in addition to neutrons, protons and
$\alpha$-particles, about 450 isotopes with charge numbers $Z$ between
10 and 32 and neutron numbers $N$ ranging from stability to neutron
drip as well as 20 heavier nuclei from the Zr and Pb region are
included.  To account for excited states, nuclear level densities have
been constructed based on HF potentials using the Grand Partition
Function approach \citep{huizenga_72,wolff_80}.  For densities above
$3 \times 10^{12}$ g/cm$^3$, the EoS is computed in the SNA, see
Sec.~\ref{sec_nse_sna}, using the thermal HF method. The nuclear
interaction is of the Skyrme type using the parameter set SKa
\cite{koehler_76}.

\paragraph{LS}
\label{sec:ls}
The EoS by \citet{Lattimer:1991nc} considers nucleons,
$\alpha$-particles and heavy nuclei in the SNA, see
Sec.~\ref{sec_nse_sna}, as degrees of freedom. The latter are
described with a medium-dependent liquid-drop model. For nucleons,
non-relativistic Fermi-Dirac statistics is used, and a
simplified momentum-independent nucleon-nucleon
interaction is employed which results in constant effective nucleon masses equal
to the applied vacuum masses. Interactions between the gas
of nucleons, $\alpha$ particles and heavy nuclei are
taken into account through an excluded volume
mechanism. $\alpha$-particles are treated as hard spheres of
volume $v_{\alpha} = 24$~fm${}^{3}$ forming an ideal Boltzmann gas,
neglecting excited states. As the density increases, nuclei undergo
geometrical shape deformations, until they dissolve in favor of
homogeneous nuclear matter above approximately saturation density. The
formation of non-spherical nuclei and bubble phases 
is described by modifying the
Coulomb and surface energies of nuclei. The phase transition to bulk
nuclear matter is treated by a Maxwell construction between the two
phases. 

The LS EoS exists for three different parameterizations of the
nucleonic interaction, which are usually denoted according to their
value of the incompressibility $K$ of 180, 220, and 375~MeV. Nowadays
the version with $K=220$~MeV is considered the most relevant of the
three, since it is the best compatible with the various constraints on
the EoS, see Sec.~\ref{sec:compatibility}.

\paragraph{STOS}
\label{sec:stos}
The EoS by \citet{shen_98,Shen:1998by,Shen:2011qu} is another
widely used general purpose EoS. It assumes the same degrees of freedom as the LS EoS:
neutrons, protons and $\alpha$-particles as well as one heavy nucleus
in the SNA. 
For nucleons a RMF model with nonlinear meson self-interactions is used
with the parameterization
TM1 \cite{Sugahara_94}. $\alpha$-particles are again described as an
ideal Maxwell-Boltzmann gas with excluded-volume corrections. Excited
states of $\alpha$-particles are neglected. The properties of the
representative heavy nucleus are obtained from Wigner-Seitz cell
calculations within the Thomas-Fermi approximation for parameterized
density distributions of nucleons and $\alpha$-particles. The
translational energy and entropy contribution of heavy nuclei is not taken into account.

\citet{Zhang:2014uma} investigated the accuracy of the parameterized
density distributions of STOS in
comparison with fully self-consistent Thomas-Fermi calculations. The
authors concluded that overall there are only small differences. In
detail it was found that the free energies of the original STOS EoS
are slightly too low compared to the self-consistent solutions. 
This and other differences were related to a too
small value of the coefficient of the (surface) gradient energy
of the density
distribution used in STOS for the description of nuclei, which is
not consistent with the employed TM1 interaction. In addition,
\citet{Zhang:2014uma} studied the effect of 
a possible bubble phase 
for the transition to uniform nuclear matter and found
that the transition can be shifted to slightly higher densities.

\begin{table}
  \caption{\label{tab:nucl_int} Nuclear matter properties of the 
    parameterizations for the nuclear
    interaction used in the general purpose EoS of Table \ref{tab:eos3d}.
    Listed are the  saturation density $n_{\rm sat}$, binding energy $B_{\rm sat}$, 
    incompressibility $K$, skewness coefficient $Q=-K'$, symmetry energy $J$,
    and symmetry energy slope coefficient $L$ at saturation density at 
    zero temperature}.\label{tab:eos_para}
\begin{center}
\begin{tabular}{c| c c c c c c c c c c c l}
\hline 
\hline 
Nuclear & $n_{\rm sat}$ & $B_{\rm sat}$ 
& $K$ & $Q$ & $J$ & $L$ \\
Interaction & (fm$^{-3}$) & (MeV) & (MeV) & (MeV) & (MeV) & (MeV) \\
\hline
SKa          & 0.155 & 16.0 & 263 & -300 & 32.9 &  74.6 \\
LS180        & 0.155 & 16.0 & 180 & -451 & 28.6\footnote{The value for the symmetry energy $J$ given here is different from the value of
  29.3~MeV in \citet{Lattimer:1991nc}. They computed $J$ as the
  energy difference between neutron and nuclear matter whereas we are
  calculating $J$ as the second derivative with respect to $Y_q$ for symmetric matter at $n_{\rm sat}$, see also \citet{Steiner_12}.}
 &  73.8 \\
LS220        & 0.155 & 16.0 & 220 & -411 & 28.6$^a$ &  73.8 \\
LS375        & 0.155 & 16.0 & 375 &  176 & 28.6$^a$ &  73.8 \\
TM1          & 0.145 & 16.3 & 281 & -285 & 36.9 & 110.8 \\
TMA          & 0.147 & 16.0 & 318 & -572 & 30.7 &  90.1 \\
NL3          & 0.148 & 16.2 & 272 &  203 & 37.3 & 118.2 \\
FSUgold      & 0.148 & 16.3 & 230 & -524 & 32.6 &  60.5 \\
FSUgold2.1   & 0.148 & 16.3 & 230 & -524 & 32.6 &  60.5 \\
IUFSU        & 0.155 & 16.4 & 231 & -290 & 31.3 &  47.2 \\
DD2          & 0.149 & 16.0 & 243 &  169 & 31.7 &  55.0 \\
SFHo         & 0.158 & 16.2 & 245 & -468 & 31.6 &  47.1 \\
SFHx         & 0.160 & 16.2 & 239 & -457 & 28.7 &  23.2 \\
\hline  
\hline
\end{tabular}
\end{center}
\end{table}

\paragraph{FYSS}
\label{sec_furu}

The EoS of \citet{furusawa11, furusawa13} can be seen as an
extension of the STOS EoS at subsaturation densities. The same RMF
parameterization TM1 is employed for the nuclear interaction as in
the STOS EoS. A distribution of various light nuclei and heavy nuclei
up to $Z\sim 1000$ is included. Heavy nuclei are not described by the
Thomas-Fermi approximation as in STOS but by a
liquid-drop type formulation with temperature-dependent bulk
energies. Shell effects are incorporated by extracting the difference
of the liquid-drop binding energies compared to experimental
\cite{audi_03} and theoretical values \cite{koura_05}. A
phenomenological density-dependence of the shell effects is
introduced, assuming that these vanish at $n_{\rm sat}$. For light
nuclei, it incorporates the Pauli-blocking
shifts of \citet{Typel:2009sy}. Furthermore, light nuclei
receive self-energy shifts originating from the
mesonic mean-fields. As an additional phenomenological interaction,
excluded-volume effects are applied for nucleons, light nuclei and
heavy nuclei. In addition to standard spherical nuclei, also a
bubble phase with low-density holes in matter of higher density is considered.
The FYSS EoS has been used to explore the
effect of light nuclei in CCSN simulations
\cite{furusawa13b}.

\paragraph{HS}
\label{sec_hs}
The basic model of the HS EoS \cite{Hempel09}
belongs to the class of extended NSE models
and describes matter as a ``chemical'' mixture of nuclei and
unbound nucleons in NSE. Nuclei are treated as classical Maxwell-Boltzmann
particles, nucleons with RMF
models employing different parameterizations. 
Several thousands of nuclei are considered, including light
ones. Binding energies are either taken from experimental
measurements \citep{audi_03} or from various theoretical nuclear
structure calculations \citep{moeller_95,lalazissis_99,geng_05}. The
latter are chosen such that they were calculated for the
same RMF parameterization as the one
applied to nucleons if available, 
otherwise the data from \citet{moeller_95}  are used. The following medium
modifications are incorporated for nuclei: screening of the Coulomb
energies by the surrounding gas of electrons in Wigner-Seitz approximation, 
excited states in the
form of an internal partition function using the level density of
\citet{fai_82}, which is  adapted from the NSE model of
\citet{ishizuka03}, and excluded-volume effects. Note that further explicit 
medium modifications of nuclei are not considered in HS. Since the
description of heavy nuclei is based on experimental nuclear masses, the
HS EoS includes the correct shell effects of nuclei 
in vacuum. On the other hand, the use of
nuclear mass tables limits the maximum mass
and charge numbers of nuclei, see \citet{Buyukcizmeci_13}.

The first version \citep{Hempel09} used the RMF 
parameterization TMA \citep{toki_95}. A few aspects of the model have been
changed in the later versions \cite{Hempel_11a}, namely a cut-off for 
the highest excitation energy of nuclei is introduced, experimental
nucleon masses
are used, and only nuclei left of the neutron dripline are considered.
At present, EoS tables are available for the following RMF parameterizations: 
TM1 \citep{Sugahara_94,Hempel_11a}, TMA
\citep{toki_95,Hempel09,Hempel_11a}, FSUgold
\citep{ToddRutel:2005zz,Hempel_11a}, NL3
\citep{lalazissis_97,Fischer_13}, DD2 \citep{Typel:2009sy,Fischer_13}, 
and IUFSU \citep{fattoyev_10,Fischer_13}.  We will denote them by
``HS(x)'', where ``x'' indicates the nuclear interaction employed.
 
\paragraph{SFHo and SFHx}
Two additional EoSs based on HS were published 
by \citet{Steiner_12}, SFHo (``o''
for optimal) and SFHx (``x'' for extreme) with new RMF 
parameterization that were fitted to some NS 
radius determinations. These two EoSs 
result in rather compact NSs and have moderately, respectively 
very low values of the slope parameter of the symmetry energy
$L$ \citep{Steiner_12,Fischer_13,Hempel:2015yma}.

\paragraph{SHT(NL3), SHO(FSU), and SHO(FSU2.1)}
The EoSs of G.~\citet{shen_11_nl3,shen_11_fsu} are based on different
underlying physical descriptions in different regimes of density and
temperature. Uniform nuclear matter at high densities and temperatures
is described by a RMF model. At intermediate densities, the same RMF
model is used within calculations of non-uniform matter,
generating a representative heavy nucleus and unbound nucleons, but no
light nuclei. At low densities and temperatures, a special form of the
virial EoS is used which includes virial coefficients up to second
order among nucleons and $\alpha$-particles.  The virial EoS
is not using Fermi-Dirac statistics for nucleons, but only
incorporates corrections for it as part of the virial coefficients. 
Nuclei with mass numbers $A=2$ and $3$
are not considered as explicit degrees of freedom, but
8980 nuclei with mass number $A\geq 12$ are included. The contribution
of heavy nuclei in NSE is modeled as a non-interacting
Maxwell-Boltzmann gas without considering excluded volume effects.
Coulomb screening is included for heavy nuclei, but not for 
$\alpha$-particles.  Note that $\alpha$-particles are only present in
the virial part of the EoS, which is completely independent of the RMF
interaction. The three different prescriptions are merged to a single
table by minimizing the free energy. In addition, a smoothing and
interpolation procedure is applied \cite{shen_11_nl3,shen_11_fsu}.

The EoSs of G.\ Shen {\em et al}.\ are available for two different
RMF interactions, NL3 \citep{shen_11_nl3} (SHT(NL3)) and FSUgold
\citep{shen_11_fsu}. 
A density dependence of the scalar meson-nucleon coupling was introduced
below $5 \times 10^{-3}$~fm$^{-3}$ in case of the NL3 interaction in order
to match the energy per nucleon of a unitary neutron gas \cite{Shen:2010pu}.
Since the FSUgold parameterization leads to a
maximum NS mass of only 1.7 M$_\odot$, an additional
phenomenological pressure contribution was introduced 
for densities above 0.2 fm$^{-3}$, leading to a
sufficiently high maximum mass of 2.1 M$_\odot$. This EoS was called
``FSU2.1'' and we abbreviate it SHO(FSU2.1), and the EoS with the
unmodified FSUgold parameterization SHO(FSU).

\subsubsection{Including additional degrees of freedom}
\label{sec:sneos_non_uniform_additional}

\begin{figure*}
\centering
\includegraphics[width = .6\columnwidth]{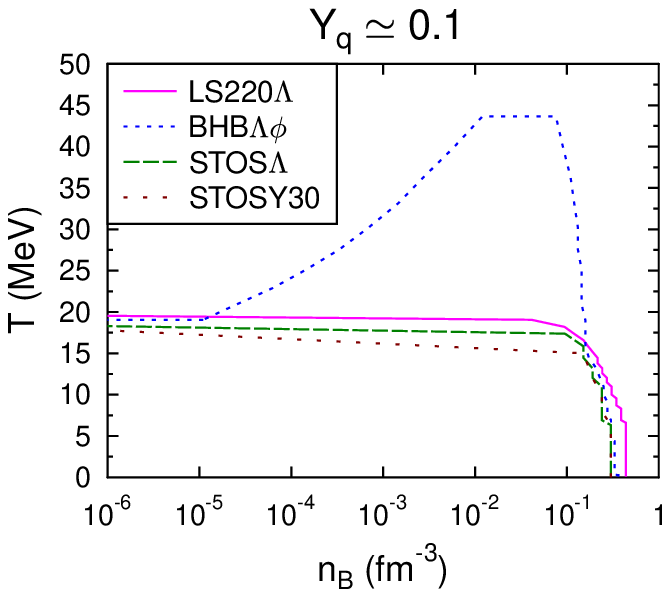}\hfill
\includegraphics[width = .6\columnwidth]{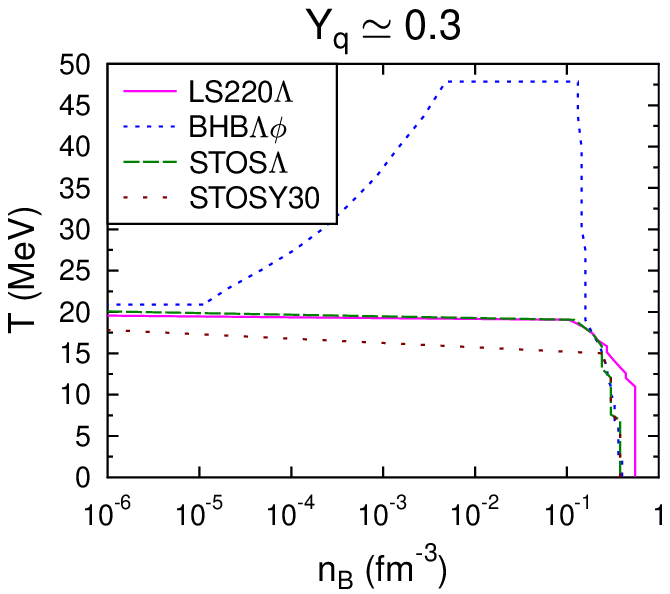}\hfill
\includegraphics[width = .6\columnwidth]{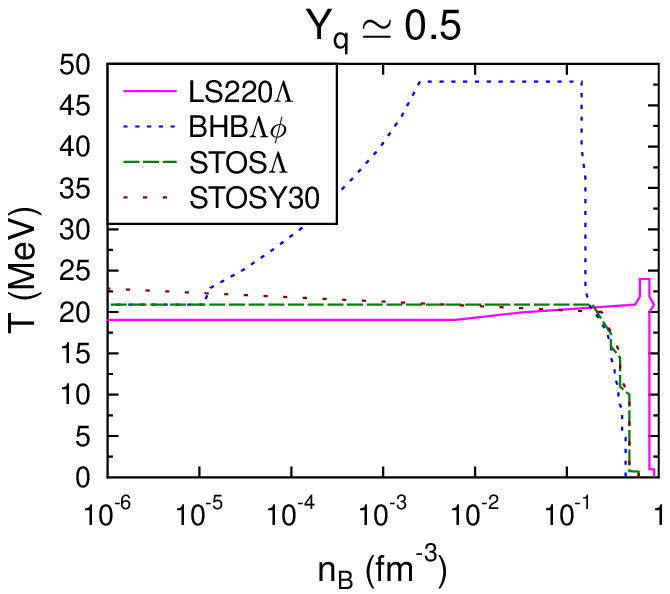}\\
\caption{The lines delimit regions where the number density fractions
  of $\Lambda$ hyperons exceed $10^{-4}$ for the different models
  discussed in the text. The figures 
  correspond
  from left to right to a fixed hadronic charge fraction $Y_q = 0.1,
  0.3,0.5$ and show the relevant regions as functions of temperature
  and baryon number density. $\Lambda$ hyperons appear at high
  densities and temperatures, i.e., in the upper right part of the
  figure. 
\label{fig:pd}
}
\end{figure*}

An EoS covering the whole thermodynamic parameter range relevant for
CCSNe and NS mergers should of course be able to correctly describe
cold $\beta$-equilibrated NSs.  As discussed in
Sec.~\ref{sec:hyppuzzle}, it might turn out that only nucleonic matter
is present in cold NSs. This does, however, not mean that additional
degrees of freedom could not occur in stellar core-collapse events and
NS mergers, where matter is strongly heated in addition to being
compressed to densities above nuclear matter saturation density.  The
temperatures and densities reached can become so high that a
traditional description in terms of electrons, nuclei, and nucleons is
no longer adequate. Compared with the cold NS EoS, temperature effects
favor the appearance of additional particles such as pions and
hyperons and they become abundant in this regime. A transition to
quark matter can also not be excluded.  In recent years, some models
have been developed extending existing purely nuclear models as
discussed in the previous section, including pions, hyperons, or
quarks.

Let us start with the models including hyperons. In
\citet{ishizuka_08} and \citet{Oertel:2012qd} the whole baryon octet
is considered. The former EoS is an extension of the STOS EoS by
\citet{shen_98} and the latter of the LS220
EoS~\cite{Lattimer:1991nc}, see the previous section. 
In \citet{ishizuka_08}, the hyperonic
interactions are fixed following a standard procedure 
for RMF models. For the vector couplings, symmetry
constraints are imposed, assuming $SU(6)$ flavor symmetry
following~\citet{Schaffner:1995th} in the isoscalar sector and isospin
symmetry in the isovector one. The remaining scalar couplings of
hyperons to nucleons are adjusted to reproduce standard values of the
single particle hyperonic potentials in symmetric nuclear matter at saturation density, extracted from
hypernuclear data, see Sec.~\ref{sec:exdatafewbody}. 
These single particle potentials are given by, see Sec.~\ref{sec:mf_models},
\begin{equation}
U_j = V_j - S_j~,
\end{equation}
involving scalar and vector self-energies. Standard values for
  $U_\Lambda = -30$ MeV and $U_\Xi = -15$ MeV are assumed, whereas
  the situation for $U_\Sigma$ the situation is ambiguous and
  \citet{ishizuka_08} have presented several versions of the EoS with
  $U_{\Sigma} = -30,0,30,90$ MeV (``STOSYxxx'', where  
``xxx'' indicates the value of the potential, and a prepended ``A'' if it is attractive).  
In the following discussion, we will keep the
version with $U_\Sigma = +30$ MeV.  In
\citet{Oertel:2012qd} hyperons are added by extending the model by
\citet{Balberg97} to finite temperature. This model is a
non-relativistic potential model similar to the one in
\citet{Lattimer:1991nc} for the nuclear part. The hyperonic couplings
are readjusted to remain compatible with the single particle hyperonic
potentials in nuclear matter, but, at the same time, predict maximum
NS masses in approximate agreement with the observation
in~\citet{demorest_10}.

\citet{ishizuka_08} and \citet{Oertel:2012qd} present models, 
in which pions are also included. 
The former will be denoted by
``STOSY$\pi$xxx''. Pions are treated as an ideal Bose gas.  Obviously,
without interactions, $\pi^-$ will form a Bose condensate below some
critical temperature, depending on the density, as discussed
extensively \cite{Migdal:1990vm,Glendenning:1997wn}.  It is now commonly
assumed, that there is an $s$-wave $\pi N$ repulsion, preventing
probably pions from condensing. \citet{ishizuka_08} show that indeed,
adding an effective $\pi N$ interaction, the domain in temperature and
density where pions condense is strongly reduced. The main effect of
pions on the EoS occurs, however, at high temperature and the ideal
gas should be a good approximation in this regime. A simplified
version, including only pions in the LS220 EoS is used in
\citet{Peres_13} (``LS220$\pi$'').  \citet{nakazato08} extend the STOS
EoS in the same way (``STOS$\pi$'').

Subsequently different models including only $\Lambda$-hyperons have
been developed. 
The first one is the work by \citet{Shen:2011qu} (``STOS$\Lambda$''),
which is very similar to the work by \citet{ishizuka_08}, except that 
slightly different hyperon interactions are employed. The
motivation is that the $\Lambda$ represents, together with the
$\Sigma^-$-hyperon, probably the most important hyperonic degree of
freedom in hot dense supernova matter. Thus, including the $\Lambda$
allows for discussing general features of the effects coming from the
hyperonic degrees of freedom without the necessity of resolving the
complicated particle composition in the presence of many different
hyperons. In addition, the $\Lambda N$ and $\Lambda\Lambda$
interactions are the best constrained from experimental data. 
And, since less degrees of freedom are
populated, the NS maximum mass is less reduced from that of purely nucleonic EoSs.
An extended version of the LS220 EoS including only
$\Lambda$-hyperons (``LS220$\Lambda$'') is discussed
in~\citet{Gulminelli:2013qr, Peres_13}. It has the feature that
a strangeness driven first order phase transition occurs at the onset of
hyperons 
\cite{SchaffnerBielich:2000wj,SchaffnerBielich:2002ki,Gulminelli:2012iq,Oertel:2016xsn}, 
see Sec.~\ref{sec:PT}. 
In \citet{Banik:2014qja} two different extended versions of the
HS statistical model are constructed
including $\Lambda$-hyperons. The density dependent RMF
parameterization DD2 is employed and the setup for the hyperonic
couplings is similar to~\citet{ishizuka_08}, i.e., assuming SU(6) symmetries
for the vector couplings. The value of the $\Lambda$ single
particle potential in symmetric nuclear matter determines the remaining
scalar couplings. The basic model is denoted as ``BHB$\Lambda$''
and the version including additional short-range repulsion in the
$\Lambda\Lambda$-channel by ``BHB$\Lambda\phi$''. 

\begin{figure*}
\centering
\includegraphics[width = 0.8\textwidth]{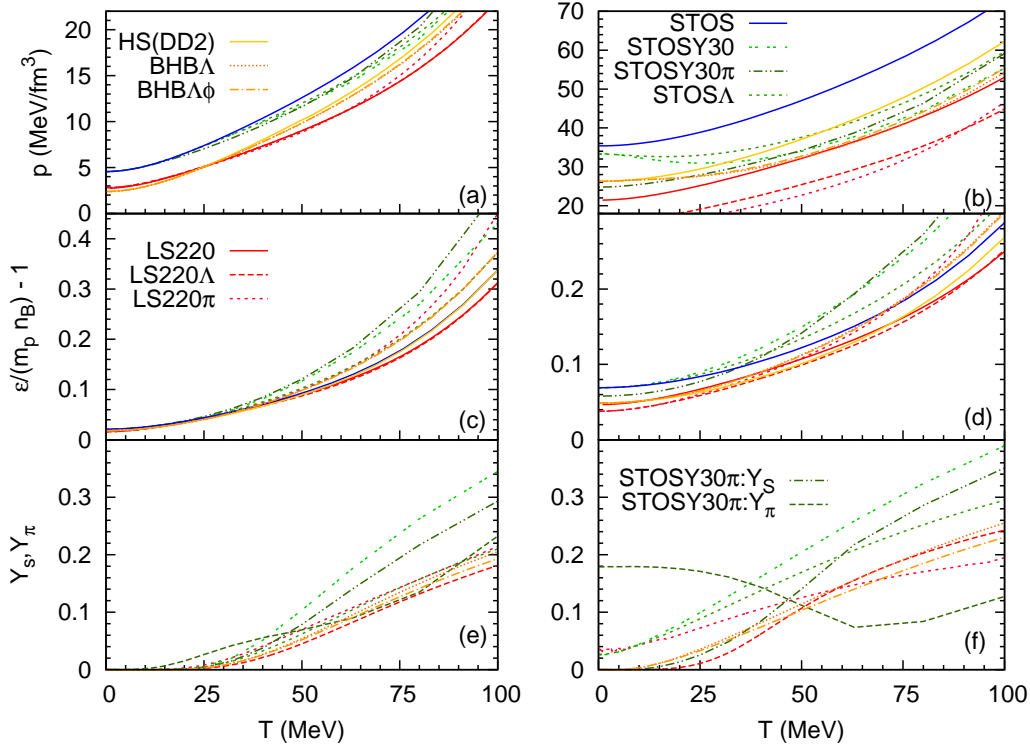}\hfill
\caption{Thermodynamic quantities as functions of temperature for $n_B
  = 0.15\ \mathrm{fm}^{-3}$ (left) and $n_B = 0.3\ \mathrm{fm}^{-3}$
  (right), corresponding roughly to  one and two times nuclear matter
  saturation density, and a charge fraction of $Y_q = 0.1$. The upper
  panels show the pressure, the middle ones the scaled internal energy per
  baryon with respect to the proton mass and the lower ones the number
  fractions of additional particles. These are: $\Lambda$-hyperons for
  ``STOS$\Lambda$'', ``LS220$\Lambda$'', ``BHB$\Lambda$'', 
  and ``BHB$\Lambda\phi$''; all hyperons for ``STOSY30''
  and ``STOSY30$\pi$'';  the $\pi^-$ fraction for
  ``LS220$\pi$''.  See text for explanation of the acronyms.}
\label{fig:thermofunct}
\end{figure*}

The onset density for hyperons lies between two and three times
$n_{\rm sat}$ at low temperatures.  As expected, upon increasing the
temperature, the density domain where hyperons appear is enlarged, in
particular above $15-20$~MeV, see Fig.~\ref{fig:pd}.  We note that due
to the presence of light nuclei in the BHB$\Lambda\phi$
EoS~\cite{Banik:2014qja}, hyperons appear at much higher temperature
in a large density domain than in the other models. Pions become more
abundant at high temperatures, too.  This can also be seen from
Fig.~\ref{fig:thermofunct}.  In the bottom panels, the different
particle number fractions are shown as a function of temperature for
$n_B = 0.15$ (e) and $0.3~\mathrm{fm}^{-3}$ (f) and for $Y_q = 0.1$.
The influence of $Y_q$ on the appearance of (neutral)
$\Lambda$-hyperons is less important than for charged
particles. The more asymmetric the matter, the higher is the charge
chemical potential, the higher the abundance of charged particles. In
neutron-rich matter the charge chemical potential is negative,
favoring negatively charged particles. This is the reason why in NS
matter, $\Sigma^-$ or $\Xi^-$ can become enhanced with respect to
$\Lambda$-hyperons, even if they have a higher mass. Thermal effects
alleviate the influence of the chemical potential, see
\citet{ishizuka_08, Oertel:2012qd}.  If the baryon number density
remains constant, the overall hyperon fractions decrease with
increasing $Y_q$ \citep{Prakash:1996xs}.

Concerning the influence on thermodynamic properties, pressure and
energy density are shown as functions of temperature for different
models in the upper and middle panel of Fig.~\ref{fig:thermofunct}.  
The STOS~\cite{shen_98}, LS~\cite{Lattimer:1991nc}, and HS(DD2) EoSs are compared
with their corresponding versions including 
hyperons~\cite{ishizuka_08,Shen:2011qu, Gulminelli:2013qr,Banik:2014qja} and/or
pions~\cite{ishizuka_08, nakazato08,Peres_13}. As can be seen from
Fig.~\ref{fig:thermofunct}, 
the effect of the additional particles on
the thermodynamic quantities is not negligible for high density and
temperature. The main effect is a reduction of pressure due to the
additional degrees of freedom. 

There exist EoSs in which the nuclear model of \citet{shen_98} has
been supplemented with a phase transition to quark matter at high
density and temperature,too. In \citet{nakazato08,nakazato13} and the work by 
\citet{sagert_09,fischer_11,sagert_12,fischer_14} the MIT bag
model \cite{Chodos74,Farhi:1984qu} is applied to the quark matter phase. The
transition from hadronic matter to quark matter is described by a
Gibbs construction in both cases, see Sec.~\ref{sec:PT}. In addition
to the nuclear interaction, the parameters of the model are the bag
constant $B$ and the strange quark mass, and possibly the 
coupling constant of strong interaction corrections. These parameters determine
the densities for the onset of the quark-hadron
phase coexistence region and the pure quark phase. It can be observed that the
onset density for the mixed phase is noticeably lowered with
decreasing charge fraction. In particular, this means
that the critical density in asymmetric CCSNe 
or NS matter can be lower than in symmetric
matter. Furthermore, it was found that the critical density is 
significantly reduced due to weak equilibrium with respect to strangeness. 
A value close to nuclear saturation or even
below is thus not in contradiction with any terrestrial experiment from
HICs. This fact is exploited in the model of
\citet{sagert_09}, where the bag constant has been chosen such that
the strongly asymmetric matter in compact stars leads to almost
pure quark stars with only a thin hadronic layer. These models are
labeled as ``STOSQxxxs'',
where ``xxx'' indicates the value of the bag
constant $B^{1/4}$ in MeV, and ``s'' stands for Sagert et al. The models of
\citet{nakazato08} lie in a parameter range ($B^{1/4} = 209$ MeV)
where the critical density is much higher, such that the resulting
NSs have only a small quark core. Strong interaction corrections are not
used in this model. If a thermal pion gas is
included in the hadronic phase, the transition to quark matter occurs
at considerably higher densities~\cite{nakazato08} due to the
softening of the hadronic EoS by the pions. The latter models are
labeled ``STOSQxxxn'' for the one with quarks and ``STOS$\pi$Qxxxn'' for
the one with quarks and pions. 
\citet{nakazato13} calculated additional hybrid EOS tables for $B^{1/4} = 162$ 
and 184~MeV. Because the corresponding maximum masses of 1.54 and 1.36~M$_\odot$
are well below the observed pulsar masses, in the following we consider only 
the table with $B^{1/4} = 209$ MeV as an representative example of the hybrid 
EoSs of \citet{nakazato08,nakazato13}.
The thermodynamic properties are 
strongly influenced by the possible existence of quark matter at high
densities and temperatures, and in particular the phase transition can
have an important effect on the dynamics of CCSNe, see Sec.~\ref{sec:ccdynamics}.

\subsubsection{Compatibility with experimental and observational constraints}
\label{sec:compatibility}

\begin{figure}
\includegraphics[width=0.8\columnwidth]{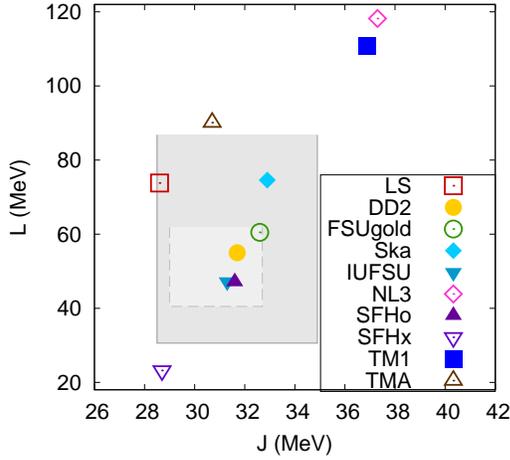}
\caption{(color online)  Slope parameter of the symmetry energy $L$ versus 
value of the symmetry energy $J$ at normal nuclear matter density. 
The dark gray region is the constraint of \citet{Lattimer:2012xj}, the light
gray region is taken from Fig.\ \ref{fig:JL-prob}.
The different symbols show the 
values of the nucleon interactions of Table~\ref{tab:eos_para}
that are applied in the general-purpose EoSs of Table~\ref{tab:eos3d}. FSU2.1 gives the same value as FSUgold.
\label{fig:jl_sneos}}
\end{figure}

\begin{figure}
\centerline{\includegraphics[width=0.5\textwidth]{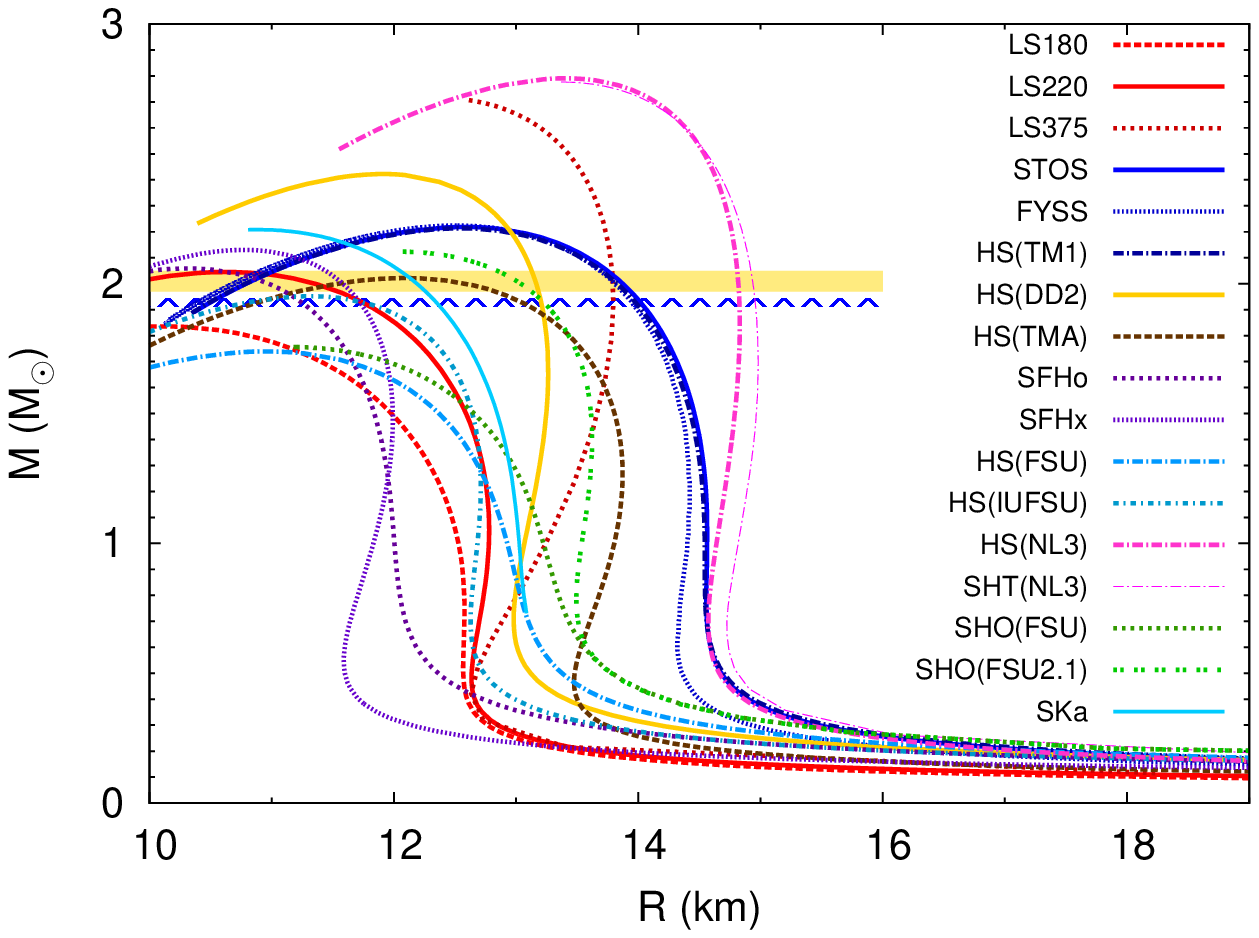} }
\caption{Mass-radius relations of spherically symmetric NSs
  for the different EoSs that cover the full thermodynamic parameter range and include
 only nucleonic degrees of freedom, cf.\ Table \ref{tab:eos3d}. 
The two horizontal bars indicate the two recent precise NS mass determinations, PSRJ1614-2230~\cite{demorest_10} 
(hatched blue) and PSR J0348+0432~\cite{Antoniadis_13} 
(yellow). The curve labeled ``SkA'', although based on the same nuclear interaction, does not represent the H\&W EoS, but the model by \citet{Gulminelli:2015csa}.
\label{fig:eos3d}}
\end{figure}

\begin{figure}
\centerline{\includegraphics[width=0.5\textwidth]{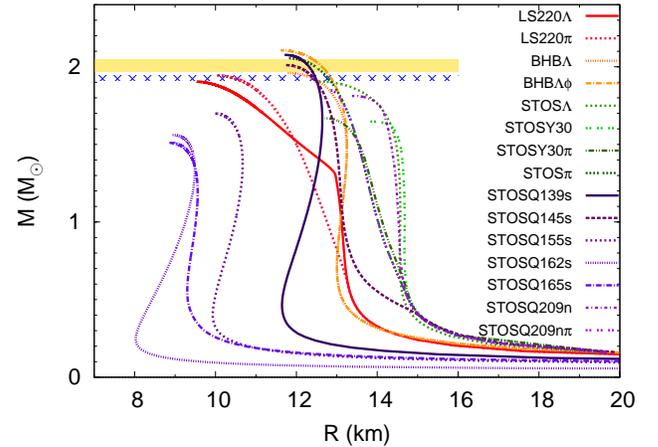} }
\caption{Same as Fig.~\ref{fig:eos3d} for EoS models including additional degrees of freedom. The onset of  additional degrees of freedom is visible as a change in the slope.
 \label{fig:hyp3d}}
\end{figure}

\begin{figure*}
\centering
\includegraphics[width=1.0\textwidth]{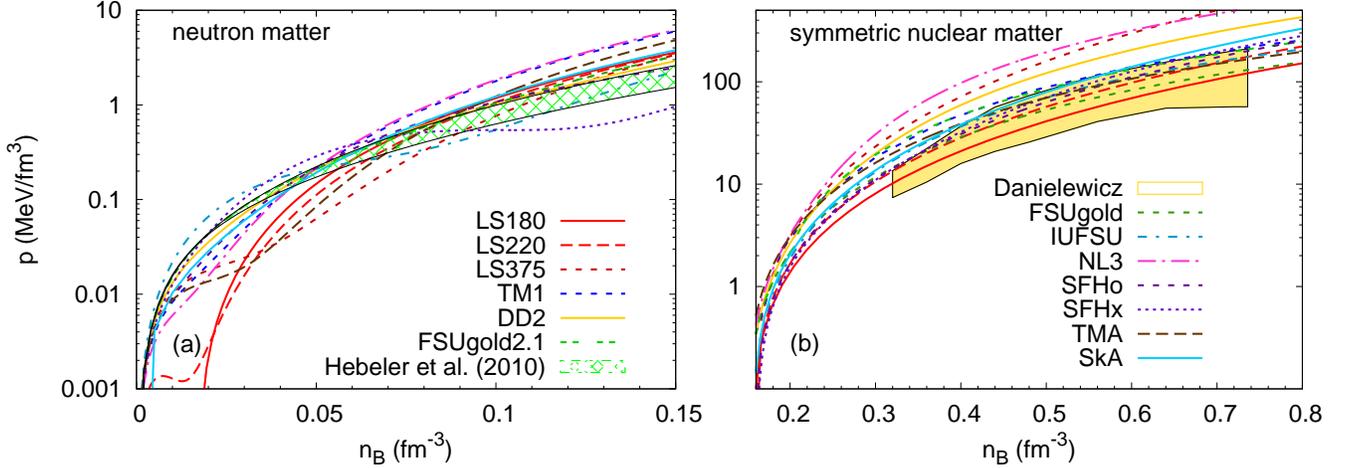}
\caption{(color online) Pressure as a function of baryon density within 
  the EoS models listed in Table~\ref{tab:eos3d} for symmetric nuclear matter (right) 
  and pure neutron matter (left). The results are compared with the theoretical 
  calculation for neutron matter from~\citet{Hebeler:2013nza} 
  and the experimental flow constraint from \citet{Danielewicz_02} for symmetric matter.
  \label{fig:Danielewiczeos+} }
\end{figure*}

In this subsection we compare the results of the general purpose EoSs presented above
with several constraints, that have been introduced
in Sec.~\ref{sec:constraints}.
As can be seen from Table~\ref{tab:nucl_int}, most of the employed
interactions give reasonable properties for compressibility, saturation density and binding energy of symmetric
matter, except that LS180 has a compressibility at the lower end of
the allowed range and LS375 and TMA have values above the allowed
range. However, some models give symmetry energies and slopes far off
the best present constraints, see Fig.~\ref{fig:jl_sneos}, i.e., the
dependence of the EoS on $Y_q$  is probably not correctly
described. 
DD2, SFHo, IUFSU and FSUgold(2.1) are  in best
agreement.

NS masses, see Fig.~\ref{fig:eos3d}, Fig.~\ref{fig:hyp3d}, and
Table~\ref{tab:eos3d}, probably represent the presently
most reliable observational constraint
on the compact star EoS.
LS180 and the models based on FSUgold give too low masses and HS(IUFSU) is
only marginally compatible. It should be noted that the maximum mass
does depend only very little on the treatment of the
inhomogeneous part of the EoS, such that all models with the same
nuclear interaction give essentially the same maximum mass. This
is not the case for the radii of intermediate mass NSs,
which are more sensitive to the treatment of the crust. Here slight
differences can be observed, e.g., between HS(TM1), STOS and FYSS,
which are all based on the TM1 interaction, see Fig.~\ref{fig:eos3d}
and Table~\ref{tab:eos3d}. Since the H\&W is not publicly available,
in Fig.~\ref{fig:eos3d} the results of the model by 
\citet{Gulminelli:2015csa} are plotted. This model is based on the same Skyrme
interaction SkA. Note that the radius of a 1.4~M$_\odot$ NS
of \citet{Gulminelli:2015csa}
differs by about 1~km compared to the value of H\&W as given in Table~\ref{tab:eos3d}.
In Sec.~\ref{sec:constraints} we have outlined the difficulties in
determining reliable NS radii and that presently no
consensus on the allowed range of values can be obtained. Let us
mention, however, that if small radii for 1.4~M$_\odot$ NSs of the
order 10-12 km were confirmed, see, e.g., \citet{Ozel_15}, then some
of the hadronic models shown here could be excluded, in particular those based
on TM1, NL3 and LS375.

As can be seen from Fig.~\ref{fig:hyp3d} and the data given in
Table~\ref{tab:eos3d}, the NS maximum masses of most of the extended models
with additional non-nucleonic degrees of freedom
are not compatible with a $2$~M$_\odot$ star. BHB$\Lambda \phi$, STOS$\pi$,
STOSQ139s, and STOSQ145s are the only ones with NS maximum masses
above $1.97~ {\rm M}_\odot$. As discussed above, for hyperonic EoSs this
is related to the hyperonic interactions used.
Recent studies for cold NS EoSs overcome the maximum mass problem 
but these models for the
interaction have not yet been applied to compute a complete EoS
covering the whole range of temperature, hadronic charge fraction
and baryon density. Anyway, with increasing temperature the effect of the
interactions becomes less important and most models agree qualitatively
for the particle composition, cf.\ Figure\ \ref{fig:thermofunct},
whether or not compatible with a 2 M$_\odot$ NS. 

The situation becomes even more severe if additional constraints are included 
in the benchmarking.
 In the right panel of
Fig.~\ref{fig:Danielewiczeos+} the experimental flow constraint of
\citet{Danielewicz_02} for the pressure as a function of baryon number
density in symmetric nuclear matter is depicted. For neutron matter
(left panel), the constraint from the $\chi$EFT calculation of
\citet{Hebeler:2013nza} is shown, cf., Sec.~\ref{sec:theosmnm}. These
constraints are compared with the EoS of symmetric and neutron matter
obtained at $T=0$ from the different models. It is obvious that none of the present
models is perfectly compatible with the neutron matter
results from \citet{Hebeler:2013nza} below saturation density. 
However, the error band shown in
the left panel of Fig.~\ref{fig:Danielewiczeos+} is perhaps too small,
see Fig.~\ref{fig:neutronmatter}, rendering some of the models at
least marginally compatible, such as DD2, SFHo or FSUgold. The LS180 and
LS220 models are in reasonable agreement with the constraint for $n_B
\gsim 0.1$ fm$^{-3}$ but give too low pressures at lower densities. 
The models SFHo and SFHx have been fitted to some NS 
radius determinations giving radii around 12 km for 1.4~M$_\odot$ 
\cite{Steiner:2010fz,Steiner_12}. 
In the extreme model SFHx, it has been tried to make these as small as 
possible, within the employed class of RMF interactions. This is the reason
why they have very low (SFHx) or moderate pressure (SFHo) for neutron matter around $n_{\rm sat}$ and correspondingly low and moderate values of $L$.
Similar observations as for neutron matter can be made from the comparison of the flow
constraint with the EoS of symmetric nuclear matter: Many present
models for the general purpose EoS seem to give a too large
pressure at suprasaturation densities.

The neutron matter EoS is a crucial anchor point for the NS 
EoS, and thus also of great significance for NS mergers. However, for CCSNe,
matter is generally more symmetric and nuclear clusters are an important 
contribution to the subsaturation EoS.
In Fig.~\ref{fig:kalpha}, see Sec.~\ref{sec:HIC}, several
of the general purpose EoSs are included in a comparison with
experimental data for cluster formation.
The LS220 EoS  shows a notable underproduction of
$\alpha$-particles, and SHT(NL3) and SHO(FSU2.1) an overproduction at
high temperatures. The other general purpose EoSs 
FYSS, SFHo, STOS and HS(DD2), are more or less in reasonable 
agreement with the constraint. The current experimental data 
do not allow to make 
further judgements about details of the medium modifications of nuclear 
clusters, e.g., to distinguish classical excluded volume effects
from quantum statistical Pauli blocking. For a further discussion, 
see \citet{Hempel:2015yma}.

To conclude, there is not a single general purpose EoS that is compatible with all
constraints, even though we considered only a few of them.
However, from the purely nucleonic models SFHo, HS(DD2), and
SHO(FSU2.1) are at least approximately consistent. 
From the EoS with additional degrees of freedom, only BHB$\Lambda
\phi$, which is also based on DD2, would be
acceptable.  Nevertheless these EoS have
further drawbacks and weaknesses: in the models based on HS (SFHo,
HS(DD2), BHB$\Lambda \phi$) the treatment of light nuclei is not as
advanced as in the quantum statistical model (Sec.~\ref{sec:qs}) or
generalized relativistic density functional (Sec.~\ref{sec:grdf}),
and it does not include an explicit medium dependence of the nuclear
binding energies of heavy nuclei, which one could extract, e.g.,
from nucleons-in-cell-calculations (Sec.~\ref{sec:nuc_in_cell}). 
Other EoSs (e.g.\ FYSS or SHO/SHT) are
more advanced in some of these aspects, but they do not fulfill,
e.g., constraints for the maximum mass or $L$. The SHO(FSU2.1) is
compatible with the maximum mass constraint only because of an
ad-hoc modification of the pressure at high densities. Furthermore,
of all possible light nuclei only the $\alpha$ particle is included
in this model. Detailed nucleons-in-cell-calculations are used
for heavy nuclei, but only at intermediate to high densities, and
light nuclei are not taken into account in this regime at all. The
usage of different prescriptions in different regimes of $T$ and $n_{B}$
can also lead to problems in the thermodynamically consistent construction of transitions.  
The state of the art in modeling the general purpose EoS is thus not
really satisfactory. There is still need
for new general purpose EoSs, that employ modern EDFs (or even
beyond) with good nuclear matter properties,  that 
tackle the problem of additional
degrees of freedom at high densities and temperatures, and that
give a detailed description of clustering at subsaturation densities.

\section{Applications in Astrophysics}
\label{sec_application_in_astro}

\subsection{Binaries and binary mergers}
\label{sec:binaries}
Coalescing relativistic binary systems containing compact objects,
either NSs or BHs, are very interesting in the context of the
EoS of dense and hot matter. They are likely to be important sources
of detectable gravitational waves (GW) by advanced LIGO/VIRGO and
KAGRA, possibly before 2020.  
NS-NS and NS-BH mergers are believed to produce short gamma-ray-bursts
(sGRB). In addition, they may represent the major source for the main
component of heavy r-process elements in the universe, see, e.g., the
recent reviews of~\citet{Shibata_11,Faber_12,Rosswog_15} and
references therein. All three, the GW signal, the sGRBs, and the
r-process abundances, contain information on the EoS.

\begin{figure}
\centering
\includegraphics[width=1.0\columnwidth]{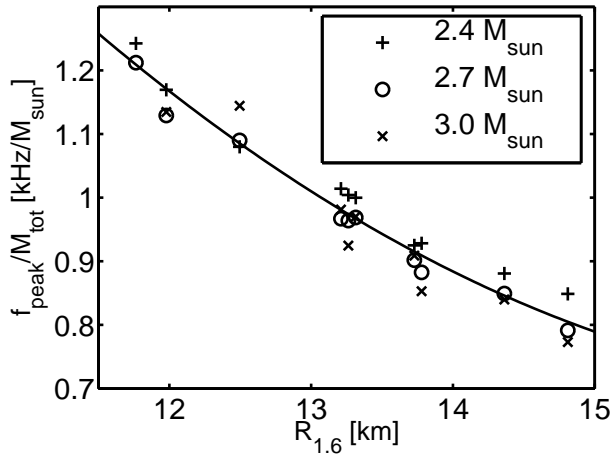}
\caption{Relation between the dominant GW frequency in the post-merger
  phase of a NS-NS binary merger event and the radius of a
  non-rotating NS with a gravitational mass of $M = 1.6$~M$_\odot$. The
  frequency has been normalized with respect to the total mass of the
  system. Different symbols refer to different total masses, where equal-mass binaries are assumed.
  Fig.\ taken from \citet{Bauswein15b} with kind permission of The European Physical Journal (EPJ).
\label{fig:fgw}}
\end{figure}

During the late inspiral phase of both, NS-NS and NS-BH systems, NSs
become tidally deformed to an extent that depends on the
underlying EoS. Numerical models suggest that the GW frequency is very
sensitive to the tidal deformation and thus to the underlying
EoS~\cite{Shibata_11,Faber_12}. 
However, the rate of NS-BH inspirals is very uncertain (to date no such system has
been observed), and the tidal effects from these systems are
probably not visible for next-generation detectors since they occur
at too high frequencies outside the range where the detectors are most
sensitive~\cite{Pannarale_11}. 
On the contrary, after the first
  detection of GW emission from a BH-BH merger~\cite{Abbott16}, there
appears to be a good chance for binary NS mergers to be detected
in the near future, and the tidal deformability has probably a
strong enough impact on the GW signal~\cite{Read_13}. Additional
information on the EoS can be obtained from the post-merger phase, in
cases where the EoS supports the formation of a hypermassive NS. The
frequencies of NS normal modes after the merger are sensitive to the
EoS and visible in the GW signal, see,
e.g.,~\citet{Sekiguchi_11,Bauswein_12,Takami_14}.  Measurements of their
frequencies could tightly constrain NS masses and radii since they are
strongly correlated. Fig.~\ref{fig:fgw} illustrates the correlation between
  the dominant GW frequency in the post merger phase, normalized to
  the total mass of the binary system, and the radius of a cold
  nonrotating NS with $M = 1.6$~M$_\odot$. It could even
be possible to give an estimate for the NS maximum mass
\cite{Bauswein:2013jpa,Bauswein:2014qla,bauswein15}.

\citet{Fryer_15} and \citet{Lawrence:2015oka} propose another test of
the EoS in binary NS mergers. Numerical models suggest that a sGRB is
produced in such a merger only if a BH forms sufficiently fast.
The fate of the core of the remnant, and thus the BH formation time,
depends on the maximum mass supported, and thus on the EoS.  Although
other factors influence the maximum mass, for
instance, the spin rate or the angular momentum distribution inside
the core~\cite{kastaun2015}, the authors suggest that it is possible
to relate the existence of a sGRB to the maximum mass of a cold
non-rotating NS. Assuming NS binary mergers to be the dominant source
of sGRB, a combined analysis of the observed burst rate and the total
merger rate with GW detectors, would then allow for constraining the
EoS.

Compact binary mergers eject initially extremely neutron rich matter,
and have therefore already very early been identified as possible
source of r-process elements in the universe~\cite{LattimerSchramm1974}. 
Recent calculations 
show that the conditions are favorable for producing a robust
r-process abundance pattern of heavy nuclei that is close to the
solar, see, e.g., \citet{Rosswog_15}.  The r-process
production rates, the final abundances, and the amount of ejected
material depend on the chemical composition and the
thermodynamic conditions in the ejecta, and thus on the EoS, see,
e.g., \citet{Bauswein_13,Wanajo:2014wha,sekiguchi15}.  The radioactive decay of
the freshly produced r-process elements should produce an
electromagnetic transient, called a kilonova/macronova. Recently, a
first candidate event has been reported,
associated with the sGRB 130603B~\cite{tanvir13}. Assuming an almost
equal mass NS-NS merger as source, \citet{Hotokezaka13} show that an
EoS giving $R_{1.35} \lsim 13.5$ km is preferred in order to match the
inferred relatively large values of ejecta masses and velocities. This
result, however, depends strongly on the initial mass ratio. Although
expected to occur much less frequently, a compact binary merger, NS-NS
or NS-BH, with an initial mass ratio substantially different from
unity, naturally produces high ejecta masses and
velocities~\cite{Oechslin_06}, without giving any particular
constraint on the EoS. 
It might be possible to obtain reliable information about the underlying EoS from
the final nucleosynthesis outcome with more observations and more detailed simulations.

\subsection{Core-collapse Supernov{\ae}}
\subsubsection{Dynamics}
\label{sec:ccdynamics}
  The dynamics of CCSNe results  from a complex interplay
  between hydrodynamics, neutrino transport, weak interactions and the
  EoS. The general expectation, called {\it Mazurek's law}, is that
  due to the strong feedback, a small modification of one of the
  ingredients does not qualitatively change the
  dynamics~\cite{lattimer00}. However, the quantitative differences
  induced by different EoS can be large enough to govern,
  e.g., the presence or absence of an explosion~\cite{janka_12,Suwa_12}.

  Since in the early phase electron pressure dominates and later on
  the collapse proceeds homologously, the dynamics of the infall
  epoch has only a mild direct dependence on the baryonic part of the
  EoS. It is sensitive to the electron fraction $Y_e$ and the
  entropy. Changes of $Y_e$ result from electron captures (EC) on nuclei 
  and free protons and  therefore depend on
  the composition, in particular the
  abundances and the mass and charge of the appearing nuclei. 
  The mass of the inner core at bounce, $M_{\mathit{ic}}$, is
  roughly proportional to $\langle Y_{L^{(e)}}^2\rangle$, the mean fraction of
  trapped leptons squared \cite{BurrowsLattimerYahil}
  which is fixed and given by $Y_e$ at the moment where neutrino trapping sets in.

\begin{figure}
\includegraphics[width=0.8\columnwidth]{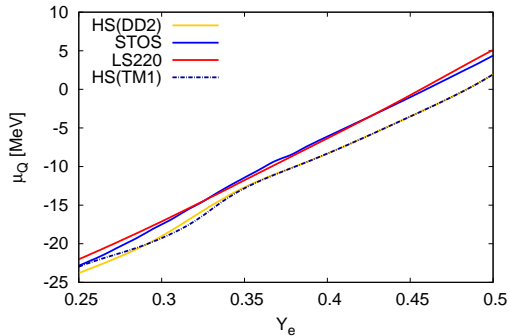}
\caption{A comparison of the values of charge chemical potential $\mu_Q$ for different EoSs and typical thermodynamic conditions in the infall epoch. The proton fraction has been varied at constant entropy per baryon $s = 1$ and constant baryon number density, $n_B = 10^{-3}\ \mathrm{fm}^{-3}$. 
\label{fig:muhat}}
\end{figure}

  Many studies, including
  those employing a statistical model for the EoS, use the single
  nucleus approximation within the simplified EC rates from
  \cite{Bruenn1985} in order to determine the evolution of $Y_e$. In
  this simple model, the reaction $Q$-value, determining the phase
  space available for the capture reaction, is approximated by the
  difference in proton and neutron chemical potentials,
\begin{equation}
\hat\mu \equiv -\mu_Q =  \mu_n - \mu_p~,
\end{equation} 
which strongly depends on the EoS. Roughly, the larger $\mu_Q$ the
  larger the electron capture rate. This quantity is illustrated for
  different EoSs in Fig. \ref{fig:muhat}. An entropy per baryon of $s
  = 1$ has been chosen, corresponding to a typical value before shock
  formation and a baryon number density of $n_B = 10^{-3}\ \mathrm{fm}^{-3}$. 
  It is evident that not the
  saturation properties of cold matter within the EoS are relevant to
  determine $\mu_Q$, but the treatment of inhomogeneous matter,
  i.e., nuclei. For instance, the difference between the model HS(TM1)
  and STOS, using the same interaction and having the same saturation
  properties, is much larger than between HS(DD2) and HS(TM1) which have
  very different nuclear properties but share the same treatment of
  nuclei. This has already been pointed out in \citet{lattimer00}
  where it was shown within a simple analytical model that $\mu_Q$ was
  much more sensitive to surface energies than to the bulk symmetry
  energy. 

  The EC rates also depend on entropy. 
  Already at progenitor level small differences
  in entropy arise between the EoS employed for
  core-collapse and the progenitor model, in general based on a
  nuclear reaction network. In addition, in STOS, the entropy
  contribution of the thermal motion of heavy nuclei is missing, see
  Sec.~\ref{sec:stos}, reducing the entropy and thus underestimating
  the deleptonization.

 Another effect is that within the \cite{Bruenn1985} description, the EC rate
  decreases strongly with the neutron number $N$.  Since in
  general the distribution of nuclear abundances is large in
  statistical models, the average heavy nucleus can be very different
  from the single heavy nucleus in the SNA, see Sec.~\ref{sec_nse_sna}
  and Fig.\ \ref{fig:nse_dist_new}. In HS(TM1), for instance,
  $N$ is in general smaller than in STOS, where the
  average mass of heavy nuclei is overestimated (related to the
  Thomas-Fermi approximation and the used value for the gradient
  energy coefficient, see Sec.~\ref{sec:stos}), leading to a higher
  EC rate in the domain where EC on nuclei is
  dominant~\cite{Hempel_11a}.

A remark of caution is in order here. It has been demonstrated
clearly that it is not sufficient to use one average heavy nucleus 
in order to determine the global EC rate on nuclei,
because of shell effects on individual rates which are not
correctly reproduced within this approximation, see,
e.g., \citet{langanke_03,juodagalvis_10} 
Better calculations for individual rates,
attenuating in particular the strong suppression of EC on neutron
rich nuclei present in the  \cite{Bruenn1985} prescription, change the infall
evolution and the bounce properties, too. The individual EC
rates can change by an order of magnitude from one nucleus to
another which has an important effect on the final $Y_L$, see,
e.g., \citet{Hix03,furusawa13b,Sullivan15,Raduta15}. 

  \citet{Hempel_11a} performed a detailed analysis of the
  different stages during the infall epoch comparing the LS180, STOS,
  and HS(TM1) EoS confirming within a simulation the different effects
  discussed above. At the time of bounce, only small differences can
  be observed between the different EoSs. A low central $Y_e$ 
  correlated with a low core mass,
  see, e.g., \citet{sumiyoshi2005,janka_12,Suwa_12,Steiner_12}. 

  A smaller core mass at bounce leads to a weaker shock forming
  closer to the center. Naively, this would lead to a situation less
  favorable for an explosion. In addition, more mass overlays the core in
  this case rendering an explosion still more difficult. However,
  different effects compete. For instance, a stronger deleptonization
  leads to a higher neutrino luminosity which in turn 
  heats the shock more strongly. Therefore no clear statement is possible and most
  studies in spherical symmetry conclude on a minor effect of the EoS
  on the overall
  dynamics \cite{sumiyoshi2005,Hempel_11a,Suwa_12,janka_12,Steiner_12,Fischer_13,Togashi:2014aga}. 
  In particular, it is difficult to relate single nuclear matter
  parameters of the EoS to CCSN dynamics, unless one compares EoS that show very
  pronounced differences. The fact that the existing general purpose EoS often differ in 
  several properties, also because of correlations among different parameters,
  see Sec.~\ref{sec:constraints}, makes systematic investigations difficult.
  In addition, as illustrated above, the treatment of inhomogeneous matter, 
  properties of nuclei and thermal effects are found to be equally important
  as the very neutron rich and dense part of the EoS \cite{Fischer_13}. 
  This might be   different for BH formation, where the PNS maximum mass is 
  decisive, cf.\ Sec.~\ref{sec_bh_formation}.

\begin{figure}
\includegraphics[width=0.8\columnwidth]{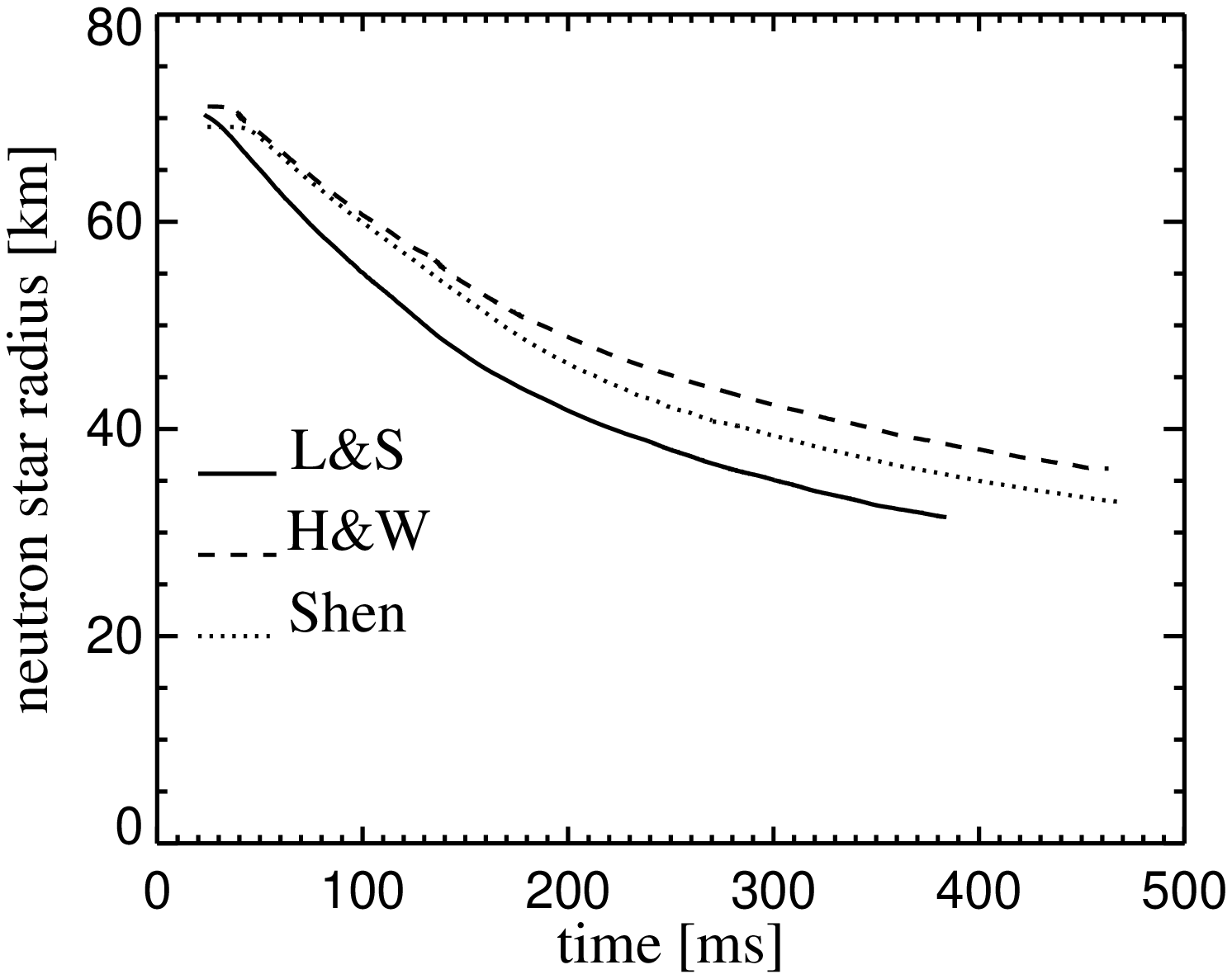}\hfill
\includegraphics[width=0.8\columnwidth]{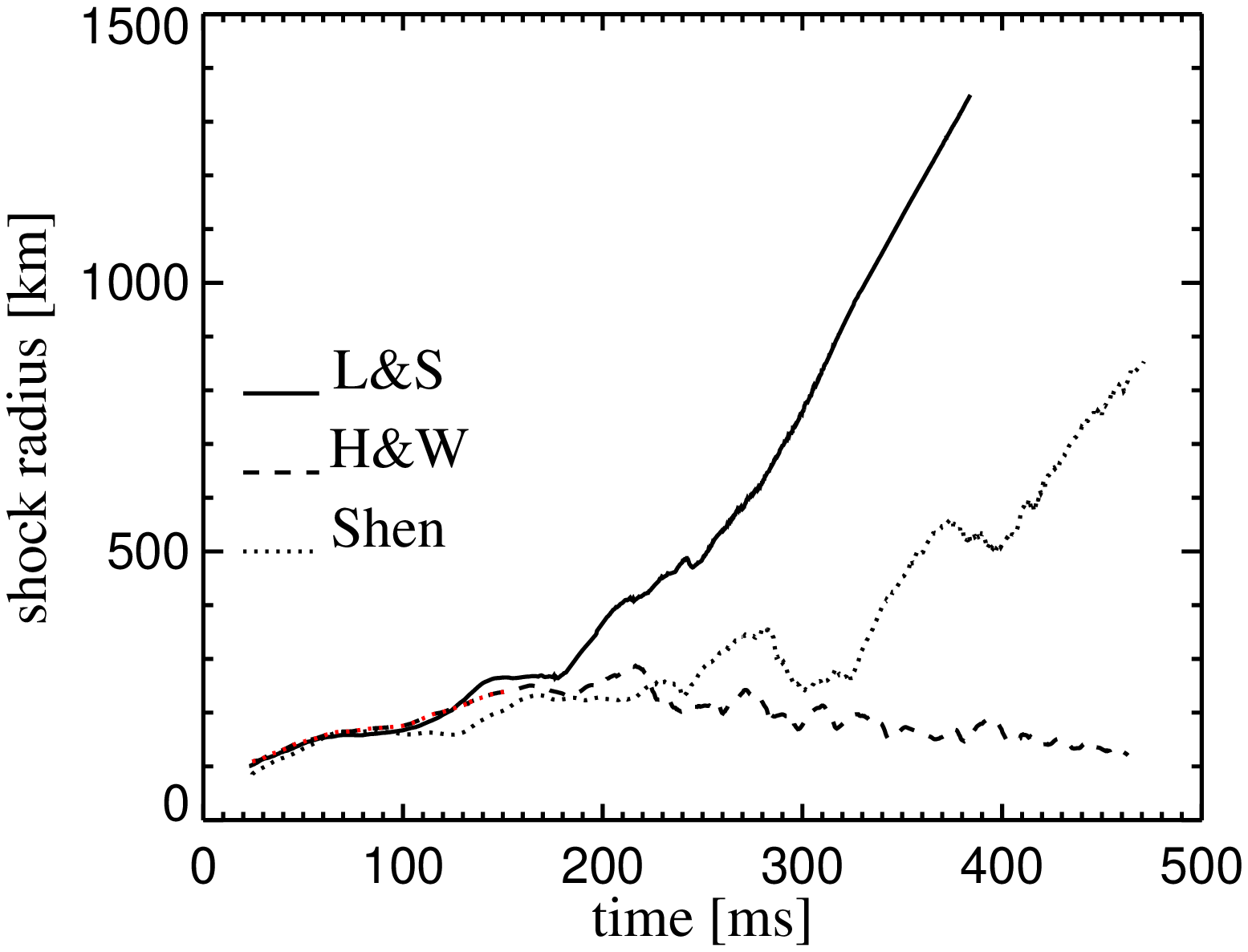}
\caption{Figures taken from \citet{janka_12arxiv}. The top panel shows
the evolution of the PNS's radius for a two-dimensional CCSN
simulation employing the LS180, the H\&W, and the STOS EoSs. The bottom panel
shows the evolution of the shock radius.
\label{fig:janka_12_eos}}
\end{figure}

   A more compact and more rapidly contracting PNS resulting
    from a ``softer'' EoS~\footnote{Please note that the terms ``soft''
    and ``stiff'' for the EoS are not necessarily related to the
    incompressibility of cold symmetric nuclear matter or any other
    nuclear matter parameter due to thermal effects. It means here
    simply that the overall PNS is more compressible for a ``soft''
    EoS.} generally seems to be favorable for explosions in multi-dimensional 
    simulations.  In particular, neutrinos are
    emitted with higher fluxes and higher energies, see, e.g., \citet{marek09}.
  This not
  only enhances neutrino cooling in 2D, but favors the formation of
  more violent hydrodynamical instabilities and stronger
  convection~\cite{Suwa_12,janka_12}. This is illustrated in
  Fig.~\ref{fig:janka_12_eos}, where the evolution of the PNS radius
  (upper panel) and the shock radius (lower panel) is shown for three
  different EoSs within a 2D simulation by the Garching group. It is
  evident that in this case the EoS decides upon the explosion. 

 There might be an imprint of the EoS on the neutrino and GW signal. 
  For instance, the faster
  deleptonization during the infall epoch leads to an enhancement in
  the neutrino luminosity at early times. The higher neutrino fluxes
  and higher energies of the emitted neutrinos from a more compact PNS
  should lead to differences in the neutrino spectra, too. From a
  galactic supernova, these differences should indeed be observable
  with present detectors, see
  ~\citet{sumiyoshi2005,marek09,Suwa_12}. A more compact PNS leads to
  higher frequencies and larger amplitudes for the emitted GW,
  too~\cite{marek09,scheidegger_10}. However, the EoS is not the only varying parameter, for example the
  unknown progenitor structure or the treatment of neutrino transport
  can induce modifications in the evolution of the CCSN, such that it
  seems very difficult to identify unambiguously one particular effect.

  Recent 1D studies of CCSNe with non-nucleonic degrees of freedom  focus on BH formation, see
  Sec.~\ref{sec_bh_formation}, where sufficiently high temperatures and densities
  are reached so that these degrees of freedom are expected to have a notable effect on the dynamics.
 An exceptional case for regular supernova explosions, i.e., without BH formation, 
 could be the onset of non-nucleonic degrees of freedom via a strong first order phase 
 transition occurring close to saturation density. As pointed out by \citet{gentile93} and 
 confirmed by \citet{sagert_09} with detailed Boltzmann neutrino transport, in spherically 
 symmetric simulations a second shock can be formed 
 as a direct consequence of a phase transition to quark matter.
This second shock was found to be strong enough to unbind the outer layers once it
merges with the standing accretion shock~\cite{sagert_09}. In this way, a CCSN
explosion is triggered due to the phase
transition to quark matter. Furthermore, the passage of the second
shock leads to a second neutrino burst that is dominated by electron
anti-neutrinos, measurable with present-day detectors
\citep{dasgupta10}. 

However, the NS maximum masses of the EoSs applied by
\citet{sagert_09} (STOSQ162s, STOSQ165s) are well below 2~M$_{\odot}$,
and thus ruled out by NS observations. In the subsequent works
exploring this scenario, see, e.g.,
\citet{sagert10,fischer_11,fischer12,nakazato13,Fischer_13},
explosions could not be obtained if the maximum mass of the employed
EoS is sufficiently high.  It is clear that the required stiffening in
the quark phase to reach 2~M$_{\odot}$ typically does not allow for a
strong first-order phase transition that seems to be necessary in this
scenario to trigger explosions.  This can be related to the so-called
masquerade problem~\cite{Alford:2004pf}, known for the mass-radius
relation of NSs where quark matter could behave very similar to
hadronic matter. On the other hand, it has not been shown yet that
a SN explosion induced by a phase transition is ruled out by the
latest pulsar mass measurements.

Overall, there are still many uncertainties and open questions about 
the CCSN explosion mechanism, such as the dependency on the progenitor, 
effects of magnetic fields and rotation, numerical convergence, the 
strength and scale of intrinsic multi-dimensional hydrodynamic effects 
(turbulence, convection, SASI, etc.), or an accurate treatment of 
neutrino interactions and their transport, see, e.g., 
\cite{mezzacappa05,Janka_06,kotake06,ott09,janka_12,burrows13}.
The EoS is one of them. Its role is not yet fully understood, 
partly due to the high complexity of the system and the interplay of all the
aforementioned aspects.

\subsubsection{Proto-neutron stars, neutrino-driven winds \& nucleosynthesis}
A neutrino-driven wind (NDW) is the emission of a low
density, high entropy baryonic gas from the surface of a newly born
PNS in a CCSN. It is driven by energy deposition of neutrinos emitted
from deeper layers and sets in after the launch of the SN
explosion. It remains active in the first seconds up to minutes. The
NDW is of great importance for the nucleosynthesis of heavy elements,
as it has been considered as one of the most promising sites for the
r-process, see for example the review by \citet{arcones_13}. However,
previous sophisticated long-term simulations of CCSNe
\citep{fischer_10,huedepohl_10} have shown that the matter emitted in
the NDW is generally proton-rich, allowing only for the so-called
$\nu$p-process \cite{froehlich06a,froehlich06b,arcones07,roberts_10,arcones_13}, 
which is not able to produce the most heavy nuclei.

\citet{MartinezPinedo:2012rb} and \citet{roberts12a,Roberts:2012um}
realized that these long-term simulations of the PNS deleptonization
phase and the NDW neglected the effect of nuclear interactions in the
charged-current (CC) interaction rates of neutrinos
with unbound nucleons. As outlined in Sec.~\ref{sec:pheno}, the
  nucleon single particle energies within mean field models can be
  written as the sum of a kinetic part which has the form of a free
  gas, depending on an effective mass, and an interaction potential
  $V_i$,
\begin{equation}
E_i = E^{\mathit{kin}}(m_i^*) + V_i  ~.
\end{equation}
The difference of neutron and proton energies, which determines the
energy available for (anti)neutrinos from CC reactions, depends on
$\Delta V = V_p - V_n$. In asymmetric matter within the hot PNS,
$\Delta V$ can be as large as 50 MeV. 
It does not affect neutral current reactions, unless $\Delta V$
carries an additional energy dependence and one has an inelastic reaction.

The simulations of \citet{MartinezPinedo:2012rb}
and \citet{roberts12a,Roberts:2012um} found that including 
$V_i$ correctly modifies the evolution of neutrino spectra and the deleptonization of
the PNS. $\Delta V$ induces an increase of the anti-neutrino
energies and a decrease of the neutrino energies. This leads
to slightly neutron-rich conditions in the NDW. Because $\Delta
V$ is related to the potential part of the symmetry energy \citep{Hempel:2015yma},
(anti)neutrino spectra and the conditions in the NDW are sensitive to the 
isospin dependence of the EoS.

Although the correct treatment of mean-field effects in the CC reactions 
favors less proton rich conditions in the NDW than previously determined, 
the values of $0.46 < Y_e < 0.5$\footnote{Note that the $Y_e$-values are
different compared with the values reported by \citet{Roberts:2012um},
due to a previous computational error which was later corrected
(L.\ Roberts, presentation at the MICRA workshop in Trento, 2013).}
  obtained in the simulations of \citet{Roberts:2012um,Martinez-Pinedo:2013jna} 
  with different EoSs are still not low enough for a robust r-process nucleosynthesis.
Similar  conclusions have been obtained within the QCD phase transition
  scenario, see Sec.~\ref{sec:ccdynamics}, where again a slightly
  proton rich NDW is produced, leading to a weak
  r-process~\cite{Nishimura_11}. However, there are still many open
questions, e.g., the employed approximations in the neutrino transport
and reaction rates, a possible progenitor dependence or the role of
light nuclei in the envelope of PNSs.

For instance, \citet{Arcones08} found that the
envelope composition is dominated by nucleons, deuterons, tritons and
$\alpha$-particles, in agreement with other works, see, e.g.,
\citet{Sumiyoshi08a,Hempel_11a,Fischer_13}. \citet{Arcones08} showed 
that light nuclei 
other than the $\alpha$-particle (mainly deuterons and tritons)
lead to a small reduction of the
average energy of the emitted electron neutrinos. 

The long-term deleptonization
and cooling of the PNS, see, e.g., 
\citet{Prakash:1996xs} and \citet{pons_01}, 
contains further interesting aspects related to the EoS. 
Within the delayed BH formation scenario, the loss of thermal energy 
and in particular the deleptonization destabilizes the PNS, cf.\ Sec.~\ref{sec_bh_formation}. 
For an exploding CCSN, convection in the subsequently contracting and cooling PNS 
is very sensitive to the EoS, as in the early postbounce phase. 
Since convection depends strongly on 
the variation of pressure with lepton fraction $Y_{L^{(e)}}$
at constant $n_B$, the EoS dependence is mainly characterized 
by the symmetry energy~\cite{Roberts:2011yw}.  The authors showed that the symmetry energy
leaves an imprint in the neutrino count rates of present-day neutrino
detectors for a galactic CCSN. 

In the simulation of \citet{Suwa:2013mva} the long-term evolution of the PNS was 
followed in a self-consistent manner starting from its birth in the CCSN up to 
$\sim$70~s. It is the first time that such a simulation entered the regime 
of conditions where the formation of the crust is expected.

\subsubsection{Black hole formation}
\label{sec_bh_formation}
In a stellar core-collapse event a NS is formed if the exploding star
successfully unbinds the ejected material after bounce. In a so-called
failed CCSN, the outcome  may
equally be a stellar mass black hole (BH) if the expanding
shock is not able to break through the infalling material and
accretion pushes the PNS over its mass-limit on the timescale of
seconds.  Alternatively, there can be a delayed BH formation
process, where either the cooling PNS becomes unstable or the fall
back of ejecta causes the collapse to a BH in the minutes
following the bounce. Numerical studies of BH formation in core
collapse have a long history, see, e.g., \citet{O'Connor:2010tk} and
references therein.

All scenarios have in common that the formation of an apparent
  horizon is accompanied by a significant drop in neutrino luminosity
  since most of the neutrino emitting material is swallowed up by the BH. 
  The GW signal could be interesting, too, in this
  context, being sensitive to oscillations in the hot
  PNS~\cite{Cerda-Duran:2013swa} and thus to the EoS.  
  In \citet{pons_01,Nakazato10b} it has been demonstrated that for
  a galactic event, the time between bounce and BH formation,
  $t_{\mathrm{BH}}$, is possibly observable from the neutrino signal
  in the Super-Kamiokande detector. However, the evolution of the core
  collapse and $t_{\mathrm{BH}}$ crucially 
  depends also on the progenitor structure. \citet{oconnor13} showed 
  that it might be possible to get information about the compactness of 
  the core of the progenitor star from the neutrino spectra and luminosities, 
  which would allow to disentangle the effects of the progenitor and of the EoS
  to some extent.
  Rotation can
  strongly change not only the time until BH formation, but also the
  neutrino signal itself~\cite{Sekiguchi:2010ja}. In particular,
  neutrino emission continues on a reduced level well after BH
  formation from the newly formed accretion disk, rendering the
  interpretation of the neutrino signal less obvious. 

  The EoS, as well, strongly influences the time until BH
  formation, since it determines the maximum mass supported by the hot
  PNS. There are two different physical mechanisms leading to the final 
  gravitational instability: either the collapse is accretion-induced or 
  due to deleptonization and/or cooling.

  In the latter case, it is not necessarily the reduced thermal
  pressure which destabilizes the cooling PNS, but deleptonization. In
  the hot PNS, due to the presence of trapped neutrinos, matter is
  very lepton rich and $Y_e$ can be as high as 0.4 (see,
  e.g., \citet{Prakash:1996xs,Pons99}). This leads to a suppression
  of additional degrees of freedom containing
  strangeness~\cite{prakash_95}, such as hyperons, a kaon condensed
  phase and/or a delayed phase transition to quark matter, see also
  Sec.~\ref{sec:general_purpose}. Consequently, meta-stable PNSs could
  exist, whose maximum mass is above that of cold,
  $\beta$-equilibrated NS \citep{Prakash:1996xs}. During the deleptonization of the hot PNS,
  the fraction of hyperons or quarks increases, eventually inducing a
  loss of stability and a collapse to a BH, see, e.g.,
  \citet{Keil95,Baumgarte96,pons_01}. 

  For an accretion induced collapse in a failed CCSN,
  $t_{\mathrm{BH}}$  is too short for considerable
  deleptonization. Here, the PNS cannot support the additionally
  accreted mass. The sensitivity of $t_{\mathrm{BH}}$ to the EoS has
  been demonstrated in many studies, see,
  e.g., \citet{sumiyoshi_07, fischer_09, O'Connor:2010tk,
    Suwa_12}.
  However, there is no straight forward relation between any
  property of a given EoS, for instance the maximum mass of a cold
  $\beta$-equilibrated NS, and the PNS mass at the onset of BH
  collapse. The reason is that BH formation is a dynamical process,
  and the temperature, density and $Y_e$ distribution in the hot PNS
  depends on many factors and is in particular very different from
  that of a cold and $\beta$-equilibrated NS. In \citet{Hempel_11a,
    Steiner_12} an interesting ansatz has been proposed: employing an
  extensive set of nuclear EoSs in simulations with a 40 M$_\odot$
  solar metallicity~\cite{woosley95} progenitor, it was shown that for
  the given setup, $t_{\mathrm{BH}}$ can be correlated with the
  maximum mass of a $\beta$-equilibrated isentropic PNS at $s = 4$.
  However, the evolution of a CCSN does not only depend on the EoS, 
  but on many other factors (as discussed above) which makes it 
  difficult to unambiguously relate $t_{\mathrm{BH}}$ to the EoS.

  Failed CCSNe have larger accretion rates
  than their exploding counterparts, such that higher temperatures and
  densities are reached within the PNS. As discussed in
  Sec.~\ref{sec:general_purpose}, this could lead to an sustained
  production of additional degrees of freedom such as quarks or
  hyperons. Subsequently, the EoS is softened, supporting less mass
  and reducing $t_{\mathrm{BH}}$ compared with a purely nuclear EoS,
  see,  e.g., \citet{ishizuka_08,sumiyoshi_09,Nakazato10a,nakazato_12,Peres_13,char2015}. Since
  these non-nucleonic degrees of freedom appear only deep inside the
  PNS, apart from $t_{\mathrm{BH}}$ no considerable difference in the
  neutrino signal is to be expected with respect to a purely nuclear
  EoS, except if the appearance of additional particles is accompanied by
  a phase transition~\cite{Nakazato10a,Peres_13}. 

Thus, although it is a promising field, there is still work needed
before we can conclude from the neutrino signal on the EoS. We
emphasize again that such difficulties are rather typical for
observables of CCSNe; cf.\ Sec.~\ref{sec:ccdynamics}.

\section{Summary and Conclusions}

Describing properties of matter in compact stars, their formation and
merger processes is a very challenging task.  The wide range of
densities, temperatures and charge fractions to be covered includes
extreme values, out of reach in terrestrial experiments.
Therefore one has to rely on theoretical modeling.  However, dense
hadronic and quark matter is difficult to describe since the many-body
problem with strongly interacting particles has to be solved.  In this
review, we have discussed theoretical and phenomenological approaches to
address these difficulties.

In addition, we reviewed constraints on the EoS that have been
obtained from experiments, astrophysical observations and {\em ab
  initio} calculations.  Let us mention here some particularly
important constraints.  First, the recent observation of two NSs with
precisely and reliably determined masses of about $2$~M$_\odot$ has
triggered intensive discussions on the composition of matter in the
central part of NSs. These results put strong constraints
on the high-density and low-temperature part of the EoS.  Secondly,
considerable progress has been made in recent years concerning
theoretical {\em ab initio} calculations of pure neutron matter 
up to
roughly saturation density, thus constraining the neutron rich part of
the EoS in this density regime. Thirdly, laboratory experiments are beginning 
to converge to a common prediction for the symmetry energy and its slope
around the saturation density.

There exist plenty of EoSs for cold NSs.  To a lesser extent this
still holds for EoSs for homogeneous hot matter in PNSs.  In this
review, the emphasis has been put on EoSs that cover the entire range of
thermodynamic variables, which is relevant for simulations of CCSNe and compact binary
mergers. They are much more rare, although in recent years much effort has
been devoted to the development of new models, focussing on two
aspects. First, the treatment of cluster formation and inhomogeneous
matter at low densities and temperatures has been considerably
improved.  It was realized that light nuclei, which were ignored
previously, can be important.  Subsequent CCSN simulations have shown
that differences in the cluster description 
induce differences in the dynamical evolution which are as
important as those arising from different nuclear interaction models.
Secondly, improved interactions and additional particles have been
considered for the high density and temperature part, such as
hyperons, mesons, and quarks. These non-nucleonic degrees of freedom
influence in particular black hole formation, and NS-NS and NS-BH
mergers.

Despite all efforts, there is much room for improvement. 
The cluster treatment is often based on a purely phenomenological 
description with several approximations and simplifications, 
see Sec.~\ref{sec:clusteredmatter}. 
The interaction models employed cannot be considered 
definite. For instance, no presently existing model is consistent
with all available constraints. 
However, it is clear that some of the
constraints have to be regarded with care. Not all of them have the
same reliability as the $2$~M$_{\odot}$ NS mass measurements,
see the discussion in Sec.~\ref{sec:constraints}.

The quality of constraints is expected to improve in the future. For
example, the current efforts to determine NS radii with an
unprecedented 5\% precision by projects such as ATHENA+, NICER, LOFT
and others promise rich information regarding the inner NS structure.
GW astronomy has the potential to give new and
completely independent insights about compact stars and their
underlying EoS. New laboratory
experiments and experimental facilities such as
RIKEN, FRIB, FAIR or NICA will provide new constraints for high density matter.
We emphasize that all available
``general purpose'' EoSs are based on phenomenological approaches due
to the computational and conceptual complexity of more microscopic
methods.  In the future, the increase in computational power is likely
to allow the latter to provide EoSs suitable for astrophysical simulations, too.

To conclude this review, let us mention, without claiming to be
exhaustive, some important questions, which have to be addressed to
develop a more ``realistic'' EoS.  Due to the large range of variables
covered in simulations, very different domains are encountered.
\begin{itemize}
\item{
Can we obtain a reliable description of all basic baryonic few-body interactions?}
\item{How and under which conditions do non-nucleonic
degrees of freedom appear?}
\item{When does nuclear matter deconfine?}
\item{Can we develop a QCD-based framework that covers the relevant range  
of variables?}
\item{How to better treat spatially inhomogeneous matter and cluster formation?}
\item{How to describe phase transitions consistently?}
\end{itemize}

Answers to any of these problems will result in better models for the
astrophysical EoS and will help to understand various fundamental
phenomena such as the composition and dynamics of NSs, the explosion
mechanism of CCSNe, the threshold to BH formation, the nature of
gamma-ray bursts, or the origin of heavy elements and the related
galactical chemical evolution.  In turn, these astrophysical insights
are potentially relevant for the analysis of (ultra-) relativistic
HICs and possibly for the search of QCD phase transitions.

\appendix

\section{Resources}

\subsection{EoS Data Bases}
\label{app:databases}
Here we present a list of publicly available EoS data bases.
A number of authors provide their EoSs online,
among them many of those presented in  Sec.~\ref{sec:general_purpose}:

\begin{itemize}

\item \citet{ls_web}

{\small
\href{http://www.astro.sunysb.edu/dswesty/lseos.html}{http://www.astro.sunysb.edu/dswesty/lseos.html}}

The original LS EoSs is available in form of a computer 
program. Several authors have generated tabulated versions from it, that can be found in
the other databases. The various tables
can differ in the range of thermodynamic variables covered and the details of 
the underlying calculation.

{\small
\href{http://www.astro.sunysb.edu/lattimer/EOS/main.html}{http://www.astro.sunysb.edu/lattimer/EOS/main.html}}

Four unpublished general purpose EoSs are
(as far as we know) tabulated  at this website, 
where, according to their names, 
three of them are based on the Skyrme interactions SKI', SKa, and SKM*, and 
the fourth has an incompressibility of 370~MeV.

\item \citet{shentable}

{\small
\href{http://user.numazu-ct.ac.jp/~sumi/eos/index.html}{http://user.numazu-ct.ac.jp/\midtilde sumi/eos/index.html}}

(hosts data of \citet{ishizuka_08} as well)

\item \citet{gshen_web}

{\small
\href{http://cecelia.physics.indiana.edu/gang_shen_eos/}{http://cecelia.physics.indiana.edu/gang\_shen\_eos/}}

\item \citet{mhhomepage}

{\small
\href{http://phys-merger.physik.unibas.ch/~hempel/eos.html}{http://phys-merger.physik.unibas.ch/\midtilde hempel/eos.html}}

\end{itemize}

The resources listed 
above provide data in the full space of temperature, density and asymmetry.
Websites that offer EoS tables for NS matter are

\begin{itemize}

\item \citet{ioffe_web}

{\small
\href{http://www.ioffe.ru/astro/NSG/nseoslist.html}{http://www.ioffe.ru/astro/NSG/nseoslist.html}}

\end{itemize}

Different groups maintain online EoS collections with additional features.
Here, we provide the links 
with a small synopsis as found at the corresponding web sites:

\begin{itemize}

\item \textsc{STELLARCOLLAPSE.ORG}

\cite{oo_web}

{\small
\href{http://www.stellarcollapse.org}{http://www.stellarcollapse.org}}

{\em  ``..., a website aimed at providing resources supporting research in stellar collapse, core-collapse supernovae, neutron stars, and gamma-ray bursts.''}

 \textsc{STELLARCOLLAPSE.ORG} provides not only tabulated EoS data but
hosts valuable resources, information and freely
available open source code for stellar collapse and related phenomena.
The available open source codes are listed in section \ref{sec:opensource}.

\item \textsc{CompOSE} 

\cite{Typel:2013rza,compose}

{\small
\href{http://compose.obspm.fr}{http://compose.obspm.fr}}

{\em  ``The online service CompOSE provides data tables for different state of the art equations of state (EoS) ready for further usage in astrophysical applications, nuclear physics and beyond.''}

\textsc{CompOSE} has been developed by the authors of this review with
support of the ESF-funded network CompStar, and the successive COST
Action MP1304, NewCompStar.  The community behind these research
networks consists of EoS developers and users.  A driving idea behind
\textsc{CompOSE} is not only to host a wide range of EoS data but
to provide it in a flexible, multiple purpose and reusable data format
applicable for, e.g., NS and CCSN EoS.  The \textsc{CompOSE} team
encourages and supports the authors of free simulation software in the
development of interfaces for the \textsc{CompOSE} data format.  The
\textsc{Lorene} library~\cite{Lorene} is \textsc{CompOSE} compatible.

\item \textsc{EOSDB}

\cite{ishizuka14}

{\small
\href{http://aspht1.ph.noda.tus.ac.jp/eos/index.html}{http://aspht1.ph.noda.tus.ac.jp/eos/index.html}}

{\em  ``Our aim is to summarize and share the current information on nuclear EoS which is available today from theoretical/experimental/observational studies of nuclei and dense matter.''}

\textsc{EOSDB} offers the possibility to search, compare and graphically represent
nuclear matter EoS and related quantities online.

\end{itemize}
 
\subsection{Open Source Simulation Software}
\label{sec:opensource}
EoSs are a crucial input to many astrophysical simulations.
We present a list of publicly available codes treating problems related to this review.

\begin{itemize}

\item \textsc{Lorene}

\cite{Lorene}

{\small
\href{http://www.lorene.obspm.fr}{http://www.lorene.obspm.fr}}

{\em  ``... a set of C++ classes to solve various problems arising in numerical relativity, and more generally in computational astrophysics.''}

\item \textsc{RNS}

\cite{RNS}

{\small
\href{http://www.gravity.phys.uwm.edu/rns/}{http://www.gravity.phys.uwm.edu/rns/}}

{\em  ``...  constructs models of rapidly rotating, relativistic, compact stars using tabulated equations of state which are supplied by the user.''}

\item \textsc{STELLARCOLLAPSE.ORG}

\cite{oo_web}

{\small
\href{http://www.stellarcollapse.org}{http://www.stellarcollapse.org}}

offers several codes:
\begin{itemize}
\item \textsc{GR1D}\\
{\em  Spherically-Symmetric Code for Stellar Collapse to Neutron Stars and Black Holes. }
\item \textsc{GR1Dv2}\\
{\em  Spherically-Symmetric Neutrino Radiation-Hydrodynamics.}
\item \textsc{SNEC}\\
{\em  The SuperNova Explosion Code}
\item \textsc{CCSNMultivar}\\
{\em  Multivariate Regression Analysis of Gravitational Waves from Rotating Core Collapse }
\end{itemize}

\item \textsc{AGILE-IDSA}

\cite{agile_idsa_web}

{\small
\href{https://physik.unibas.ch/~liebend/download}{https://physik.unibas.ch/\midtilde liebend/download}}

{\em  ``...  provides tools to run a rudimentary and approximate model of a core-collapse supernova with neutrino transport in spherical symmetry through the phases of stellar collapse, bounce, and early postbounce evolution.''}

\end{itemize}

\begin{acknowledgments}
We thank S.~Furusawa, C.~Ishizuka, K.~Nakazato, and K.~Sumiyoshi for
providing the data of their EoS, M.~Fortin for providing
Fig.~\ref{fig:nsradii}, and A.~Perego for providing
Fig.~\ref{fig:phasediagram_nsmerger}. We are grateful to A.~Raduta,
N.~Buyukcizmeci, I.~Mishustin, and A.~Botvina for providing us with
the data for Fig.~\ref{fig:nse_dist_new}.  We appreciate the
thoughtful and encouraging comments of T.~Fischer, T. Gaitanos,
C.~Miller, G.~R{\"o}pke, J.~Schaffner-Bielich and H.~Wolter when
confronted with early versions of this review.  This work has been
partially funded by the SN2NS project ANR-10-BLAN-0503 and by
NewCompStar, COST Action MP1304.  M.H.\ is supported by the Swiss
National Science Foundation.  T.K.\ is grateful for support by the
Polish National Science Center (NCN) under grant number
UMO-2013/09/B/ST2/01560.  S.T.\ is supported by the Helmholtz
Association (HGF) through the Nuclear Astrophysics Virtual Institute
(VH-VI-417) and by the DFG through grant No.~SFB1245.

\end{acknowledgments}

\bibliography{rmp_eosrev}

\end{document}